# Symmetries of Analytic Paths

Christian Fleischhack[*]

*Institut für Mathematik*
*Universität Paderborn*
*Warburger Straße 100*
*33098 Paderborn*
*Germany*

March 21, 2015

## Abstract

The symmetries of paths in a manifold $M$ are classified with respect to a given pointwise proper action of a Lie group $G$ on $M$. Here, paths are embeddings of a compact interval into $M$. There are at least two types of symmetries: Firstly, paths that are parts of an integral curve of a fundamental vector field on $M$ (continuous symmetry). Secondly, paths that can be decomposed into finitely many pieces, each of which is the translate of some free segment, where possibly the translate is cut at the two ends of the paths (discrete symmetry). Here, a free segment is a path $e$ whose $G$-translates either equal $e$ or intersect it in at most finitely many points. Note that all the statements above are understood up to the parametrization of the paths. We will show, for the category of analytic manifolds, that each path is of exactly one of either types.

For the proof, we use that the overlap of a path $\gamma$ with one of its translates is encoded uniquely in a mapping between subsets of $\mathrm{dom}\,\gamma$. Running over all translates, these mappings form the so-called reparametrization set to $\gamma$. It will turn out that, up to conjugation with a diffeomorphism, any such set is given by the action of a Lie subgroup of $O(2)$ on $S^1$, restricted in domain and range to some compact interval on $S^1$. Now, the infinite subgroups correspond to the continuous symmetry above, finite ones to the discrete symmetry.

## 1 Introduction

### 1.1 Background

Connections in fibre bundles is one of the cores of geometry. It generalizes the classical differential calculus and provides us the optimal framework to study fundamental geometric entities like curvature or geodesics. Its applications range from algebraic topology with its famous index theorems over the de Rham decomposition classifying irreducible Riemannian manifolds, to theoretical physics where connections correspond to fundamental fields. Sometimes, one is interested in particular connections like the Levi-Civita covariant derivative in Riemannian or Lerentzian manifolds; sometimes, one is studying the space $\mathcal{A}$ of all connections, possibly after factorizing by the group of vertical isomorphisms. Also intermediate stages, like flat or symmetric connections have been under investigation. On the other hands, in some situations it may be advisable not to consider just smooth connections, but more general ones. For instance, in the compact case, the group of vertical isomorphisms turns into a Hilbert-Lie transformation group on the space of connections only after admitting also Sobolev structures [8]. Even more, in loop quantization one is naturally lead to the inclusion of distributional connections [2].

---

[*]e-mail: `fleischh@math.upb.de`

Intimitely related to the concept of connections is that of parallel transport. Recall that connections in principal fibre bundles can be defined in two equivalent ways: either geometrically as horizontal distributions compatible with the action of the structure group $S$; or analytically as $\mathfrak{s}$-valued 1-forms of type Ad that reduce the fundamental $S$-vector fields to their respective generators. The presence of a connection allows now to lift paths in the base manifold to horizontal paths in the bundle $P$. This way one assigns to the path a morphism between the fibres over the endpoints of the path. As the concatenation of paths corresponds to the concatenation of fibre morphisms, one can regard parallel transports as (in some sense smooth) homomorphisms from the groupoid $\mathcal{P}$ of all paths (modulo parametrization) in the base manifold $M$ to the fibre morphisms $\operatorname{Mor} P$ in $P$. Remarkably, the parallel transports uniquely determine the connection.

This relation between connections and parallel transports forms the basis of many results on the structures mentioned in the beginning. For instance, the Ambrose-Singer theorem allows to identify [3] the moduli space of flat connections modulo vertical automorphisms over a compact 2-surface with the space of all homomorphisms from the homotopy group $\pi_1(M)$ to $S$ modulo conjugation. Or, for the rise of distributional connections, consider the abelian $C^*$-algebra of all matrix functions of parallel transports along all the paths as a subalgebra of the bounded functions on $\mathcal{A}$, provided $S$ is compact. Its spectrum turns out to be isomorphic to $\overline{\mathcal{A}} := \operatorname{Hom}(\mathcal{P}, \operatorname{Mor} P)$ serving as the home of distributional connections. Note that $\mathcal{A}$ is a proper, but dense subset of $\overline{\mathcal{A}}$; moreover, $\overline{\mathcal{A}}$ contains *all* homomorphisms, not just smooth ones. The theory of such distributional connections also triggered the present paper, more precisely, the quest for symmetric elements in $\overline{\mathcal{A}}$.

Symmetries of smooth connections have already been investigated for quite a long time. Here, symmetry means invariance w.r.t. the action of some Lie group $G$ of automorphisms on the underlying principal fibre bundle. Wang [11] showed that for fibre-transitive actions the space of symmetric connections is one-to-one with a subset of $L(\mathfrak{g}, \mathfrak{s})$, determined by two algebraic conditions that involve $G$-stabilizers. Later, Harnad et al. [7] claimed a generalization to actions with conjugate stabilizers. Unfortunately, the restrictions are still too tight to incorporate, e.g., spherical symmetry on full $\mathbb{R}^n$. To cure this, Hanusch [5] recently developed a general algebraic characterization of symmetric connections. His results have turned out to be crucial for the loop-inspired quantization of cosmological models [1], i.e., of symmetry-reduced gravity.

Symmetries of generalized connections have been studied only very recently. For this, one has to lift the action of $G$ on $\mathcal{A}$ to an action of $\overline{\mathcal{A}}$. The main idea for this is to use the natural action of $G$ on parallel transports and the induced action on the algebra of their matrix functions [6, 4]. As the $C^*$-completion of this algebra is $G$-invariant, the action of $G$ is getting lifted. At the end, one sees that $h \in \overline{\mathcal{A}}$ is invariant under the action of $G$ iff

$$h(\varphi_g \circ \gamma) = \varphi_g \circ h(\gamma) \quad \text{for all paths } \gamma \text{ in } M \text{ and all } g \in G.$$

Here, $\varphi_g$ denotes both the induced action of $g \in G$ on $M$ (left-hand side) and on $\operatorname{Mor} P$ (right-hand side). It is now obvious that the main obstacle for identifying homomorphisms are relations between $\varphi_g \circ \gamma$ and $\gamma$. We know from smooth connections that their parallel transports along two paths are independent as soon as these paths intersect in at least finitely many points. Thus, it seems natural to focus in the symmetric context primarily on situations where the intersection of $\varphi_g \circ \gamma$ and $\gamma$ has at least an accumulation point. In the $C^\infty$ category of paths this does not give much further information, but if the paths are analytic[1] this already implies that they share a full segment. Indeed, now the independence of the non-symmetric analytic level above transfers

---

[1]Indeed, in loop quantization one usually assumes to be given (piecewise) analytic paths, mainly because then the set of finite graphs is directed, i.e., any two finite graphs are always subgraphs of a third one. Obviously, this is not given in the $C^\infty$ category. The directedness is crucial for the core of loop quantum gravity, namely measure theory on $\overline{\mathcal{A}}$. In fact, $\overline{\mathcal{A}}$ is a projective limit of powers of $S$, indexed by the *directed* set of finite graphs, making measure theory feasible.

to the symmetric distributional level.

To illustrate this, let us be more specific and let us assume that $\gamma = \gamma_1 \gamma_2$ for subpaths $\gamma_1$ and $\gamma_2$, where $\gamma_2$ is the $s$-translate of $\gamma_1$. Then, by homomorphy and invariance, $h(\gamma) = h(\gamma_1) \circ h(\gamma_2) = h(\gamma_1) \circ \varphi_g \circ h(\gamma_1)$, whence $h(\gamma)$ and $h(\gamma_1)$ are not independent. On the other hand, if there is no $g \neq \mathbf{1}$, such that $\gamma$ and $\varphi_g \circ \gamma$ have more than just isolated intersections, then there are no relations between the values of $h$ on $\gamma$ and on some of its proper subpaths [6]. This brings us to the goal of the present paper:

*Classify analytic paths according to their symmetries.*

Here, symmetries refer not only to the invariance of a given path under a certain subgroup of the symmetry group, but more general to a non-trivial overlapping of the paths with some of its translates.

## 1.2 Ideas and Assumptions

Throughout the whole paper, let us be given an analytic manifold $M$, a Lie group $G$ and an analytic left action $\varphi$ of $G$ on $M$. The main theorem of our present paper is going to classify analytic paths $\gamma$ in $M$ w.r.t. their symmetry. For this, we have to investigate how $\gamma$ can intersect its own translates $g\gamma$ by $G$. Of course, isolated intersections are hard to classify. Therefore, let us look for occurrences of at least accumulation points of intersections. By analyticity, this already means that both paths overlap with nontrivial translates along full subpaths. Are there prototypes for paths in $M$ that share full segments with their translates? Indeed, there are. As Hanusch [6] pointed out, there are at least two types, namely

1. integral curves of fundamental vector fields induced by the Lie algebra of $G$ on $M$;
2. concatenations of translates of a free segment.

Here, a free segment is some path $\gamma$ whose translates either coincide with $\gamma$ or have no (nontrivial) overlap with $\gamma$. In particular, partial overlaps are not allowed for them. As we see, paths from the first class exhibit a continuous, infinite symmetry, whereas those from the second one only show a discrete, finite symmetry. The former ones will be called **Lie paths**, the latter ones[2] **brick paths**. The big challenge has been to decide whether any path is in one of the two classes or not.

In general, the answer is negative. For instance, consider $\mathbb{R}$ acting on $\mathbb{R}^2$ by $\varphi_\lambda(x) \longmapsto e^\lambda x$. Obviously, the Lie paths are given by all the radial straight lines not passing the origin, i.e., the straight paths connecting $ax$ and $bx$ for $0 \neq [a,b]$ and $x \neq 0$. The straight path $\delta$ connecting $x$ and $-x$, however, is not a Lie path. Neither it is a brick path. In fact, let $\gamma$ be a free segment for $\delta$. As $\gamma$ is free iff any of its translates is so, we can assume that $\gamma$ is a subpath of $\delta$. If $\gamma$ did not touch 0, none of its translates does, whence they cannot cover $\delta$. Hence, $\gamma$ touches 0. But, now, any dilation by $e^\lambda$ overlaps parts of $\gamma$, whence $\gamma$ cannot be free. The deeper reason for that behaviour is that the dilations do not act pointwise properly.

Thus, let us assume pointwise properness from now on. Within this framework, Hanusch [6] was able to show that indeed there are no other paths possible than those above, provided the action is transitive or proper, and, moreover, if it admits only stabilizers that are normal subgroups of $G$. For the particular situation [6] of Euclidean subgroups acting on $\mathbb{R}^3$, we have, e.g.,

| | acting group | transitive | proper | normal stabilizers | condition met |
|---:|---:|:---:|:---:|:---:|:---:|
| translations | $\mathbb{R}^3$ | yes | yes | yes | yes |
| rotations around 0 | $SO(3)$ | no | yes | no | no |
| all motions | $\mathbb{R}^3 \rtimes SO(3)$ | yes | yes | no | no |

---

[2]Hanusch called them "free paths". We refrain from that notion as free paths need not be unrelated to any of its nontrivial translates. Nevertheless, we keep Hanusch's notion "free segment".

Thus, in the latter cases it had remained open whether there are further types of paths or not.

In our present paper, we will show that indeed in all cases above there are Lie paths and brick paths only. Even more, we will be able to generalize Hanusch's classification results to any pointwise proper action. The key idea is to focus on the set $\mathbf{P}_\gamma$ of so-called essential reparametrizations $\varrho_g$ for $\gamma$. These are certain analytic diffeomorphisms between subsets of $\mathrm{dom}\,\gamma \subseteq \mathbb{R}$ that fulfill $\varphi_g \circ \gamma = \gamma \circ \varrho_g$ on the domain of $\varrho_g$. Obviously, they characterize the overlapping behaviour of $\varphi_g \circ \gamma$ and $\gamma$. For our classification result, however, we will forget these particular definitions. Rather, we will transfer the properties of the action to properties of sets $\mathbf{P}$ consisting of appropriate homeomorphisms between subsets of intervals. Then we will investigate the second, abstract problem. It will turn out that Lie paths correspond to infinite $\mathbf{P}_\gamma$, whereas brick paths correspond to finite ones. The most important statement is that a subset of any infinite $\mathbf{P}_\gamma$ is a local semi-subgroup isomorphic to $[0, T)$ for some $T > 0$.

Beyond that classification result, we will also show for brick paths how to choose a subpath as free segment $\gamma$ and which group elements are necessary to cover the full path by translates. We only have to distinguish between two cases, namely whether $\mathbf{P}_\gamma$ containes only locally increasing functions or not.

1. In the former case, we will show that there is a $g \in G$, such that the translates of $\gamma$ by $\mathbf{1}, g, g^2, \ldots, g^k$ cover $\gamma$ by concatenation completely (possibly we have to cut at the boundary). Note that this behaviour somewhat resembles that of Lie curves, however, with $\mathbb{R}$ replaced by $\mathbb{Z}$. Indeed, if $g = \mathrm{e}^A$ for some $A \in \mathfrak{g}$, then we have $g = \mathrm{e}^{tA}$, $t \in \mathbb{Z}$, as acting elements to cover $\gamma$; for Lie curves, we admit $t \in \mathbb{R}$.

2. In the latter case, we will need a further element $h \in G$. It is a sort of reflection for which $h^2$ fixes $\gamma$. Now, the translates of $\gamma$ by $h, \mathbf{1}, gh, g, g^2h, g^2, \ldots, g^k$ cover $\gamma$ (here, possibly skipping parts of the last two paths).

Before we are now going into the theory, let us first fix the very basic notions, give explicit examples for the two types of curves and finally present the main theorem and strategy.

## 1.3 Curves and Paths

**Definition 1.1**
- A map between subsets of real-analytic manifolds is called **analytic** iff it is the restriction of a real-analytic function between open subsets.
- A map between subsets of real-analytic manifolds is called **diffeomorphism** iff it is the bijective restriction of a diffeomorphism between open subsets.

In the following all manifolds are assumed real-analytic. Moreover, in the following definition, an interval is always understood to contain at least two points.

**Definition 1.2**
- A **curve** in $M$ is any analytic mapping from an interval to $M$.
- A **path** in $M$ is an analytic embedding of a compact interval into $M$.
- An **open path** in $M$ is an analytic embedding of an open interval into $M$.

**Definition 1.3** Two curves $\gamma_1$ and $\gamma_2$ **coincide up to the parametrization** iff $\gamma_1 = \gamma_2 \circ \varrho$ for some analytic diffeomorphism $\varrho : \mathrm{dom}\,\varrho_1 \longrightarrow \mathrm{dom}\,\varrho_2$ with $\dot\varrho > 0$.

**Definition 1.4** A **subpath** of a path $\gamma$ is any path $\gamma \circ \varrho$, where $\varrho$ is an analytic diffeomorphism whose range is a nontrivial interval in the domain of $\gamma$.

Note that, in our definition, the orientation of the subpath does not matter. In particular, this means that among the subpaths of $\gamma$ is the inverted[3] path $\gamma^-$ which is the path that runs along $\gamma$ in the opposite direction, i.e., $\gamma^-(t) = \gamma(b + a - t)$ for $\mathrm{dom}\,\gamma^- := \mathrm{dom}\,\gamma = [a,b]$.

Now, let us investigate the two main types of analytic paths. For this, let $\varphi$ be an analytic left action of a Lie group $G$ on the manifold $M$.

---
[3]In order to avoid confusion with the preimage mapping $\gamma^{-1}$, we will use the notation $\gamma^-$.

## 1.4 Lie Paths

First, let us consider the integral curves of the fundamental vector field that is induced by $\varphi$.

**Definition 1.5** A **Lie curve** is a mapping $\delta : t \longmapsto e^{tA}x$ for some $A \in \mathfrak{g}$ and $x \in M$.

Here, $\mathfrak{g}$ is the Lie algebra of $G$. Of course, any Lie curve $\delta$ is defined on full $\mathbb{R}$. With $L_s : \mathbb{R} \longrightarrow \mathbb{R}$ denoting the right-shift by $s$, we have $\delta(L_s(t)) \equiv \delta(s+t) = e^{sA}e^{tA}x = e^{sA}\delta(t)$, hence

$$\delta \circ L_s \;=\; \varphi_{e^{sA}} \circ \delta \qquad \text{for all } s \in \mathbb{R}.$$

**Lemma 1.1** Any Lie curve is either constant or injective or periodic.

Moreover, it is an immersion unless it is constant. We give the standard proof for completeness.

**Proof** Assume that the Lie curve $\delta : t \longmapsto e^{tA}x$ is not an immersion. Then there is some $t$ with $\dot\delta(t) = 0$. Since $\dot\delta(s) = (\varphi_{e^{sA}})'(\dot\delta(0))$, we see that then $\dot\delta \equiv 0$, whence $\delta$ is constant. Thus, let us now assume that $\delta$ is an immersion, hence locally an embedding. Collect in $I$ all $s \in \mathbb{R}$ with $e^{sA}x = x$. Obviously, $I$ is a closed subgroup of $\mathbb{R}$. However, as $\delta$ is locally injective, there is an $\varepsilon > 0$, such that $0$ is the only element in $I \cap (-\varepsilon, \varepsilon)$. Hence, $I = \{0\}$, i.e., $\delta$ is injective, or $I = c\mathbb{Z}$ for some $0 < c \in \mathbb{R}$, i.e., $\delta$ is periodic. **qed**

Typically, we are interested in paths with compact domains only, whence we define

**Definition 1.6**
- A **partial Lie curve** is an injective restriction of an open Lie path to a compact interval.
- A **Lie path** is a partial Lie curve up to parametrization.

## 1.5 Brick Paths

Another type – indeed, the other, as we are going to prove – is represented by the sine-curve example mentioned in the introduction. It is merely the concatenation of $G$-translates of a single path that has no symmetry at all.

**Definition 1.7** A curve $\delta$ is called **free segment** [6] iff for each $g \in G$, the curves $\delta$ and $\varphi_g \circ \delta$ either coincide up to parametrization or do not share a subpath.

Obviously, any $G$-translate of a free segment is a free segment again.

**Definition 1.8**
- A curve $\gamma$ is called **brick curve** iff there is some free segment $\delta$, such that $\gamma$ is the concatenation of curves that are – at least after some possibly orientation-reversing reparametrizations – $G$-translates of $\delta$.
- The respective free segment $\delta$ is said to **generate** $\gamma$.
- A **brick path** is an injective restriction of a brick curve to a compact interval.

For example, re-consider the sine curve. More precisely, let $M = \mathbb{R}^2$ and let $\delta(t) := (t, \sin t)$ describe the graph of the sine function. Let us now study the following subgroups $G$ of the Euclidean group $\mathbb{R}^2 \rtimes O(2)$ of $\mathbb{R}^2$ and their natural actions:

1. $G$ is the trivial subgroup. {**1**}

    Then $\delta$ itself is a free segment. Indeed, any curve is a free segment if the acting group is trivial. Note that the same behaviour occurs if we let $G$ equal the translation group generated by a shift that is not parallel to the $x_1$-axis.

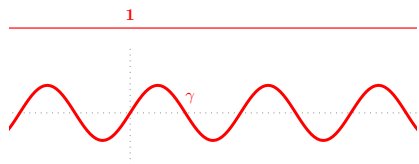

2. $G$ is the subgroup generated by the rotation by $\pi$ around the origin.  $\{\mathbf{1}, -\mathbf{1}\}$

   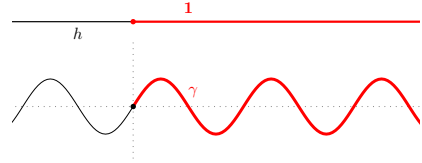

   Then the restriction $\gamma$ of $\delta$ to $\mathbb{R}_+$ is a free segment. Setting $h := (\mathbf{0}, -\mathbf{1})$, then $\delta$ is the concatenation
   $$(\varphi_h \circ \gamma)^- \cdot \gamma.$$
   Recall that $\delta^-$ describes the path or curve that is inverse to $\delta$.

3. $G$ is the group generated by the translations on $\mathbb{R}^2$.  $\mathbb{R}^2$

   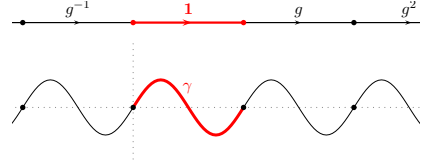

   Then the restriction of $\delta$ to any interval having at most length $2\pi$ is a free segment. Any other restrictions are brick curves w.r.t. the free segment $\gamma := \delta|_{[0,2\pi]}$. In particular, with $g := ((2\pi, 0), \mathbf{1})$, the sine curve $\delta$ is the concatenation
   $$\cdots (\varphi_{g^{-1}} \circ \gamma) \cdot (\varphi_{g^0} \circ \gamma) \cdot (\varphi_{g^1} \circ \gamma) \cdot (\varphi_{g^2} \circ \gamma) \cdots$$

4. $G$ is the connected component of the unit in the Euclidean group.  $\mathbb{R}^2 \rtimes SO(2)$

   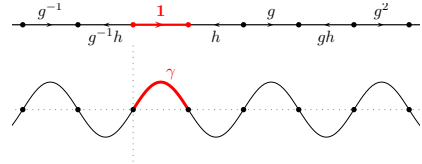

   Then the free segments are just the curves that do not contain a zero in their interior. More precisely, these curves are restrictions of $\delta$ to any interval containing elements of $\pi\mathbb{Z}$ at most at their boundary like $[3\pi, 4\pi]$. Any other restrictions are brick curves w.r.t. the free segment $\gamma := \delta|_{[0,\pi]}$. In fact, define
   $$\begin{aligned} g &:= \big((2\pi, 0),\ \mathbf{1}\big) & \ldots & \quad \text{shift by } 2\pi \text{ parallel to the } x_1\text{-axis} \\ h &:= \big((2\pi, 0), -\mathbf{1}\big) & \ldots & \quad \text{rotation by } \pi \text{ around } (\pi, 0) \end{aligned}$$

   Then the sine curve $\delta$ is the concatenation
   $$\cdots (\varphi_{g^{-1}} \circ \gamma) \cdot (\varphi_{g^{-1}h} \circ \gamma)^- \cdot (\varphi_{g^0} \circ \gamma) \cdot (\varphi_{g^0 h} \circ \gamma)^- \cdot (\varphi_{g^1} \circ \gamma) \cdot (\varphi_{g^1 h} \circ \gamma)^- \cdot (\varphi_{g^2} \circ \gamma) \cdots$$

5. $G$ is the full Euclidean group.  $\mathbb{R}^2 \rtimes O(2)$

   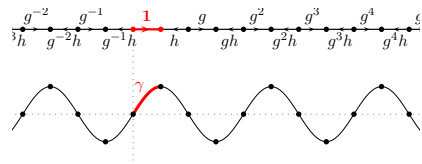

   There the situation is similar to the previous case. Simply define $\gamma$ to be the restriction of $\delta$ to $[0, \frac{\pi}{2}]$, and let
   $$\begin{aligned} g &:= \big((\pi, 0), \operatorname{diag}(1, -1)\big) \\ h &:= \big((\pi, 0), \operatorname{diag}(-1, 1)\big) \end{aligned}$$
   $g$ shifts by $\pi$ parallel to $x_1$-axis and then reflects thereon; $h$ is the reflection at vertical line $x_1 = \frac{\pi}{2}$. The decomposition is then exactly as above.

Note that in none of the cases above, we had to reparametrize the shifted paths. This, of course, is not a general behaviour and even non-geometric. In fact, for $G = \mathbb{R}^2$, we could consider a reparametrized sine curve $\delta \circ \varrho$ for some analytic diffeomorphism $\varrho : \mathbb{R} \longrightarrow \mathbb{R}$, where $\dot\varrho$ is not $2\pi$-periodic. Then it is clear that the corresponding free segment is not mapped to a subpath of $\delta \circ \varrho$ by the $2\pi$-shift, unless we apply a further reparametrization.

Another interesting observation concerns the products used in the concatenation. There, for the free segment $\gamma$ having compact domain, only translates by

$$\ldots, g^{-1}, g^0, g^1, g^2, \ldots \qquad \text{or} \qquad \ldots, g^{-1}, g^{-1}h, g^0, g^0h, g^1, g^1h, g^2, g^2h, \ldots$$

had been necessary; moreover, the paths that are moved using $h$ get inverted. This behaviour will even turn out to be true in general. Indeed, such an $h$ as above occurs if $G$ contains elements that flip $\delta$, i.e., elements that fix some point in $\operatorname{im}\delta$, but not full $\delta$. Otherwise, we can always find some $g$ that generates full $\delta$ out of a single free segment. The only situation not covered above is the occurrence of periods. Indeed, it may happen that $g^n$ acts trivially. Take, e.g., the unit circle line $\delta(t) := (\cos t, \sin t)$ in $\mathbb{R}^2$ and let $G = \{\mathbf{1}, -\mathbf{1}\}$. Then, of course, the restriction of $\delta$ to $[0, \pi]$ is a free segment, and $\delta$ the concatenation

$$\cdots \gamma \cdot (\varphi_g \circ \gamma) \cdot \gamma \cdot (\varphi_g \circ \gamma) \cdot \gamma \cdots.$$

Finally, note that later we will consider paths instead of general curves, i.e., mappings on a compact interval. Then, of course, any decomposition of such a path will only need a subsequence of the group elements above. Moreover, we have to take care of the two ends of the path; there we have to cut some segments appropriately.

## 1.6 Strategy

The major goal of our present paper is to prove

**Theorem 1.2** Let $\varphi$ be a pointwise proper analytic left action of a Lie group $G$ on an analytic manifold $M$. Then any analytic path in $M$ is either a Lie path or a brick path.

Additionally, we will explicitly determine the free segments that constitute brick curves.

**Remark** Note that the notion of being a Lie path or a brick path is *not* an *intrinsic* notion of $M$, but refers to the action $\varphi$ of $G$ on $M$. In fact, any Lie path in $M$ w.r.t. a nontrivial action of $G$ is a brick path w.r.t. to the trivial action.

To get a clue of the main strategy, recall the examples above. If $\gamma$ overlaps its $g$-translate nontrivially, then $\varphi_g \circ \delta$ will equal $\delta \circ \varrho_g$ for some diffeomorphism $\varrho_g$. Indeed, although $g = (2\pi, 0)$ moves the full-sine-function graph onto itself, $\varphi_g \circ \delta$ and $\delta$ do not coincide, but rather $\varphi_g \circ \delta = \delta \circ \varrho_g$ with $\varrho_g(t) := t + 2\pi$. For paths, we have to additionally take care of the fact that the corresponding $\varrho_g$ possibly has a domain that is smaller than that of $\delta$. Our strategy is now to classify the possible sets of reparametrization functions $\varrho_g$ for paths. It will turn out that a path is a Lie path iff there are infinitely many nontrivial $\varrho_g$. We will, however, do this classification in a rather abstract manner, almost without any reference to their origin in reparametrizations between overlaps of translates.

Overall, the present paper is divided into three parts.

1. Section 2 reduces the intersection behaviour of a path with its translates to properties of the respective reparametrization functions. They form a so-called motion set.

2. Sections 4 to 11 classify all motion sets abstractly.

3. Section 3 uses the classification of motion sets to prove the main theorem.

Observe that the sequence of sections as presented here, does not follow the logic. However, as we do not need the full classification of motion sets for proving Theorem 1.2, we will prepone the relevant statements from the later sections and postpone only their proofs.

# 2 Intersections and Reparametrizations

Let us start with the reduction of the symmetry problem to overlaps of paths. First, in Subsection 2.1, we will discuss this for any two paths, before we study the intersection of a path $\gamma$ with its translates in Subsection 2.2. Their overlaps will be encoded in so-called essential reparametrization functions, i.e., diffeomorphisms between subsets of the domain of $\gamma$. Running over all translates, we can collect these functions getting a so-called reparametrization set in Subsection 2.3. We close with the example of subgroups of $O(2)$ acting on $S^1$ in Subsection 2.4. This will turn out very important as we will see at the end of the paper that any reparametrization set is conjugate to precisely such an example. Even more, the proof of the main theorem will crucially rely on that classification result.

In the following, any path will have domain $[a, b]$, unless we explicitly specify another domain.

## 2.1 Intersection of Two Analytic Paths

**Definition 2.1**
- A **local reparametrization** between open paths $\delta_1$ and $\delta_2$ is an analytic diffeomorphism $\varrho$, whose domain is an open interval with $\delta_1 = \delta_2 \circ \varrho$ thereon[4].
- A local reparametrization is said to be **around** $A$ iff $A \subseteq \delta_1(\text{dom}\,\varrho)$.
- A local reparametrization is said to be **maximal** iff it does not have a proper extension that is a local reparametrization again.

We will shortly write local reparametrizations around $x$ instead of some around $\{x\}$.

**Lemma 2.1** Let $\varrho_\iota$ be local reparametrizations between $\delta_1$ and $\delta_2$.
If the union of their domains forms an interval, then
$$\varrho(t) \;:=\; \varrho_\iota(t) \qquad \text{for } t \in \text{dom}\,\varrho_\iota$$
defines a local reparametrization with $\text{dom}\,\varrho = \bigcup_\iota \text{dom}\,\varrho_\iota$.

**Proof**
- $\delta_2$ injective $\implies \varrho$ well defined.
  In fact, $\varrho_\iota$ and $\varrho_{\iota'}$ coincide on the intersection of their domains as $\delta_2 \circ \varrho_\iota = \delta_1 = \delta_2 \circ \varrho_{\iota'}$.
- $\delta_1$ injective $\implies \varrho$ injective.
  In fact, $\varrho_\iota(t) = \varrho(t) = \varrho(t') = \varrho_{\iota'}(t')$ implies $\delta_1(t) = [\delta_2 \circ \varrho_\iota](t) = [\delta_2 \circ \varrho_{\iota'}](t') = \delta_1(t')$.
- $\varrho_\iota$ analytic $\implies \varrho$ analytic.
  In fact, $\varrho = \varrho_\iota$ on $\text{dom}\,\varrho_\iota$. This domain is open; and all of them span the domain of $\varrho$.
- $\varrho_\iota$ analytic diffeomorphisms $\implies \varrho$ analytic diffeomorphism.
  In fact, $\varrho$ is injective and each $\varrho_\iota$ is non-singular. **qed**

**Corollary 2.2** Any local reparametrization can be extended to a unique maximal one.

Note that the maximal one is around $A$ if the former one is so as well.

**Proof** Let $\varrho : J_1 \longrightarrow J_2$ be a local reparametrization and $x \in \delta_1(J_1) = \delta_2(J_2)$. Moreover, let $\{\varrho_\iota\}$ be the family of all local reparametrizations around $x$. Use these $\varrho_\iota$ to define a local reparametrization $\overline{\varrho}$ on $\bigcup_\iota \text{dom}\,\varrho_\iota$, by Lemma 2.1. Obviously, $\overline{\varrho}$ is around $x$ again. It even extends $\varrho$ as $\varrho$ equals $\varrho_\iota$ for some $\iota$. Moreover, if $\varrho'$ extends $\overline{\varrho}$, then $\varrho'$ is around $x$, hence equal to some $\varrho_{\iota'}$, whence $\overline{\varrho}$ extends $\varrho'$ by construction. Hence, $\overline{\varrho}$ is maximal. Uniqueness follows immediately. **qed**

**Proposition 2.3** Let $\delta_1$ and $\delta_2$ be open paths, and let $S$ be the intersection of their images.
Then there is a unique maximal local reparametrization around any accumulation point within $S$.

---
[4]This includes $\text{dom}\,\varrho \subseteq \text{dom}\,\delta_1$ and $\text{im}\,\varrho \subseteq \text{dom}\,\delta_2$.

This proposition extends Lemma 5.3/1 in [6].

**Proof** Let $x$ be some accumulation point within $S$.
- As $x \in S$, there are $t_i \in \operatorname{dom} \delta_i$ with $\delta_1(t_1) = x = \delta_2(t_2)$.
- Shrinking the domains of $\delta_i$ to open $J_i$, if necessary, we may assume that boths paths live in a single chart $(U, \kappa)$ of $M$ with $\delta_1$ mapping to the $\mathbf{e}_1$-axis with $\kappa(x) = \mathbf{0}$.
- Let $\kappa_k : U \longrightarrow \mathbb{R}$ denote the $k$-th component of $\kappa$. As, by assumption, $\delta_2$ meets the $\mathbf{e}_1$-axis locally infinitely often, $\kappa_k \circ \delta_2$ vanishes at infinitely many points for each $k \geq 2$. By analyticity, it vanishes everywhere. This shows that $\delta_2$ maps to the $\mathbf{e}_1$-axis.
- Consequently, there are analytic embeddings $\varrho_i : J_i \longrightarrow \mathbb{R}$ with $\kappa \circ \delta_i = \varrho_i \cdot \mathbf{e}_1$. Shrinking them in range, we get analytic diffeomorphisms $\varrho_i : J_i \longrightarrow \varrho_i(J_i)$.
- Now, the analytic diffeomorphism $\varrho := \varrho_2^{-1} \circ \varrho_1 : J_1 \longrightarrow J_2$ fulfills
$$\kappa(\delta_1(t)) \;=\; \varrho_1(t)\mathbf{e}_1 \;=\; \varrho_2(\varrho(t))\mathbf{e}_1 \;=\; \kappa(\delta_2(\varrho(t)))$$
for $t \in J_1$, hence $\delta_1 = \delta_2 \circ \varrho$ on $J_1$.
- Finally, we extend $\varrho$ to the unique maximal local reparametrization around $x$ by Corollary 2.2. **qed**

**Corollary 2.4** Let $\delta_1$ and $\delta_2$ be open paths, and let $S$ be the intersection of their images. Moreover, let $A$ be a connected component of $S$ consisting of at least two points. Then there is a unique maximal local reparametrization $\varrho$ around $A$. Moreover, the domain $I$ of $\varrho$ is open and $A$ equals $\delta_1(I)$.

**Proof**
- As $\delta_1$ is an embedding, $\delta_1^{-1}(S) \subseteq \operatorname{dom} \delta_1$ is homeomorphic to $S$ and its connected components are intervals. Unless a connected component contains just a single point, it is a genuine interval, i.e., consists of accumulation points. Again using the embedding property of $\delta_1$, we see that $A = \delta_1(I)$ consists of accumulation points only.
- Now taking local reparametrizations around each $\delta_1(t) \in A$, we get a family $\{\varrho_t\}$ of local reparametrizations, whose domains form some interval containing $I$. By Lemma 2.1, they give rise to a local reparametrization $\varrho$ around $A$. If, however, the domain of $\varrho$ was larger than $I$, we would get a contradiction to the assumption that $I$ is a connected component in $\delta_1^{-1}(S)$. Therefore, the domain of $\varrho$ is exactly $I$. In particular, $\delta_1(I) = A$ and $\varrho$ is maximal. Therefore, finally, $I$ is open. **qed**

**Lemma 2.5** Let $\gamma_1$ and $\gamma_2$ be paths in $M$, and let $S$ be the intersection of their images. Moreover, let $K$ be a connected component of $\gamma_1^{-1}(S)$ that does not consist of a single point. Then $K$ is a compact interval whose boundary points are mapped by $\gamma_1$ to the vertex set $V := \{\gamma_1(a), \gamma_1(b), \gamma_2(a), \gamma_2(b)\}$.

**Proof**
- As each $\gamma_i$ is continuous with compact domain, $S$ is compact as well and so does $\gamma_1^{-1}(S)$. Its connected components are compact intervals.
- Let $\delta_i$ be the open paths given by $\gamma_i$ restricted to $(a, b)$. Denote the intersection of their images by $T$.
- Let now $K$ be a nonsingular connected component in $\gamma_1^{-1}(S)$ and assume $\gamma_1(r) \notin V$ for some $r \in K$. Then, in particular, $r \in \operatorname{dom} \delta_1$. As $\gamma_1(K) \setminus V \subseteq S \setminus V \subseteq T$ consists of accumulation points only, $\delta_1(r) = \gamma_1(r)$ is an accumulation point within $T$. Hence, there is some local reparametrization $\varrho$ around $\delta_1(r)$ between $\delta_1$ and $\delta_2$. In particular, its domain is contained in that of $\delta_1$. Therefore, $\gamma_1(\operatorname{dom} \varrho) = \delta_1(\operatorname{dom} \varrho) \subseteq T \subseteq S$, hence $\operatorname{dom} \varrho \subseteq K$ by connectivity. As $\operatorname{dom} \varrho$ is open, but $K$ is compact, $r$ cannot be a boundary point of $K$. **qed**

**Corollary 2.6** Let $\gamma_1$ and $\gamma_2$ be paths, and let $S$ be the intersection of their images. Then $\gamma_i^{-1}(S)$ is the union of at most finitely many singleta and at most two compact intervals.

This corollary coincides with Lemma 5.3/5 in [6].

**Proof** Decompose $\gamma_i^{-1}(S)$ into connected components, each being a compact interval.
- *Nontrivial Intervals*[5]
  If we would have $n \leq \infty$ nonsingular intervals, then their endpoints get mapped to $2n$ vertices. As there are at most four vertices at all, there can be at most two compact nonsingular intervals.
- *Trivial Intervals*
  Assume now that there are infinitely many singleta. As $\gamma_i$ has compact domain, these singleta have to cluster somewhere, say at $x$. As $S$ is compact, we have $x = \gamma_1(t) \in S$. Thus, we may assume that $x_n \to x$ for some $x_n = \gamma_1(t_n) \neq x$, each of them forming a singlet. Moreover, $t_n \to t$.
  Let us next extend $\gamma_i$ to open paths $\overline{\gamma}_i$. As, of course, $x_n$ and $x$ are contained in the images of both extended paths as well, there is a local reparametrization $\varrho$ around $x$. As its domain is open and connected, it contains $[t, t_n]$ for some $n$. Now, $[t, t_n]$ is even contained in the domain of $\gamma_1$ as its boundary points are. On the other hand, from $\overline{\gamma}_2(\varrho(t)) = \overline{\gamma}_1(t) = x = \gamma_2(t') = \overline{\gamma}_2(t')$ for some $t' \in \mathrm{dom}\,\gamma_2$, we get $\varrho(t) \in \mathrm{dom}\,\gamma_2$ by injectivity of $\overline{\gamma}$. The same applies to $t_n$ instead of $t$, whence $\varrho(t_n)$ and $\varrho(t)$ are in the domain of $\gamma_2$, as well as the compact interval connecting them.
  Altogether, $\gamma_1$ and $\gamma_2 \circ \varrho$, not only their extensions, coincide on the interval between $t_n$ and $t$. Therefore, $\{x_n\}$ is not a connected component of $S$. Contradiction. **qed**

Let us collect our findings in the following

**Proposition 2.7** Let $\gamma_1$ and $\gamma_2$ be paths in $M$, and let $S$ be the intersection of their images. Then there is a unique analytic diffeomorphism
$$\varrho \;:\; \gamma_1^{-1}(S) \longrightarrow \gamma_2^{-1}(S)$$
that fulfills
$$\gamma_1 \;=\; \gamma_2 \circ \varrho \qquad \text{on the domain of } \varrho \tag{1}$$
Moreover, $\mathrm{dom}\,\varrho$ and $\mathrm{im}\,\varrho$ is the union of
- at most finitely many points, and
- at most two nonsingular compact intervals,

whereas $a$ or $b$ is contained in $\{t, \varrho(t)\}$ for any endpoint $t$ of such an interval.

**Proof** We already know that $\gamma_1^{-1}(S)$ as well as $\gamma_2^{-1}(S)$ consist of at most finitely many points and at most two nonsingular compact intervals with the properties above. Define $\varrho := \gamma_2^{-1} \circ \gamma_1 : \gamma_1^{-1}(S) \longrightarrow \gamma_2^{-1}(S)$. By injectivity of $\gamma_2$, this map is well defined and bijective. We only have to show that it is analytic on each of the at most two intervals mentioned above. Denote such an interval by $K$ and extend $\gamma_i$ to open paths $\overline{\gamma}_i$. Obviously, $\gamma_1(K)$ lies in the intersection $T$ of the images of $\overline{\gamma}_1$ and $\overline{\gamma}_2$. As $K$ is nonsingular, there is a local reparametrization $\widetilde{\varrho}$ around $\gamma_1(K)$ between $\overline{\gamma}_1$ and $\overline{\gamma}_2$. Now, for any $t \in K$,
$$\overline{\gamma}_2(\varrho(t)) \;=\; \gamma_2(\varrho(t)) \;=\; \gamma_1(t) \;=\; \overline{\gamma}_1(t) \;=\; \overline{\gamma}_2(\widetilde{\varrho}(t)),$$
whence $\varrho$ coincides on $K$ with the analytic $\widetilde{\varrho}$. **qed**

The intersection of two paths in an isolated point is rather accidental and does not appear to give us structural information. That along a full nontrivial subpath is different. Therefore, we will concentrate on the "essential", i.e., nondiscrete parts of $\varrho$ in the following.

---

[5] Intervals are **nontrivial** iff they consist of at least two points.

**Definition 2.2** Let $\gamma_1$ and $\gamma_2$ be paths in $M$, and denote by $K$ the (possibly empty) union of the nonsingular compact intervals in $\gamma_1^{-1}(S)$ according to Proposition 2.7.
Then the **essential reparametrization** between paths $\gamma_1$ and $\gamma_2$ is an analytic diffeomorphism $\varrho : K \longrightarrow \varrho(K)$ with $\gamma_1 = \gamma_2 \circ \varrho$ on $K$.

Proposition 2.7 tells us that the intersection of paths along nontrivial subpaths is completely encoded in the respective essential reparametrization. Later, for paths related by a group action, we will classify the possible essential reparametrizations in a much more abstract way. Before, however, we will shortly review how a path can intersect its own translates.

## 2.2 Group Actions

In the previous subsection, we have seen how any two analytic paths may intersect. Now, we are going to study a more special situation. Namely, let $\varphi$ be some (analytic, left) action of a Lie group $G$ on $M$. As usual, we define $\varphi_g : M \longrightarrow M$ by $\varphi_g(x) := \varphi(g,x) \equiv gx$. Moreover, fix some path $\gamma$ in $M$. Now, we are interested how the translates $g\gamma \equiv \varphi_g \circ \gamma$ intersect $\gamma$. As we have seen above, the crucial entity to be investigated is the essential reparametrization between $g\gamma$ and $\gamma$.

**Definition 2.3** We denote by $\varrho_g$ the essential reparametrization between $\varphi_g \circ \gamma$ and $\gamma$, i.e.,

$$\varphi_g \circ \gamma \;=\; \gamma \circ \varrho_g \qquad \text{on } \operatorname{dom} \varrho_g.$$

The set of all nontrivial essential reparametrizations between $\gamma$ and its $G$-translates is denoted by $\mathbf{P}_\gamma$.

Note that $\varrho_g$ may be trivial, i.e., may have empty domain.

Naively, one might think that $g \longmapsto \varrho_g$ is a homomorphism as it is the case for $g \longmapsto \varphi_g$. However, this is not the case, in general. Indeed, we have to see that $\varphi_g \circ \gamma = \gamma \circ \varrho_g$ only holds on the domain of $\varrho_g$. Nevertheless, there are some properties that look similar to homomorphy as we are now going to derive.

**Proposition 2.8** We have for all $g, h \in G$

$$\begin{aligned}
\varrho_{\mathbf{1}} &= \mathbf{1} \\
\varrho_{g^{-1}} &= \varrho_g^{-1} \\
\varrho_{gh} &= \varrho_g \circ \varrho_h \qquad \text{on the non-discrete part of } \varrho_h^{-1}(\operatorname{dom} \varrho_g)
\end{aligned}$$

Note, e.g., that $\varrho_{g^{-1}} \circ \varrho_g$ does not equal the identity $\mathbf{1}$ on $[a,b]$, in general. Rather it is given by the identity on the image of $\varrho_g$.

**Lemma 2.9** The domains of $\varrho_g^{-1}$ and $\varrho_{g^{-1}}$ coincide.

**Proof** Denote the intersection of the images of $\gamma$ and $g^{-1}\gamma$ by $S$, and its non-discrete part by $S_0$. Thus, the domain of $\varrho_{g^{-1}}$ is just $[\varphi_{g^{-1}} \circ \gamma]^{-1}(S_0) = \gamma^{-1}(\varphi_g(S_0))$. On the other hand, the intersection of the images of $g\gamma$ and $\gamma = gg^{-1}\gamma$ is now given by $gS \equiv \varphi_g(S)$. As well, its non-discrete part is $\varphi_g(S_0)$. Now, the image of $\varrho_g$ is $\gamma^{-1}(\varphi_g(S_0))$. The proof follows as the image of $\varrho_g$ is the domain of $\varrho_g^{-1}$. **qed**

**Proof Proposition 2.8**
- The first equation is trivial.
- For the second observe that we have $\varphi_g \circ \gamma = \gamma \circ \varrho_g$ on $\operatorname{dom} \varrho_g$, hence $\gamma \circ \varrho_g^{-1} = \varphi_g^{-1} \circ \gamma$ on $\operatorname{dom} \varrho_g^{-1}$. On the other hand, $\varphi_{g^{-1}} \circ \gamma = \gamma \circ \varrho_{g^{-1}}$ on $\operatorname{dom} \varrho_{g^{-1}}$. By Lemma 2.9, both domains coincide, whence $\varrho_g^{-1} = \varrho_{g^{-1}}$ thereon, by injectivity of $\gamma$.

- For the third equation observe $\varphi_{gh} \circ \gamma \equiv \varphi_g \circ \varphi_h \circ \gamma = \varphi_g \circ \gamma \circ \varrho_h = \gamma \circ \varrho_g \circ \varrho_h$, whereas the second equality holds on $\mathrm{dom}\,\varrho_h$ and the third one on $L := \varrho_h^{-1}(\mathrm{im}\,\varrho_h \cap \mathrm{dom}\,\varrho_g) \subseteq \mathrm{dom}\,\varrho_h$. This means that the images of $\gamma$ and $\varphi_{gh} \circ \gamma$ intersect at least in $[\varphi_{gh} \circ \gamma](L)$. Consequently, $\varphi_{gh} \circ \gamma = \gamma \circ \varrho_{gh}$ at least on the non-discrete part of $L$. The proof follows from injectivity of $\gamma$. **qed**

The analyticity of the objects under investigation yields

**Lemma 2.10** If $\varrho_g$ is the identity on some nontrivial interval, then it is the identity on full $[a, b]$.

**Proof** Denote that interval by $I$. By definition $\varphi_g \circ \gamma = \gamma$ on $I$. As both sides are analytic, they have to coincide on $[a, b]$. This gives the proof. **qed**

**Corollary 2.11** If $\varrho_g$ and $\varrho_h$ coincide on a nontrivial interval, then they are equal.

**Proof** By assumption, $\varrho_h(\mathrm{dom}\,\varrho_g)$ contains a nontrivial interval. Now, $\varrho_{gh^{-1}}$ equals $\varrho_g \circ \varrho_{h^{-1}}$ thereon by Proposition 2.8; hence it is the identity thereon. Lemma 2.10 shows that $\varrho_{gh^{-1}}$ is the identity on full $[a, b]$. Again using Proposition 2.8, $\varrho_g$ equals $\varrho_{gh^{-1}} \circ \varrho_h \equiv \varrho_h$ on $\varrho_h^{-1}(\mathrm{dom}\,\varrho_{gh^{-1}}) \equiv \mathrm{dom}\,\varrho_h$, as the latter one is nondiscrete by assumption. Exchanging the rôles of $g$ and $h$, we get the proof. **qed**

As already mentioned in the introduction, we will restrict ourselves later mostly to the case of pointwise proper actions. Recall that an action $\varphi$ is called **pointwise proper** in $x$ iff any sequence $(g_i) \subseteq G$ has a converging subsequence provided $(g_i x)$ converges. Similarly, it is pointwise proper iff it is in each $x \in M$. The most important consequence for essential reparametrizations will come from

**Proposition 2.12** Let $\varphi$ be pointwise proper or isometric, $(g_i)$ a sequence in $G$ and $s, t \in [a, b]$. If now both $(\varrho_{g_i}(s))$ and $(\varrho_{g_i}(t))$ are well defined and converging, then
$$\lim \varrho_{g_i}(s) = \lim \varrho_{g_i}(t) \iff s = t\,.$$

**Proof** We only have to show the $\Longrightarrow$ direction.
- $\varphi$ is pointwise proper.
  By assumption, $\gamma(\varrho_{g_i}(s)) = g_i \gamma(s)$ is converging as well as $g_i \gamma(t)$. Consequently, there is a subsequence $(g_j)$ of $(g_i)$ converging to some $g$. It fulfills
  $$g\gamma(s) = \lim g_j \gamma(s) = \lim g_j \gamma(t) = g\gamma(t)\,.$$
  Acting with $g^{-1}$ on both sides, this implies $s = t$ by injectivity of $\gamma$.
- $\varphi$ is isometric.
  $$\begin{aligned} d(\gamma(s), \gamma(t)) &= d(g_i \gamma(s), g_i \gamma(t)) = d(\gamma(\varrho_{g_i}(s)), \gamma(\varrho_{g_i}(t))) \\ &\to d(\gamma(\lim \varrho_{g_i}(s)), \gamma(\lim \varrho_{g_i}(t))) = 0. \end{aligned}$$
  Again, the proof follows from injectivity of $\gamma$. **qed**

This property above is sufficient to prove that $\varrho_g$ is the identity as soon as its domain is full $[a, b]$ and $\varrho_g$ is increasing. We will show this in a more abstract context in Corollary 6.8. Now, we finally transfer the notion of being a free segment to the level of essential reparametrizations.

**Proposition 2.13** Let $\varphi$ be pointwise proper.
Then an analytic path $\gamma$ with domain $I$ is a free segment iff for all $\varrho \in \mathbf{P}_\gamma$
$$\varrho^{-1}(I) \cap I \text{ nontrivial} \iff \varrho = \mathbf{1}\,.$$

**Proof** This is an easy consequence of Lemma 2.14 below, as $\varrho_g = \mathbf{1}$ iff $\varphi_g \circ \gamma = \gamma$. **qed**

**Lemma 2.14**  Let $\varphi$ be pointwise proper, and let $\gamma$ be an analytic path in $M$ having domain $I$. Then we have for all $g \in G$

$$\varphi_g \circ \gamma \text{ and } \gamma \text{ share a common subpath} \iff \varrho_g^{-1}(I) \cap I \text{ is nontrivial.}$$

**Proof** $\implies$  By assumption, there is a nontrivial interval $J \subseteq I$ and two analytic diffeomorphisms $\varrho_i : J \longrightarrow J_i \subseteq I$ with $\varphi_g \circ \gamma \circ \varrho_1 = \gamma \circ \varrho_2$. As $J_1 \subseteq I$ and also $\varphi_g(\gamma(J_1)) = \gamma(J_2) \subseteq \gamma(I)$, we have $[\varphi_g \circ \gamma](J_1) \subseteq \varphi_g(\gamma(I)) \cap \gamma(I)$, hence $J_1 \subseteq \operatorname{dom} \varrho_g$. This implies $\gamma \circ \varrho_2 = \varphi_g \circ \gamma \circ \varrho_1 = \gamma \circ \varrho_g \circ \varrho_1$, hence $\varrho_g \circ \varrho_1 = \varrho_2$. Altogether, $\varrho_g(J_1) = J_2$, giving $J_1 \subseteq \varrho_g^{-1}(I) \cap I$.

$\impliedby$  Let $J_1 \subseteq \varrho_g^{-1}(I) \cap I$ and let $J_2 := \varrho_g(J_1)$. Then $\varphi_g \circ \gamma \circ \mathbf{1} = \gamma \circ \varrho_g$ on $J_1$, hence $\varphi_g \circ \gamma$ and $\gamma$ share a common subpath. **qed**

Similarly, one can see that a path is a concatenation of translates of free segments (possibly cut at the ends) iff the sets $\varrho_g^{-1}(I)$ with $g$ running over $G$ cover $I := \operatorname{dom} \gamma$.

## 2.3  Formalization

We have learned so far that the overlap behaviour is completely encoded in the properties of the essential reparametrization functions. In the previous subsection, moreover, we have derived several properties that these function have. In the following, we are going to forget now that the mappings are essential reparametrizations; instead we consider them as appropriate mappings having the properties derived above. This will be sufficient to get significant parts of the proof of our main theorem.

Recall from Proposition 2.7 that any essential reparametrization is a homeomorphism between subsets of $[a,b]$ with a particular behaviour on the boundary of connected components: any boundary point of the domain equals $a$ or $b$ or is mapped to $a$ or $b$. As described above, this restricts the number of domain components to be two or less. As these properties will turn out to be central for the following, let us summarize them in

**Definition 2.4**  A map $\varrho$ is called **standard** iff it is a homeomorphism between compact subsets of $[a,b]$ each consisting of one or two nontrivial compact subintervals of $[a,b]$, such that each boundary point[6] of these intervals equals $a$ or $b$ or is mapped by $\varrho$ to $a$ or $b$.

Indeed, by Proposition 2.7, any essential reparametrization is an analytic standard map (unless its domain is empty). Remarkably, there are just a very few topological types of standard maps. It is a simple exercise that the types in Table 1 on page 14 will comprise all possible ones, at least up to exchange of domain with image and up to flipping the interval $[a,b]$. Here, the upper line shows the domain, the lower one the image. Moreover, the dashed lines show the mapping of the boundary points. The notion of perfect maps will become relevant later (see Definition 5.1 and Subsection 5.1).

As already indicated in the table, any standard map can be given an orientation in a natural way:

**Lemma 2.15**  A standard map is either increasing on all connected components or decreasing on all connected components.

**Proof**  Of course, any homeomorphism between single intervals is monotonous, whence the only situation we have to control is $\varrho : K_1 \longrightarrow K_2$ with $K_1$ and $K_2$ consisting of two intervals. Assume $K_1 = [a,s] \sqcup [t,b]$. By standardness and bijectivity, $\{\varrho(s), \varrho(t)\} = \{a,b\}$. If $\varrho$ is increasing on $[a,s]$, then $\varrho(s) > \varrho(a)$, hence $\varrho(s) = b$ and $\varrho(t) = a$. This means $\varrho(t) \leq \varrho(b)$, whence $\varrho$ is increasing on $[t,b]$ as well. The other constellations are completely similar. **qed**

---

[6] Here, boundary point is understood w.r.t. $\mathbb{R}$.

| positive orientation | | negative orientation | |
|---|---|---|---|
| [diagram] | perfect | [diagram] | perfect |
| [diagram] | perfect | [diagram] | perfect |
| [diagram] | perfect | [diagram] | perfect |
| [diagram] | not perfect | [diagram] | not perfect |
| [diagram] | not perfect | [diagram] | not perfect |

Table 1: All topological types of standard maps

**Definition 2.5** A homeomorphism between subsets of $[a,b]$ is called

$$\begin{aligned}\textbf{positive} &\iff \text{it increases on all connected components;}\\ \textbf{negative} &\iff \text{it decreases on all connected components.}\end{aligned}$$

Sometimes we will speak of positive orientation instead of positivity. We should emphasize that positivity of $\varrho$ does not mean that $\varrho > 0$; instead, a diffeomorphism $\varrho$ is positive iff $\dot\varrho > 0$. Neither does a positive $\varrho$ need to increase on its full domain. In fact, we do not impose any restrictions between points in different components. As we have seen in the proof above, a positive standard map defined on the union of two intervals is given pictorially by

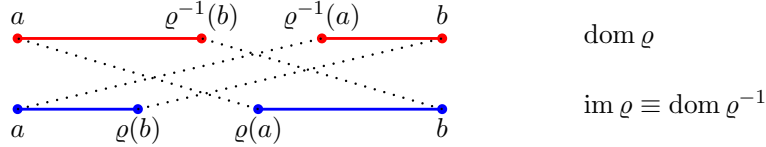

So, e.g., the right part of the domain of $\varrho$ is given by the interval $[\varrho^{-1}(a), b]$. Of course, we have assumed that $a \in \dom \varrho^{-1} \equiv \im \varrho$. Otherwise, the right part is empty, indeed. For negative standard maps, we can give similar relations. Moreover, they have an important property:

**Lemma 2.16** Any negative standard map has exactly one fixed point in the interior of each connected component of its domain.

Later we will see that any negative essential reparametrization will be characterized uniquely by its fixed point, provided the action is analytic and pointwise proper.

**Proof** Let $[s,t]$ be a connected component of the domain of the standard mapping $\sigma$ with $s < t$.
- Assume first $t < b$. As $\sigma$ is standard, $s$ must equal $a$ and $\sigma(t)$ must be $a$ or $b$. In the latter case, we would get $\sigma(s) < b = \sigma(t)$ implying that $\sigma$ is positive. Hence $\sigma(t) = a$. As $s \neq t$, we have $\sigma(s) > a = \sigma(t)$. Now, by the mean value theorem, there is some $\mathbf{x} \in [s,t]$ with $\sigma(\mathbf{x}) = \mathbf{x}$. Obviously, $\mathbf{x}$ neither equals $s$ nor $t$, hence lies in $(s,t)$. Uniqueness is clear from strict monotonicity of $\sigma$.
- Assume next $a < s$. Then the argumentation is similar.
- Assume finally $a = s < t = b$. Here, the statement is trivial. **qed**

Let us now continue with formalizing the properties of the set of essential reparametrizations.

**Definition 2.6**  A set **P** of mappings between closed subsets of $[a,b]$ is called
- **standard** $\iff$ all elements of **P** are standard maps;
- **positive** $\iff$ all elements of **P** are positive maps;
- **unital** $\iff$ the identity **1** is in **P**;
- **involutive** $\iff$ with any element also its inverse is in **P**;
- **multiplicative** $\iff$ for any $\varrho_1, \varrho_2 \in \mathbf{P}$ with $\mathrm{dom}(\varrho_1 \circ \varrho_2)$ containing a nontrivial interval $I$, there is a $\varrho_1 \bullet \varrho_2 \in \mathbf{P}$ that coincides with $\varrho_1 \circ \varrho_2$ on $I$;[7]
- **analytic** $\iff$ any $\varrho_1, \varrho_2 \in \mathbf{P}$ equal iff they coincide on a nontrivial interval;
- **pointwise proper** $\iff$ for $\varrho_i \in \mathbf{P}$ with $s,t \in \mathrm{dom}\,\varrho_i$ as well as converging $(\varrho_i(s))$ and $(\varrho_i(t))$:
$$\lim \varrho_i(s) = \lim \varrho_i(t) \iff s = t.$$

**Proposition 2.17**  Let $\varphi$ be an analytic left action of the Lie group $G$ on $M$, and let $\gamma$ be an analytic path. Then the set $\mathbf{P}_\gamma$ of all nontrivial essential reparametrizations between $\gamma$ and its $G$-translates is standard, analytic, unital, involutive, multiplicative.

If the action is even pointwise proper or isometric, then $\mathbf{P}_\gamma$ is pointwise proper.

**Definition 2.7**  A set **P** is called **reparametrization set** iff it is a standard, unital, analytic, involutive, multiplicative, pointwise proper set of mappings between closed subsets of $[a,b]$.

Thus, $\mathbf{P}_\gamma$ is a reparametrization set for any analytic path $\gamma$, provided $\varphi$ is an analytic and pointwise proper left action of some Lie group $G$ on some analytic manifold $M$. Note that it seems that we have now lost some information. In particular, we do not require the mappings in a reparametrization set to be diffeomorphisms or just differentiable. However, differentiability will re-appear later as any continuous one-parameter subgroup in a Lie group is smooth.

**Proof**  Standardness follows from Proposition 2.7 together with Definitions 2.2, 2.4 and 2.3. Analyticity is due to Corollary 2.11. Unitality and involutivity come from Proposition 2.8. For multiplicativity, let $\varrho_i = \varrho_{g_i}$ be two essential reparametrization functions with non-discrete $\mathrm{dom}\,\varrho_1 \circ \varrho_2 \equiv \varrho_2^{-1}(\mathrm{dom}\,\varrho_1)$. Again by 2.8, there is some $\varrho := \varrho_{g_1 g_2}$, such that $\varrho$ coincides with $\varrho_1 \circ \varrho_2$ on its non-discrete domain. This gives multiplicativity. Finally, pointwise properness follows from Proposition 2.12. **qed**

We are now going to solve two tasks: firstly, we shall classify all reparametrization sets and, secondly, based on this classification, we shall prove Theorem 1.2, i.e., that any $\gamma$ is either a Lie path or a brick path. Nevertheless, we will present this in the opposite order. This means, first, in Section 3, we will just motivate the needed results from the reparametrization set classification and then prove the mentioned theorem. Only afterwards, starting with Section 4, we will then perform the full classification of reparametrization sets.

## 2.4 Prototypical Examples

Above, we have already discussed the symmetries of the sine curve. There are much simpler examples that will even turn out prototypical, namely the motions on $M = S^1$ or $M = \mathbb{R}^1$.

---

[7]If we speak on $\varrho_1 \bullet \varrho_2 \in \mathbf{P}$ for some $\varrho_i$ in the following, we always comprise the assumption that there is some nontrivial interval contained in $\mathrm{dom}\,\varrho_1 \cap \mathrm{im}\,\varrho_2$ with the desired property. Moreover, given analyticity, this element $\varrho_1 \bullet \varrho_2$ is unique if existing. See also Proposition 4.5 for multiple products.

Each analytic path therein is – up to parametrization and up to orientation – just the identical mapping on a compact interval[8] $I$, hence we can even identify the path with $I$. Then the essential reparametrization of a motion $g$ is simply given by its restriction

$$\varrho_g \; : \; I \cap g^{-1}I \; \longrightarrow \; gI \cap I$$

in domain and range to $I$, provided $I \cap g^{-1}I$ does not contain singular points; otherwise, further restrict domain and image to their non-discrete parts. It is now very remarkable that any reparametrization set $\mathbf{P}$ will turn out to be (up to conjugation with some homeomorphism) the reparametrization set of an analytic path on $S^1$ under the action of a subgroup $U$ of $O(2)$. If $\mathbf{P}$ descends from an analytic and pointwise proper action of a Lie group $G$ on $M$, then $U$ is even a Lie group. This will turn out crucial for the classification of path symmetries in the general case.

Let us illustrate the properties of our prototypical example of $S^1$ in two steps. For simplicity, we restrict ourselves to orientation preserving motions, i.e. shifts only. First we consider some Lie group $U$ of shift operators on $\mathbb{R}$. As $U$ can be identified with a Lie subgroup of $\mathbb{R}$, it equals one of the groups $\mathbb{R}$ (continuous case) or $s\mathbb{Z}$ (discrete case) with fixed $s \in \mathbb{R}$. Let us take $I = [0, 3] \subseteq \mathbb{R}$ and $U$ to contain the integer shifts. Then $U \cong \mathbb{Z}$ is generated by the right-shift by 1. Its corresponding restriction to $I$ is

$$\varrho \equiv \varrho_1 \; : \; [0, 2] \; \longrightarrow \; [1, 3].$$

Similarly, the essential reparametrization $\varrho_2$ of the right shift by 2 maps $[0, 1]$ to $[2, 3]$. The shift by 3, however, is more delicate. Here, $I$ and $g^{-1}I$ share just the point 0, whence the reparametrization function has discrete domain; it maps 0 to 3. Hence, the essential reparametrization has empty domain. The same applies to any larger shifts. Discussing the inverses similarly, we see that only $\varrho_{-2}, \varrho_{-1}, \mathbf{1}, \varrho, \varrho_2$ have non-discrete domain, hence qualify for being an essential reparametrization. Thus, they comprise $\mathbf{P}_\gamma$.

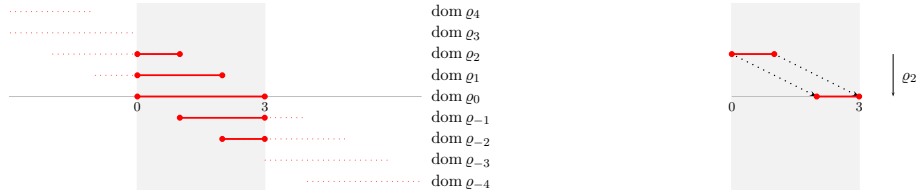

Note that product of elements in $\mathbf{P}_\gamma$ is not always defined. Indeed, we have $\varrho_1 \circ \varrho_1 = \varrho_2$, i.e., $\varrho_2$ equals $\varrho^2$, but the product $\varrho_1 \circ \varrho_2$ is not well defined within $\mathbf{P}_\gamma$. In fact, it maps 0 to 3, hence is not an essential reparametrization. However, even if the product is defined, it need not be an element in $\mathbf{P}_\gamma$ again. For instance, $\varrho_{-1} \circ \varrho_2 \equiv \varrho^{-1} \circ \varrho^2$ having domain $[0, 1]$ equals $\varrho^1$ thereon as expected, but not on the full domain of $\varrho^1$. This has been the deeper reason for introducing the notion of $\bullet$. Indeed, $\varrho^{-1} \circ \varrho^2$ extends to a unique element of $\mathbf{P}_\gamma$, namely $\varrho$, which now allows to write $\varrho^{-1} \bullet \varrho^2 = \varrho^1$. This means, the $\bullet$-product has the desired property $\varrho^x \bullet \varrho^y = \varrho^{x+y}$, if defined, i.e., if $x, y$ and $x + y$ have modulus at most 2.

Let us now come to the case of $S^1$. There, in contrast to $\mathbb{R}$, a shift operator might move parts of the domain around the circle and let it "re-enter" from the other side again. This can, of course, only happen if the interval is larger than half the circle. Now, to get nicer figures, let us assume that $S^1 = \mathbb{R}/4\mathbb{Z}$, i.e., $S^1$ can be considered as the interval $[0, 4]$ with identified endpoints. Also, the integer shifts form a group isomorphic to $\mathbb{Z}_4$. Taking $I = [0, 3] \subseteq S^1$, the restriction in domain and range to $[0, 3]$ of the unit right-shift is defined on $[0, 2] \sqcup \{3\}$. Hence, its essential reparametrization $\varrho \equiv \varrho_1$ is defined on $[0, 2]$ and coincides with the corresponding mapping in the $\mathbb{R}$-case above. This is no longer true for $\varrho_2$. In fact, $\varrho_2$ maps now $[0, 1] \sqcup [2, 3]$ to itself, while exchanging both subintervals. We see already here, that $\varrho_2$ does not coincide with $\varrho^2 = \varrho_1 \circ \varrho_1$; they equal only on $[0, 1]$ being the domain of $\varrho^2$. Again, we only have $\varrho_2 = \varrho \bullet \varrho$. Observe that

---
[8]Compact intervals on $S^1$ are defined to be compact connected proper subsets. Alternatively, they can be considered as images of compact intervals in $\mathbb{R}$ of length less than $s$ under the canonical projection $\mathbb{R} \longrightarrow \mathbb{R}/s\mathbb{Z} \cong S^1$.

$\varrho_3$ now is defined and equals $\varrho \bullet (\varrho \bullet \varrho)$. Moreover, we have $\varrho_{-1} = \varrho_3$, showing that $\mathbf{P}_\gamma$ consists of exactly four elements $\mathbf{1}, \varrho, \varrho_2, \varrho_3$. Note that if we concatenate the "generator" $\varrho$ by means of the normal product $\circ$, there is no difference between the present example on $S^1$ and the corresponding example on $\mathbb{R}$ above. It is the $\bullet$-product, that makes the difference. The careful handling of this difference will be crucial for the classification of reparametrization sets.

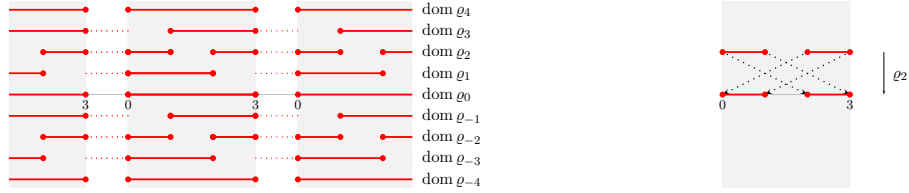

Let us close with some comments

- Any reparametrization set w.r.t. $\mathbb{R}$ can also be considered as one w.r.t. $S^1$. For this, let $I \subseteq \mathbb{R}$ be the interval corresponding to the path $\gamma$ and let $U$ be some Lie subgroup of $\mathbb{R}$. Moreover identify $S^1$ with $\mathbb{R}/s\mathbb{Z}$. Here $s$ has to fulfill two requirements: First, $s$ must be larger then the double length of $I$; this means that $I$ as an interval in $S^1$ comprises less than a half-circle of $S^1$. Second, unless $U$ is trivial, $s$ must be contained in $U$; this guarantees that the image of $U$ under the canonical projection $\mathbb{R} \longrightarrow \mathbb{R}/s\mathbb{Z}$ is discrete again for discrete $U$. It is now a simple exercise that the reparametrization set obtained this way on $S^1$ is the same as the original one on $\mathbb{R}$.

- In principle, we have had no need to restrict ourselves above to *Lie* subgroups of $\mathbb{R}$ or $S^1$. Indeed, we can formulate the theory also for non-Lie subgroups. However, as we will see pointwise proper actions, the reparametrization set will always correspond to some with a Lie subgroup acting on $S^1$.

- In the examples above only three of the five types of standard maps occur (cf. Table 1). These are exactly three types called "perfect". For $\mathbb{R}$ or $S^1$, obviously the other cases are excluded since the action is isometric. However, we will see in Proposition 5.4 that indeed any reparametrization set consists of such perfect mappings only.

## 3 Classification Theorem

In this section we are going to prove our main Theorem 1.2. Actually, we will even derive a simple and explicit criterion when $\gamma$ is a Lie or a brick path. To get an idea, consider some Lie path $\gamma(t) := e^{tA}x$. Obviously, the action of each $g = e^{sA}$ on $\gamma$ corresponds to an essential reparametrization $\varrho_g(t) = s + t$, upon restriction to the correct domain. In particular, we see that there are infinitely many group elements that perform mutually different actions on $\gamma$. Thus, one might be tempted to assume that $\gamma$ is a Lie path iff there are infinitely many different $g \in G$ leading to a nontrivial overlap. This, however, is not yet the correct idea as, e.g., for the trivial action of a compact $G$ obviously each path is a brick path. Instead, we shall use the essential reparametrizations. Note that different group elements may have identical essential reparametrizations. Indeed, if $G_\gamma$ is the stabilizer of $\gamma$, then all elements in $gG_\gamma$ lead to the same essential reparametrization, hence to the same element in $\mathbf{P}_\gamma$. The advantage of our abstract notation is to have already modded out the undesirable stabilizer and to get this way a direct classification of the paths by cardinality of their respective $\mathbf{P}_\gamma$. It is now natural to define

**Definition 3.1** Let $\gamma$ be a path and $\mathbf{P}_\gamma$ its set of nontrivial essential reparametrizations. Then

$$\gamma \text{ has } \mathbf{infinite\ symmetry} \iff \mathbf{P}_\gamma \text{ infinite}$$
$$\gamma \text{ has } \mathbf{finite\ symmetry} \iff \mathbf{P}_\gamma \text{ finite}$$

Now, we can state our main classification result for symmetries of analytic paths.

**Theorem 3.1** Let $\varphi$ be a pointwise proper analytic left action of a Lie group $G$ on an analytic manifold $M$. Then we have for all analytic paths $\gamma$ in $M$:

$$\gamma \text{ Lie path} \iff \gamma \text{ has infinite symmetry}$$
$$\gamma \text{ brick path} \iff \gamma \text{ has finite symmetry}$$

## 3.1 Road to Classification of Reparametrization Sets

As announced, the proof of Theorem 3.1 above will be reduced to the classification of reparametrization sets to be done later in this article. At the moment, let us motivate the main results we need already for the path classification. As it will turn out, that $\mathbf{P}_\gamma$ is finite iff its subset $\mathbf{P}_\gamma^+$ containing the positive elements of $\mathbf{P}_\gamma$ is finite, let us restrict ourselves for the time being to the case of positive reparametrization sets $\mathbf{P}$. Then the main ideas will be as follows:

1. *Standard implies Perfect.*
   If $\varrho \in \mathbf{P}$, then its domain is $[a,b]$ iff its range is $[a,b]$, by pointwise properness.

2. *Restriction to* $\mathbf{P}_a := \{\varrho \in \mathbf{P} \mid a \in \operatorname{dom} \varrho\}$
   This will be possible as $\varrho$ is defined at $b$ iff $\varrho^{-1}$ is defined at $a$.

3. *Fixed point criterion*
   If $\varrho \in \mathbf{P}_a$ has a fixed point, then $\varrho = \mathbf{1}$. This will mainly be due to pointwise properness.

4. *Identification of* $\mathbf{P}_a$ *with* $\mathbf{P}(a) := \{\varrho(a) \mid \varrho \in \mathbf{P}\}$
   In fact, the previous point implies that $\varrho$ is uniquely determined by its value in $a$.

5. *Pointwise limits in* $\mathbf{P}_a$ *correspond to limits in* $\mathbf{P}(a) \subseteq [a,b)$.
   We will show that $\varrho_1(a) < \varrho_2(a)$ implies $\varrho_1 < \varrho_2$ around $a$.

6. $\mathbf{P}(a)$ *either dense or finite*
   For this, we will show that $\mathbf{P}(a)$ accumulates somewhere iff it accumulates everywhere.

7. *Existence of a generator of* $\mathbf{P}$ *(finite case)*
   $\mathbf{P}_a$ has minimal element $\mu$ larger than $\mathbf{1}$. It fulfills $\mathbf{P} = \{\mu^k\}_{k \leq n}$

8. *Existence of roots in* $\mathbf{P}(a)$ *(infinite, complete case)*
   For $\mathbf{P}(a) = [a,b)$, we have roots: each $\varrho \in \mathbf{P}_a$ equals some $\sigma^k$ with $\sigma \in \mathbf{P}_a$.

9. *Local sub-semigroup* $[0,T) \longmapsto \mathbf{P}$ *(infinite, complete case)*
   Approximating $\varrho^x$ by roots for some $\varrho > \mathbf{1}$, the mapping $x \longmapsto \varrho^x$ is even an isomorphism.

These ideas will lead to the following statements to be proven later in the subsequent sections. Indeed, Proposition 3.2 is part of Proposition 6.2 (see also Definitions 5.5 and 6.1). Theorem 3.3 comes from Theorem 9.12 (see also Definition 9.1). Theorem 3.4 originates in Theorem 10.13 (see also Definition 10.2). Finally, Theorem 3.6 comes from Theorem 8.1 (finite case) and Proposition 11.17 with Corollary 11.18 (infinite full case). Observe for all cases that each reparametrization set is a motion set by Proposition 5.4.

Note that the reparametrization set is assumed positive only if explicitly given.

**Proposition 3.2** Let $\mathbf{P}$ be a *positive* reparametrization set and let $\varrho, \sigma \in \mathbf{P}_a$.
Then $\varrho(a) < \sigma(a)$ implies $\varrho < \sigma$ on any interval $[a,s]$ contained in $\operatorname{dom} \sigma$.

In particular, this includes the statement that $[a,s]$ is contained in the domain of $\varrho$.

**Theorem 3.3** Let **P** be a *finite* reparametrization set.
Then there is a nontrivial interval $I \subseteq [a, b]$, such that $\mathbf{P}^{-1}(I)$ covers $[a, b]$ and
$$I \cap \varrho^{-1}(I) \text{ nontrivial} \iff \varrho = \mathbf{1}.$$

The restriction of $\gamma$ to $I$ will become a free segment $e$ by the nontriviality condition. The condition that $\mathbf{P}^{-1}(I) \equiv \{\varrho^{-1}(I) \mid \varrho \in \mathbf{P}\}$ covers $[a, b]$ gives that $\gamma$ is indeed a concatenation of $G$-translates of $e$. Note, however, that $I$ need not always be fully contained in the image of $\varrho$ as soon as $\varrho^{-1}(I)$ contains $a$ or $b$. This corresponds to the fact that only a part of a translate of $e$ might be needed to obtain the remaining part of $\gamma$ towards its ends.

**Theorem 3.4** Let **P** be a reparametrization set and $\mathbf{P}^+ := \{\varrho \mid \varrho \in \mathbf{P} \text{ positive}\}$. Then
$$\mathbf{P} \text{ infinite} \iff \mathbf{P}^+ \text{ infinite} \implies \mathbf{P}^+(a) \text{ is dense in } [a, b].$$

Note that we cannot conclude[9] yet that an infinite reparametrization set gives full $\mathbf{P}(a) = [a, b)$. Unfortunately, we will need that assumption for the final classification of analytic paths; in fact, any Lie path has surely full $\mathbf{P}(a)$ as well as any brick path has finite. But, if **P** indeed comes from a set of essential reparametrizations, we have closedness, hence fullness in the infinite case by denseness. Remarkably, this will remain the only point where we the classification of reparametrization sets back-reacts with that of analytic paths. Therefore, we will state and prove closedness already here:

**Proposition 3.5** If $G$ acts pointwise properly, then $\mathbf{P}_\gamma^+(a)$ is closed in $[a, b)$.

Note that $\varrho(a)$ can never be $b$ for standard and positive $\varrho$.

**Proof** Write shortly $\mathbf{P} := \mathbf{P}_\gamma^+$, and let $\varrho_{g_i}(a) \to t < b$ for some $\varrho_{g_i} \in \mathbf{P}(a)$ with $g_i \in G$.
- We may assume that **P** is infinite, as otherwise the statement is trivial. Therefore, by Theorem 3.4, $\mathbf{P}(a)$ is dense in $[a, b]$. Thus, there is some $\sigma$ in $\mathbf{P}_a$ with $t < \sigma(a) < b$.
- As $\varphi_{g_i}(\gamma(a)) = \gamma(\varrho_{g_i}(a)) \to \gamma(t)$, we have $g_i \to g$ by pointwise properness (if necessary, after taking a subsequence). Choose some interval $[a, s]$ in the domain of $\sigma$. As we may assume $\varrho_{g_i}(a) < \sigma(a)$ for all $i$, the interval $[a, s]$ is even contained in the domain of any $\varrho_{g_i}$, by Proposition 3.2. As the range of $\gamma$ is compact, we have for all $r \in [a, s]$
$$[\varphi_g \circ \gamma](r) = \lim \varphi_{g_i}(\gamma(r)) = \lim \gamma(\varrho_{g_i}(r)) \in \operatorname{im} \gamma.$$
- Thus, the intersection of the images of the $\varphi_g \circ \gamma$ and $\gamma$ contains $\varphi_g(\gamma[a, s])$ at least. Obviously, this set contains an accumulation point, whence the essential reparametrization $\varrho_g$ is nontrivial and its domain contains $[a, s]$ at least. Hence,
$$\gamma(\varrho_g(r)) = \varphi_g(\gamma(r)) = \lim \varphi_{g_i}(\gamma(r)) = \lim \gamma(\varrho_{g_i}(r)) = \gamma(\lim \varrho_{g_i}(r))$$
for $r \in [a, s]$, since $\gamma$ is an embedding. Altogether $\varrho_{g_i}$ converges pointwise to $\varrho_g$ on $[a, s]$, whence $\varrho_g$ is increasing as all $\varrho_{g_i}$ do. Moreover, $\gamma(\varrho_g(a)) = \lim \varphi_{g_i}(\gamma(a)) \to \gamma(t)$. Now, by injectivity of $\gamma$, we have $t = \varrho_g(a) \in \mathbf{P}(a)$. **qed**

Let us now state the central result from the theory of reparametrization sets. It basically states that any positive **P** is isomorphic to the restriction of a subgroup of $SO(2)$ to some interval, with $SO(2)$ considered as the set of left translations on $S^1$. As we will see in a moment, it is the local commutativity of this subgroup that allows to prove that any path with infinite reparametrization set is already a Lie path.

---
[9]For instance, consider the group $\mathbb{Z}$ acting on $S^1 = \mathbb{R}/\mathbb{Z}$ by shifts by $\lambda n$ with $n \in \mathbb{Z}$ and non-rational $\lambda$. Then one easily checks that the restriction of this action to some nontrivial subinterval $I$ of $S^1$ yields a reparametrization set **P**. In particular, **P** is pointwise proper as $\mathbb{Z}$ acts by isometries (see Proposition 2.17). However, $\mathbb{Z}$ itself does not act pointwise properly, whence Proposition 3.5 is not applicable to show closedness of **P**. And indeed, assuming 0 to be contained in the interior of $I$, we have $\mathbf{P}(0) = \{[\lambda n]\} \cap I$ which is well known to be dense in $I$, but not closed.

**Theorem 3.6**  Let **P** be a *positive* reparametrization set with $\mathbf{P}_\gamma^+(a)$ closed in $[a,b]$. Then there is a local sub-semigroup **T** of $\mathbb{R}$ and, for any $t \in \mathbf{T}$, some $\varrho^t \in \mathbf{P}_a$, such that

$$\varphi: \begin{array}{rcl} \mathbf{T} & \longrightarrow & \mathbf{P}(a) \\ t & \longmapsto & \varrho^t(a) \end{array}$$

is a homeomorphism and we have
$$\begin{array}{rcl} \varrho^{t_1+t_2} & = & \varrho^{t_1} \bullet \varrho^{t_2} \quad \text{ for all } t_1, t_2, t_1 + t_2 \in \mathbf{T} \\ \varrho^0 & = & \mathbf{1}\,. \end{array}$$

In particular, $\varrho^{t_1+t_2}$ and $\varrho^{t_1} \circ \varrho^{t_2}$ coincide in a neighbourhood of $a$.

Recall [9] that semigroups are defined on $\mathbb{R}_+$ and local semigroups on $[0, T)$ for some $T > 0$. Here, we assume that **T** is a local sub-semigroup, i.e., the restriction of a Lie subgroup $U$ of $\mathbb{R}$ to its intersection with a bounded interval $[0, T) \subseteq \mathbb{R}$. In particular, the addition of $t_1$ and $t_2$ from **T** is defined as soon as the usual sum $t_1 + t_2$ in $\mathbb{R}$ is in **T** again; indeed, we identify then the sums in $\mathbb{R}$ and in **T**. Note that we could even extend this notion to subgroups $U$ that are no longer *Lie* subgroups. Indeed, it could just be any subgroup. Of course, unless this subgroup equals $x\mathbb{Z}$ for some $x \in \mathbb{R}$, it is already dense. We refrain, however, from discussing these cases as they do not appear for reparametrization sets due to closedness.

### 3.2 Finite Symmetry

Let us start with the shorter part, namely the case of finite symmetry. In Proposition 5.23 of [6], Hanusch has already shown that any path is a brick path, if it has a subpath being a free segment. Here, we will give another condition that will turn out much more general as it allows to classify the paths for *any* pointwise proper action. In particular, the existence of free segments is now a general implication, no longer an assumption.

**Theorem 3.7**  Let $G$ act pointwise proper and let $\gamma$ be a path of finite symmetry. Then there is free segment $e$ that generates $\gamma$ as a brick path.

**Proof**  As $\gamma$ is of finite symmetry, $\mathbf{P}_\gamma$ is finite. Now, choose some nontrivial interval $I$ with the properties listed in Theorem 3.3, and define $e$ to be the restriction of $\gamma$ to $I$.
- $e$ is a free segment.
  Assume that $e$ and $\varphi_g \circ e$ share a subpath, say $e|_J$. Then, of course, the essential reparametrization $\varrho_g$ is well defined, hence in $\mathbf{P}_\gamma$ with $J \subseteq \operatorname{im} \varrho_g$. This implies $J \cong \varrho_g^{-1}(J) \subseteq I \cap \varrho_g^{-1}(I)$. Theorem 3.3 gives $\varrho_g = \mathbf{1}$. Hence $e$ and $\varphi_g \circ e$ are even identical.
- $e$ generates $\gamma$.
  Consider the paths $\varphi_g \circ e$ for $\varrho_g \in \mathbf{P}_\gamma$. Any two of them either coincide up to parametrization or do not share a subpath. Thus, we only have to show that for each $t \in [a, b]$ there is some $g \in G$, such that $\gamma(t)$ is contained in $\varphi_g \circ e$. This, however, is indeed the case. In fact, for any such $t$, there is some $\varrho_h \in \mathbf{P}_\gamma$ with $t \in \varrho_h^{-1}(I)$. Thus, $[\varphi_h \circ \gamma](t) = \gamma(\varrho_h(t))$ is in $\gamma(I)$, hence in the image of the free segment $e$. **qed**

It will be a direct consequence of Proposition 9.13 that one needs at most two elements $g$ and $h$ in $G$ to generate the full path $\gamma$ by translates of a free segment $e$. Indeed, choosing $e$ appropriately, $\gamma$ is a subpath of the concatenation of the (possibly reparametrized) translates of $e$ w.r.t. the group elements

$$\mathbf{1}, g, \ldots, g^n \quad \text{or} \quad g^{-1}h,\ \mathbf{1}, h,\ g, gh,\ g^2, g^2h, \ldots, g^n, g^nh$$

for some appropriate $n$. Here, observe that the translates of $e$ that involve $h$ need to be inverted. For concrete examples, we refer to the sine-graph curve in Section 1.5. Just observe that we use now paths, i.e., do no longer admit curves with non-compact domains.

## 3.3 Infinite Symmetry

Throughout this subsection, $\gamma$ is always an analytic path and $\mathbf{P}$ equals the set $\mathbf{P}_\gamma^+$ of positive and nontrivial essential reparametrization functions. Of course, $\mathbf{P}$ is a reparametrization set again. By Theorem 3.4, $\mathbf{P}$ is infinite iff $\gamma$ has infinite symmetry.

**Definition 3.2**
$$\begin{aligned} G_\gamma &:= \{g \in G \mid \varrho_g = \mathbf{1}\} \equiv \{g \in G \mid \varphi_g \circ \gamma = \gamma\} \\ H_\gamma &:= \{h \in G \mid \varrho_h \in \mathbf{P}_a\} \\ H &:= \langle H_\gamma \rangle \end{aligned}$$

Here, $\langle A \rangle$ denotes the smallest Lie subgroup in $G$ that contains $A$. Let us compare our definition with Definition 5.18 in [6] by Hanusch.

1. For analytic and pointwise proper actions, our definition of $G_\gamma$ is equivalent to the one by Hanusch. This follows, e.g., from Lemma 5.19/1 in [6]. In our framework, this is comprised in the fact that any positive element in a reparametrization set is the identity as soon as it has full domain; see Subsection 6.1 for several statements that imply this.

2. The notion of $H_\gamma$ is related to Hanusch's $H_{\gamma,\gamma}$, but they do not coincide, in general. Usually, $H_\gamma$ is a proper subset of $H_{\gamma,\gamma}$. In fact, we restrict ourselves to positive elements of $\mathbf{P}_\gamma$ that are defined in $a$, while for $H_{\gamma,\gamma}$ all elements of $\mathbf{P}_\gamma$ are taken into account. Whereas the inclusion of all positive elements of $\mathbf{P}_\gamma$ in $H_\gamma$ would lead to the same $H$ as above, it is a priori unclear whether the same will be true if we drop the positivity condition. On the other hand, restricting ourselves to positivity allows us to exploit the full strength of Theorem 3.6 and to ultimately end with the desired classification. Indeed, in the infinite symmetry case, it is the commutativity of the multiplication in $(\mathbf{P}_\gamma)_a^+$ that will transfer to the commutativity of $H/G_\gamma$ and this way letting $\gamma$ correspond to an appropriate one-parameter subgroup.

**Lemma 3.8** $G_\gamma$ is a normal Lie subgroup in $H$.

**Proof** Obviously, $G_\gamma$ is a closed, hence Lie subgroup of $G$. Let now $g \in G_\gamma$ and $h \in H_\gamma$. By assumption, domain and image of $\varrho_h$ are nontrivial, hence of $\varrho_{h^{-1}}$ as well. Since, on the domain of $\varrho_{h^{-1}}$, we have

$$\begin{aligned} \varphi_{hgh^{-1}} \circ \gamma &= \varphi_h \circ \varphi_g \circ \varphi_{h^{-1}} \circ \gamma \\ &= \varphi_h \circ \varphi_g \circ \gamma \circ \varrho_{h^{-1}} = \varphi_h \circ \gamma \circ \varrho_{h^{-1}} = \varphi_h \circ \varphi_{h^{-1}} \circ \gamma = \gamma, \end{aligned}$$

$\varphi_{hgh^{-1}} \circ \gamma$ even equals $\gamma$ everywhere, by analyticity. Hence $hgh^{-1} \in G_\gamma$. As $H$ is generated by $H_\gamma$, we have $hG_\gamma h^{-1} \subseteq G_\gamma$ for all $h \in H$. As $H$ is closed in $G$, now $G_\gamma$ is a closed, hence Lie subgroup of $H$. **qed**

**Corollary 3.9**
1. $G_\gamma$ stabilizes all points in the $H$-orbit of $x$.
2. $G_\gamma$ is the kernel of the induced $H$-action on $Hx$, if $\mathbf{P}$ is infinite.

**Proof**
1. Obviously, $G_\gamma$ stabilizes $x = \gamma(a)$. Now, for $g \in G_\gamma$ and $h \in H$, we have $ghx = h(h^{-1}gh)x = hx$, as $h^{-1}gh \in G_\gamma$. Hence, $g$ acts trivially on $Hx$.
2. Let $h$ act trivially on $Hx$. As $\mathbf{P}$ is infinite, $\mathbf{P}(a)$ is dense in $[a,b]$, whence $Hx$ comprises at least a dense subset of $\mathrm{im}\,\gamma$. By continuity, we have $h\gamma = \gamma$, i.e., $h \in G_\gamma$. **qed**

**Lemma 3.10** The mapping
$$\begin{aligned} \Theta: \mathbf{P}_a &\longrightarrow H/G_\gamma \\ \varrho_h &\longmapsto [h] \end{aligned}$$
is well defined.

**Proof** If $\varrho_{h_1} = \varrho_{h_2}$ both having domain $I$, then we have $\varphi_{h_1} \circ \gamma = \gamma \circ \varrho_{h_1} = \gamma \circ \varrho_{h_2} = \varphi_{h_2} \circ \gamma$ on $I$. By analyticity, we even have $\varphi_{h_1} \circ \gamma = \varphi_{h_2} \circ \gamma$ everywhere. This gives the proof.

**qed**

**Lemma 3.11** Let $h_1, h_2 \in H_\gamma$ and $\varrho^{t_i} = \varrho_{h_i}$ for $t_i \in \mathbf{T}$.
Then $\varrho^{t_1+t_2} = \varrho_{h_1 h_2}$ and $h_1 h_2 \in H_\gamma$ as long as $t_1 + t_2 \in \mathbf{T}$.

**Proof** Theorem 3.6 gives $\varrho^{t_1+t_2} = \varrho_{h_1} \bullet \varrho_{h_2}$. In particular, $\varrho^{t_1+t_2}$ coincides with $\varrho := \varrho_{h_1} \circ \varrho_{h_2}$ on the non-discrete part of dom $\varrho$. Proposition 2.8 shows that also $\varrho_{h_1 h_2}$ coincides with $\varrho$ thereon. Thus, $\varrho^{t_1+t_2} = \varrho_{h_1 h_2}$ thereon. In particular, $\varrho_{h_1 h_2}$ is positive again, whence it is in $\mathbf{P}$ as so does $\varrho^{t_1+t_2}$. Corollary 2.11 shows now that $\varrho_{h_1 h_2}$ equals $\varrho^{t_1+t_2}$ everywhere.

**qed**

**Theorem 3.12** $K_\gamma := H/G_\gamma$ is an abelian Lie group for any path of infinite symmetry.

**Proof** Let $h_1, h_2 \in H_\gamma$. By Theorem 3.6, there are $2t_1, 2t_2 \in \mathbf{T}$ with $\varrho_{h_1} = \varrho^{2t_1}$ and $\varrho_{h_2} = \varrho^{2t_2}$. As $t_1$ and $t_2$ are contained in $\mathbf{T}$, there are $k_1, k_2 \in H_\gamma$ with $\varrho_{k_1} = \varrho^{t_1}$ and $\varrho_{k_2} = \varrho^{t_2}$. As, moreover, $t_1 + t_2$ is contained in $\mathbf{T}$, we have by Lemma 3.11

$$\begin{aligned}
\varrho_{k_1 k_2} &= \varrho^{t_1+t_2} = \varrho^{t_2+t_1} = \varrho_{k_2 k_1} \\
\varrho_{h_1} &= \varrho^{2t_1} = \varrho^{t_1+t_1} = \varrho_{k_1 k_1} \\
\varrho_{h_2} &= \varrho^{2t_2} = \varrho^{t_2+t_2} = \varrho_{k_2 k_2}
\end{aligned}$$

Using Lemma 3.10, we get $[k_1][k_2] = [k_1 k_2] = [k_2 k_1] = [k_2][k_1]$ from the first line, as well as $[h_1] = [k_1]^2$ and $[h_2] = [k_2]^2$ from the other ones. Consequently, also $[h_1]$ and $[h_2]$ commute. Finally, observe that $H_\gamma$ generates $H$. This gives the proof.

**qed**

Note that infinite symmetry as well as closedness have been used to conclude $t \in \mathbf{T}$ from $2t \in \mathbf{T}$.

**Lemma 3.13** Let $G$ be a Lie group that acts on itself by left translations.
Then $E(g) := \gamma(\varrho_g(a))\gamma(a)^{-1}$ is multiplicative, i.e.,

$$E(gh) = E(g)E(h) \qquad \text{whenever } \varrho_{gh} = \varrho_g \circ \varrho_h \text{ in } a \text{ is well defined.}$$

**Proof** Use $g\gamma(a) \equiv [\varphi_g \circ \gamma](a) = \gamma(\varrho_g(a))$ and Proposition 2.8 to derive

$$\gamma(\varrho_{gh}(a)) = \gamma(\varrho_g(\varrho_h(a))) = g\gamma(\varrho_h(a)) = \gamma(\varrho_g(a))\,\gamma(a)^{-1}\,\gamma(\varrho_h(a))$$

**Proposition 3.14** Let $\varphi$ be a free, transitive and pointwise proper action of $G$ on $M$.
Then any path with infinite symmetry is a Lie path.

**Proof**
- As $\varphi$ is transitive and pointwise proper, it is even proper.[10] Therefore and by freeness, there is an equivariant diffeomorphism between the action $\varphi$ of $G$ on $M$ and the left translation of $G$ on itself. Therefore, we may assume that we are in the latter situation.
- Proposition 3.5 says that $\mathbf{P}(a)$ equals $[a, b)$, and Theorem 3.6 provides us with a collection $\{\varrho^t\} \subseteq \mathbf{P}_a$, such that $\boldsymbol{\varphi}(t) := \varrho^t(a)$ is a homeomorphism $\boldsymbol{\varphi} : \mathbf{T} \longrightarrow [a, b)$ and $t \longmapsto \varrho^t$ is a local semigroup in $\mathbf{P}_a$.
- Define $F(t) := \gamma(\boldsymbol{\varphi}(t))\gamma(\boldsymbol{\varphi}(0))^{-1}$ and choose $g_i \in G$ with $\varrho^{t_i} = \varrho_{g_i}$ for $t_i \in \mathbf{T}$. If also $t_1 + t_2 \in \mathbf{T}$, we have $\varrho^{t_1+t_2} = \varrho_{g_1 g_2} \in \mathbf{P}_a$, whence $\varrho_{g_1 g_2} = \varrho_{g_1} \circ \varrho_{g_2}$ in $a$ by Theorem 3.6. Thus,

$$F(t_1 + t_2) = E(g_1 g_2) = E(g_1)E(g_2) = F(t_1)F(t_2).$$

by Lemma 3.13. As $\mathbf{T}$ is an interval containing 0 and since $F$ is continuous, it generates a continuous one-parameter subgroup of $G$. As any continuous group homomorphism

---

[10] Let $(g_i x_i)$ and $(x_i)$ converge. As $x_i = h_i x$ for some $h_i \in G$, there is a converging subsequence $(h_i)$. Consequently, again taking a subsequence, $(g_i h_i)$ converges as well. Thus, also $g_i$ converges.

between Lie groups is smooth, we get that $F(t) = e^{tA}$ for some $A \in \mathfrak{g}$. Consequently, $[\gamma \circ \boldsymbol{\varphi}](t) = e^{tA}\gamma(a)$ for $t \in \mathbf{T}$. As $\gamma$ is an analytic embedding, $\boldsymbol{\varphi}$ is even an analytic diffeomorphism. This means, that $\gamma$ is up to the parametrization a partial analytic curve. **qed**

**Theorem 3.15** Let $M$ be an analytic manifold, $G$ a Lie group and $\varphi$ a pointwise proper analytic left action of $G$ on $M$. Then we have for all analytic paths $\gamma$ in $M$:

$$\gamma \text{ is a Lie path} \iff \gamma \text{ has infinite symmetry}$$

**Proof** The implication is trivial as any Lie path is assumed to be injective. To prove the other direction, we will reduce the transformation group step-by-step to its "core" as follows:

$$(M, G) \longrightarrow (M, H) \longrightarrow (H/H_x, H) \longrightarrow (H/H_x, K_\gamma) \longrightarrow (H/H_x, K_\gamma/K_x)$$

- As $H$ is a Lie subgroup of $G$, the action $\varphi$ restricts to an analytic left action $H$ on $M$.
- Equip the orbit $Hx$ with the analytic manifold structure of the homogeneous space $H$ modulo the $H$-stabilizer $H_x$ of $x := \gamma(a)$. For general reasons [10][11], the canonical mapping $\iota : H/H_x \longrightarrow M$ with $\iota[h] := hx$ is an initial, but not necessarily embedded submanifold with range $Hx$. Thus, $\varphi$ can be restricted in range to an analytic action $h' \circ [h] = [h'h]$ of $H$ on $H/H_x$, or seen from the orbit, $h' \circ hx = (h'h)x$.
- For the third step, recall that $G_\gamma$ is a normal subgroup that equals the kernel of the action of $H$ on $H/H_x$. Thus, $h'G_\gamma \circ [h] := [h'h]$ is the resulting action of $K_\gamma = H/G_\gamma$.
- Finally, recall that $K_\gamma$ is abelian, whence the corresponding stabilizer $K_x$ of $x$ (or more precisely that of $[\mathbf{1}] = H_x \in H/H_x$) is a normal Lie subgroup of $K_\gamma$ and, moreover, equals the stabilizer of any point in the orbit of $x$. As, by construction, $K_\gamma$ acts transitively, $K_x$ is the stabilizer of any group element, hence the kernel of the action. Thus, $L_\gamma := K_\gamma/K_x$ is an abelian Lie group that acts freely and transitively on $H/H_x$ via $(h'G_\gamma)K_x \circ [h] := [h'h]$. Consequently, $L_\gamma$ is diffeomorphic to $H/H_x$.

Now, $L_\gamma$ acts pointwise properly and $\gamma$ restricts in range to a Lie path $\widetilde{\gamma}$ in $H/H_x$.

- If fact, let $h_i \in H$, such that $(h_i G_\gamma)K_x \circ [h] = [h_i h]$ is converging. Then $(h_i h)x = \iota[h_i h]$ is converging as well, i.e., $h_i h \to g$ for some subsequence and some $g \in G$. As $H$ is closed, we have $g \in H$, hence $h_i$ converges within $H$. Thus, at least a subsequence of $(h_i G_\gamma)K_x$ is converging.
- As $\gamma$ has infinite symmetry, Theorem 3.4 and Proposition 3.5 show[12] that the image of $\gamma$ lies in $Hx$. Using the initial manifold $\iota : H/H_x \longrightarrow M$, we can restrict $\gamma$ in range to an injective analytic mapping $\widetilde{\gamma} : [a, b] \longrightarrow H/H_x$. Indeed, $\widetilde{\gamma}$ is an analytic path by compactness. Of course, $\widetilde{\gamma}$ has infinite symmetry again.
- Since the action of $L_\gamma$ on $H/H_x$ fulfills the requirements of Proposition 3.14, we see that $\widetilde{\gamma}$ is a Lie path. This means, there is some $C$ in the Lie algebra of $L_\gamma$ and some analytic diffeomorphism $\boldsymbol{\varphi} : \mathbf{T} \longrightarrow [a, b)$, such that $[\widetilde{\gamma} \circ \boldsymbol{\varphi}](t) = e^{tC}\widetilde{\gamma}(a)$ for all $t \in \mathbf{T}$.

Finally, $\gamma$ is a Lie path.

- As the exponential mapping intertwines the projections to quotient Lie algebras with those to quotient Lie groups, we find some $A \in \mathfrak{h} \subseteq \mathfrak{g}$ such that $e^{tA}\iota[h] = \iota(e^{tC} \circ [h])$. Consequently,

$$[\gamma \circ \boldsymbol{\varphi}](t) \;=\; [\iota \circ \widetilde{\gamma} \circ \boldsymbol{\varphi}](t) \;=\; [\iota \circ e^{tC} \circ \widetilde{\gamma}](a) \;=\; e^{tA}\gamma(a)$$

for all $t \in \mathbf{T}$, and even for $t \in \overline{\mathbf{T}}$, as $\boldsymbol{\varphi}$ continues obviously. Since $\gamma$ is an analytic embedding, $\boldsymbol{\varphi}$ is an analytic diffeomorphism. **qed**

---

[11]The reference deals with the smooth category. However, the theory basically relies on the implicit function theorem that can be generalized to the analytic category.

[12]Actually, we still have to show that $\gamma(b)$ is in $Hx$. Thus, let $t_i \to b$ for $t_i < b$. Then $\gamma(t_i) = h_i x$ for some $h_i \in H_\gamma \subseteq H$. As $\gamma(t_i) \to \gamma(b)$, there is a subsequence of $h_i$ converging to $g \in G$. As $H$ is closed, even $g \in H$. Hence $\gamma(b) = hx \in Hx$.

## 4 Reparametrization Sets

The central object of our paper is the reparametrization set $\mathbf{P}_\gamma$ of all nontrivial essential reparametrizations of a path under the action of a Lie group. In Definition 2.6, we have transferred notions like analyticity or pointwise properness from the manifold to $\mathbf{P}_\gamma$. This has lead to the general notion of reparametrization sets. We have already seen how their properties lead to the symmetry classification of analytic paths. The remaining sections will now be devoted to the abstract theory of reparametrization sets. This way, we will derive the still missing results of the previous section.

In this section we will introduce the main notions and restrict ourselves on basic operations within reparametrization sets $\mathbf{P}$ like concatenation and inversion. Here, mostly all elements of $\mathbf{P}$ need not be standard; just homeomorphisms with non-discrete domains suffice. In particular, pointwise properness will become relevant in the subsequent sections only.

### 4.1 Extensions of Mappings

When we discussed the properties of essential reparametrizations, we have seen that the concatenation $\varrho_g \circ \varrho_h$ of two such mappings need not be such a reparametrization again; but, if it has non-discrete domain, then it coincides with $\varrho_{gh}$ thereon. Let us give this set a name:

**Definition 4.1** Let $\varrho$ be defined on some subset of $[a, b]$.

$$\mathbf{I}(\varrho) \quad \ldots \quad \text{union of all nontrivial intervals contained in } \operatorname{dom} \varrho$$

Recall that an interval is **nontrivial** iff it containes at least two elements.[13] Obviously, $\mathbf{I}(\varrho)$ is the non-discrete part of $\operatorname{dom} \varrho$. Moreover, it is clear that $\mathbf{I}(\varrho)$ always contains $\mathbf{I}(\sigma \circ \varrho)$; sloppily, the fat parts of the domain are getting smaller at most.

The concatenation of essential reparametrizations above refers not only to some subset of the domain, but also to the property that two functions coincide on this subset. So, let us take this as a motivation for

**Definition 4.2** Let $\varrho$ and $\sigma$ be defined on some subsets of $[a, b]$. Then:

$$\varrho \text{ **extends** } \sigma \quad \Longleftrightarrow \quad \varrho \text{ and } \sigma \text{ coincide on } \mathbf{I}(\sigma).$$

Note first that there may be many extensions of some $\sigma$. And second, there may be extensions that take different values on the discrete part of $\sigma$. It might even happen that the domain of $\varrho$ is smaller than that of $\sigma$. Nevertheless, we always have $\mathbf{I}(\varrho) \supseteq \mathbf{I}(\sigma)$ if $\varrho$ extends $\sigma$. Note finally, that the extension property is transitive, i.e., $\varrho$ extends $\tau$ as soon as $\varrho$ extends $\sigma$ and $\sigma$ extends $\tau$.

Transferred to essential reparametrizations, the definition above just says that $\varrho_{gh}$ extends $\varrho_g \circ \varrho_h$. More general, multiplicativity of some $\mathbf{P}$ means that for any two functions $\varrho_1$ and $\varrho_2$ in $\mathbf{P}$ having non-empty $\mathbf{I}(\varrho_1 \circ \varrho_2)$, there is some $\varrho_1 \bullet \varrho_2 \in \mathbf{P}$ extending $\varrho_1 \circ \varrho_2$. It appears now justified to call it an extension as there are at most finitely many isolated points in the domain of any concatenation of standard mappings. In fact, the domain of $\varrho_1 \circ \varrho_2$ equals $\varrho_2^{-1}(\operatorname{im} \varrho_2 \cap \operatorname{dom} \varrho_1)$, hence is homeomorphic to $\operatorname{im} \varrho_2 \cap \operatorname{dom} \varrho_1$. As now both domains and ranges of standard mappings consist of finitely many compact intervals, such an intersection does so as well. Only nontriviality of the intervals need no longer be given, but in any case there are at most finitely many isolated points.

Very important in the following is the property that the extension of mappings is compatible with the concatenation, at least if we are considering homeomorphisms. For this, let $\varrho$ and $\sigma$ as well as the corresponding indexed terms all be mappings between subsets of $[a, b]$.

---

[13] Moreover, for brevity we may sometimes denote the compact interval connecting $s$ and $t$ by $[s, t]$ even if $s > t$; in other words, we have $[s, t] \equiv [t, s]$.

**Proposition 4.1** Let $\sigma_2$ be a homeomorphism. Then we have

$$\varrho_i \text{ extends } \sigma_i \quad \text{for } i = 1, 2 \quad \Longrightarrow \quad \varrho_1 \circ \varrho_2 \text{ extends } \sigma_1 \circ \sigma_2$$

Before we prove this proposition, let us collect to further statements.

**Lemma 4.2** If $\sigma$ is a homeomorphism, then $\sigma\bigl(\mathbf{I}(\varrho \circ \sigma)\bigr)$ is contained in $\mathbf{I}(\varrho)$.

**Proof** Let $t \in \mathbf{I}(\varrho \circ \sigma) \subseteq \operatorname{dom} \sigma$, hence $t \in I \subseteq \operatorname{dom}(\varrho \circ \sigma)$ for some nontrivial interval $I$. Then $\sigma(I)$ is a nontrivial interval in $\sigma\bigl(\operatorname{dom}(\varrho \circ \sigma)\bigr) \equiv \operatorname{dom} \varrho \cap \operatorname{im} \sigma \subseteq \operatorname{dom} \varrho$. Consequently, $\sigma(t) \in \mathbf{I}(\varrho)$.
**qed**

**Corollary 4.3** Let $\varrho_1, \ldots, \varrho_k$ be homeomorphisms. Then

$$\mathbf{I}(\varrho_k \circ \ldots \circ \varrho_1) \text{ non-empty} \quad \Longrightarrow \quad \mathbf{I}(\varrho_j \circ \ldots \circ \varrho_i) \text{ non-empty for all } i \leq j$$

**Proof** We get the statement inductively. Just observe that we get the implication for
- $1 = i \leq j = k-1$, since clearly $\mathbf{I}(\varrho_{k-1} \circ \ldots \circ \varrho_1) \supseteq \mathbf{I}(\varrho_k \circ \ldots \circ \varrho_1)$;
- $2 = i \leq j = k$, since Lemma 4.2 implies $\mathbf{I}(\varrho_k \circ \ldots \circ \varrho_2) \supseteq \varrho_1\bigl(\mathbf{I}(\varrho_k \circ \ldots \circ \varrho_1)\bigr)$ with $\varrho_1$ being a homeomorphism.
**qed**

**Proof Proposition 4.1**
Let $I \subseteq \mathbf{I}(\sigma_1 \circ \sigma_2) \subseteq \mathbf{I}(\sigma_2)$ be a nontrivial interval.
- As $\varrho_2$ extends $\sigma_2$, we have $\varrho_2 = \sigma_2$ on $I$.
- Lemma 4.2 implies $\sigma_2(I) \subseteq \sigma_2\bigl(\mathbf{I}(\sigma_1 \circ \sigma_2)\bigr) \subseteq \mathbf{I}(\sigma_1)$, hence $\varrho_1 = \sigma_1$ on $\sigma_2(I)$.
- Altogether, we have $\varrho_1 \circ \varrho_2 = \varrho_1 \circ \sigma_2 = \sigma_1 \circ \sigma_2$ on $I$, hence also $I \subseteq \operatorname{dom}(\varrho_1 \circ \varrho_2)$.
**qed**

## 4.2 Definitions

Let us introduce a few further notions, complementing Definition 2.6.

**Definition 4.3** A set $\mathbf{P}$ of mappings between closed subsets of $[a, b]$ is called
- **homeomorphic** $\iff$ all elements of $\mathbf{P}$ are homeomorphisms;
- **non-discrete** $\iff$ all elements $\varrho$ of $\mathbf{P}$ have nontrivial $\mathbf{I}(\varrho)$;
- **oriented** $\iff$ all elements in $\mathbf{P}$ are positive or negative;
- **exponential** $\iff$ for any $\varrho \in \mathbf{P}$ with non-empty $\mathbf{I}(\varrho^k)$ there are $\varrho_i \in \mathbf{P}$ that extend $\varrho^i$ for all $i \leq k$;

Obviously, any standard $\mathbf{P}$ is also homeomorphic, non-discrete and oriented. Moreover,

**Lemma 4.4** If $\mathbf{P}$ is oriented, then $\mathbf{P} = \mathbf{P}^+ \sqcup \mathbf{P}^-$ with

$$\begin{aligned} \mathbf{P}^+ &:= \{\varrho \in \mathbf{P} \mid \varrho \text{ positive}\} \\ \mathbf{P}^- &:= \{\varrho \in \mathbf{P} \mid \varrho \text{ negative}\} \end{aligned}$$

**Definition 4.4** Let $\mathbf{P}$ and $\mathbf{Q}$ consist of mappings between subsets on $[a, b]$ and $[c, d]$, respectively, Then $\mathbf{P}$ and $\mathbf{Q}$ are **isomorphic** iff there is a homeomorphism $\varphi : [a, b] \longrightarrow [c, d]$, such that

$$\varphi \circ \mathbf{P} = \mathbf{Q} \circ \varphi.$$

We leave it as a trivial exercise to the reader to show that all properties of Definitions 2.6 and 4.3 are preserved under isomorphy. The same will apply to those in Definition 5.2 and 10.2 to come.

## 4.3 Multiplicativity

As we have learned already from the prototypical examples in Subsection 2.4, the concatenation of mappings is rather subtle. We have abstractly introduced the new multiplication $\bullet$ extending $\circ$ provided we are given nontrivial domains. This way, we can guarantee that the product of two standard elements is standard again (if existing). Let us now see that we can generalize this concept easily from two to finitely many factors. Again, the main lesson will be that $\bullet$-products are just the same as the respective $\circ$-concatenations as long as we are on the non-discrete domains of $\varrho_k \circ \ldots \circ \varrho_1$.

**Proposition 4.5** Let **P** be homeomorphic, multiplicative and analytic.
   Then we have for all $\varrho_k, \ldots, \varrho_1 \in \mathbf{P}$ with non-empty $\mathbf{I}(\varrho_k \circ \ldots \circ \varrho_1)$:
   - There is a unique $\varrho_k \bullet \ldots \bullet \varrho_1 \in \mathbf{P}$ extending $\varrho_k \circ \ldots \circ \varrho_1$.
   - In particular, $\bullet$ is associative and
   $$\varrho_k * \ldots * \varrho_1 \quad \text{extends} \quad \varrho_k \circ \ldots \circ \varrho_1 \,.$$
   Here, $*$ may be replaced by $\circ$ or $\bullet$ at any position freely, with $\bullet$ being a higher-order[14] operation than $\circ$.

Note that associativity is usually not guaranteed for all possible choices of the maps $\varrho_i$. Indeed, the domain assumption is crucial. For instance, let $[a,b] = [-1,1]$, let $\varrho$ be the unit right-shift, i.e., $\varrho(t) := t+1$, and let $\mathbf{P} := \{\varrho^{-1}, \mathbf{1}, \varrho\}$. We define $\varrho \bullet \varrho^{-1}$ as well as $\varrho^{-1} \bullet \varrho$ to be $\mathbf{1}$. Moreover, $\varrho \bullet \mathbf{1} := \varrho$, etc. One immediately checks that **P** is multiplicative and analytic. Observe, however, that $\varrho \bullet \varrho$ is not defined. Indeed, the domain of $\varrho \circ \varrho$ is just $\{-1\}$, whence there is no "need" for $\varrho \bullet \varrho$ to exist. In particular, $(\varrho \bullet \varrho) \bullet \varrho^{-1}$ is not defined, although $\varrho \bullet (\varrho \bullet \varrho^{-1}) = \varrho \bullet \mathbf{1} = \varrho$ is. Note that the domain of $\varrho \circ \varrho \circ \varrho^{-1}$ is $\{0\}$, hence discrete. This also shows that **P** does not form a groupoid, in general. – Now to the proof of the proposition.

**Proof** Let $1 \leq l < k$. Then, by induction,
$$\varrho_k * \ldots * \varrho_{l+1} \quad \text{extends} \quad \varrho_k \circ \ldots \circ \varrho_{l+1}$$
$$\varrho_l * \ldots * \varrho_1 \quad \text{extends} \quad \varrho_l \circ \ldots \circ \varrho_1$$

Now, since $\varrho_l \circ \ldots \circ \varrho_1$ is a homeomorphism, Proposition 4.1 implies that
$$(\varrho_k * \ldots * \varrho_{l+1}) \circ (\varrho_l * \ldots * \varrho_1) \quad \text{extends} \quad \varrho_k \circ \ldots \circ \varrho_1 \,.$$

In particular, by multiplicativity, there is some
$$(\varrho_k \bullet \ldots \bullet \varrho_{l+1}) \bullet (\varrho_l \bullet \ldots \bullet \varrho_1) \in \mathbf{P} \quad \text{extending} \quad \varrho_k \circ \ldots \circ \varrho_1.$$

Analyticity implies uniqueness, hence associativity. **qed**

As analyticity has been used only to prove uniqueness and associativity, we get

**Corollary 4.6** If **P** is homeomorphic and multiplicative, then **P** is also exponential.

Again an immediate consequence of the proposition above is

**Corollary 4.7** Let **P** be homeomorphic, multiplicative and analytic.
   If $\mathbf{I}(\varrho_k * \ldots * \varrho_1)$ is (well defined and) non-empty, then any function one gets from $\varrho_k * \ldots * \varrho_1$ by replacing one or more $\circ$ by $\bullet$ extends that $\varrho_k * \ldots * \varrho_1$.

For example, we have

---

[14]This means that, e.g., $\varrho_1 \bullet \varrho_2 \circ \varrho_3$ always reads $(\varrho_1 \bullet \varrho_2) \circ \varrho_3$, but not necessarily equals $\varrho_1 \bullet (\varrho_2 \circ \varrho_3)$. In fact, $\bullet$ is only an operation on **P**, and $\varrho_2 \circ \varrho_3$ might fail to be in **P**, although both $\varrho_2$ and $\varrho_3$ are in **P**.

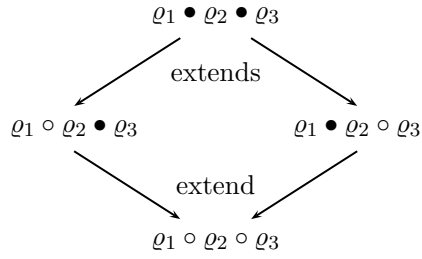

## 4.4 Involutivity

The inversion of standard elements is less problematic than their multiplication. Nevertheless, the concatenation of $\varrho^{-1}$ and $\varrho$ is usually not $\mathbf{1}$ as discussed above. It is just $\mathbf{1}$ restricted to the domain of $\varrho$. Here, however, the situation is quite comfortable. Assuming multiplicativity and that $\mathrm{dom}\,\varrho$ is non-discrete, any $\varrho^{-1}\bullet\varrho$ is in $\mathbf{P}$ again and equals $\mathbf{1}$ on $\mathrm{dom}\,\varrho$. Now, analyticity and unitality give $\varrho^{-1}\bullet\varrho = \mathbf{1}$. This makes the following lemma obvious.

**Lemma 4.8** Let $\mathbf{P}$ be homeomorphic, non-discrete, multiplicative, involutive, unital, analytic. Then we have
$$\varrho^{-1}\bullet\varrho \;=\; \mathbf{1} \;=\; \varrho\bullet\varrho^{-1} \qquad \text{for all } \varrho\in\mathbf{P}$$

**Corollary 4.9** Let $\mathbf{P}$ be homeomorphic, non-discrete, multiplicative, involutive, unital, analytic. Then we have for all $\varrho_1, \varrho_2 \in \mathbf{P}$ with $\varrho_1 \bullet \varrho_2 \in \mathbf{P}$:
$$\varrho_1 \bullet \varrho_2 = \mathbf{1} \iff \varrho_1 = \varrho_2^{-1}$$

**Proof** Since $\mathrm{dom}[\varrho_1 \circ \varrho_2 \circ \varrho_2^{-1}] \cong \mathrm{im}\,\varrho_2^{-1} \cap \mathrm{dom}[\varrho_1 \circ \varrho_2] = \mathrm{dom}\,\varrho_2 \cap \mathrm{dom}[\varrho_1\circ\varrho_2] = \mathrm{dom}[\varrho_1 \circ \varrho_2]$ contains a nontrivial interval, we have
$$\varrho_1 \;=\; \varrho_1\bullet(\varrho_2\bullet\varrho_2^{-1}) \;=\; (\varrho_1\bullet\varrho_2)\bullet\varrho_2^{-1} \;=\; \mathbf{1}\bullet\varrho_2^{-1} \;=\; \varrho_2^{-1}.$$
**qed**

**Lemma 4.10** Let $\mathbf{P}$ be homeomorphic, non-discrete, multiplicative, involutive, unital, analytic. If now $\mathbf{I}(\varrho_k \circ \ldots \circ \varrho_1)$ is non-empty, then
$$[\varrho_k \bullet \ldots \bullet \varrho_1]^{-1} \;=\; \varrho_1^{-1} \bullet \ldots \bullet \varrho_k^{-1}.$$

**Proof** If $I \subseteq \mathbf{I}(\varrho_k\circ\ldots\circ\varrho_1)$ is a nontrivial interval, then $[\varrho_k\circ\ldots\circ\varrho_1](I)$ is a nontrivial interval contained in $\mathrm{im}[\varrho_k\circ\ldots\circ\varrho_1] = \mathrm{dom}[\varrho_k\circ\ldots\circ\varrho_1]^{-1}$. Now use Proposition 4.5 and Corollary 4.9. **qed**

Altogether, we see that all group-like stuff (concatenation, inversion, unit element) transfers to reparametrization sets as long as we stick to intervals contained in the domain of the product and/or inverses. In particular, we may freely replace $\circ$ by $\bullet$ thereon.

**Lemma 4.11** Let $\mathbf{P}$ be homeomorphic, non-discrete, multiplicative, involutive, unital, analytic. Then we have for all $\varrho,\sigma \in \mathbf{P}$ with nonempty $\mathbf{I}(\varrho\circ\sigma\circ\varrho^{-1})$
$$\varrho\bullet\sigma\bullet\varrho^{-1} = \mathbf{1} \iff \sigma = \mathbf{1}$$

**Proof** ($\Longrightarrow$ only) If $I \subseteq \mathbf{I}(\varrho\circ\sigma\circ\varrho^{-1})$ is an interval, then $I \subseteq \mathrm{dom}\,\varrho^{-1}$, hence $\varrho^{-1}(I)$ is an interval again, by homeomorphy. Now, $[\varrho\circ\sigma\circ\varrho^{-1}](t) = t$ for $t \in I$ implies $\sigma(\varrho^{-1}(t)) = \varrho^{-1}(t)$. Consequently, $\sigma$ is $\mathbf{1}$ on the interval $\varrho^{-1}(I)$, whence everywhere by analyticity. **qed**

## 4.5 $\mathbb{Z}_2$-Grading

**Lemma 4.12** Let **P** be homeomorphic, multiplicative, analytic and oriented.
The orientation induces a $\mathbb{Z}_2$ grading w.r.t. the $\bullet$-multiplication. More precisely,

$$\mathbf{P}^+ \bullet \mathbf{P}^+ \text{ and } \mathbf{P}^- \bullet \mathbf{P}^- \quad \text{are contained in} \quad \mathbf{P}^+;$$
$$\mathbf{P}^+ \bullet \mathbf{P}^- \text{ and } \mathbf{P}^- \bullet \mathbf{P}^+ \quad \text{are contained in} \quad \mathbf{P}^-.$$

**Proof** If $\varrho = \varrho_1 \bullet \varrho_2$, then $\varrho_1 \circ \varrho_2$ is defined on a nontrivial interval. Now, the statement is obvious since, e.g., the concatenation of a decreasing and an increasing function is decreasing. **qed**

In particular, the lemma above implies that any property of **P** we have defined so far, is inherited by $\mathbf{P}^+$. In particular, if **P** is a reparametrization set, then $\mathbf{P}^+$ is a reparametrization set.

## 5 Perfect Maps

As we have already seen in the introduction, pointwise properness of the acting transformation group will turn out an important assumption, when we study the intersection behaviour of an analytic path with its translates. Transferred to the level of standard maps, pointwise properness states that different points cannot converge to the same point if we act on them by the same sequence of standard maps. This is obviously given for any prototypical reparametrization set as we have seen in Subsection 2.4. In fact, there any shift preserves the distance between two points (unless one or both are pushed out of the interval). Of course, such a simple description is, in general, not available yet, just as we can by no means guarantee that each $\varrho \in \mathbf{P}$ preserves distance. Nevertheless, pointwise properness will strongly restrict the possible behaviour of $t \in \text{dom}\,\varrho$, when one successively applies $\varrho$. Indeed, as we will see, given any point $t$ in any interval $I \subseteq \text{dom}\,\varrho$, either a positive $\varrho$ fixes $t$ or some $\varrho^k$ pushes it out of $I$ for finite $k$ (e.g., as $t$ is not contained in $\text{dom}\,\varrho^k$). This means, in particular, that $(\varrho^k(t))$ cannot converge for positive standard $\varrho$ unless $t$ is a fixed point. This, on the other hand, will imply that such a $\varrho$ cannot have full domain and non-full image. In particular, this excludes some of the topological types of standard mappings to appear within reparametrization sets. The remaining types will called perfect.

In this section, we will first see in which respect pointwise properness is responsible for the rise of such perfect maps. Then we study some domain issues and end with the behaviour of perfect mappings under inversion and concatenations.

### 5.1 Appearance of Perfect Maps

**Proposition 5.1** Let **P** be homeomorphic, exponential, positive, pointwise proper.
Then we have for all $\varrho \in \mathbf{P}$, all nontrivial compact intervals $I \subseteq \text{dom}\,\varrho$ and all $t \in I$.

$$\varrho(t) = t \iff \varrho^k(t) \in I \quad \text{for all } k$$

We will deduce the proposition from

**Lemma 5.2** Let **P** be homeomorphic.
Let $\varrho \in \mathbf{P}$, let $n \in \mathbb{N}$, and let $I \subseteq \text{dom}\,\varrho$ be an interval. Then:[15]

$$\varrho^k(t) \in I \text{ for all } k \leq n \implies [t, \varrho(t)] \subseteq \text{dom}\,\varrho^k \text{ for all } k \leq n$$

**Proof** Let $J := [t, \varrho(t)]$. Since $J \subseteq \text{dom}\,\varrho$ and since $\varrho$ is continuous and monotonous on intervals, $\varrho(J)$ equals $[\varrho(t), \varrho^2(t)] \subseteq I$. Thus, $J \subseteq \varrho^{-1}(\text{dom}\,\varrho \cap \text{im}\,\varrho) = \text{dom}(\varrho \circ \varrho)$. Inductively, $J \subseteq \text{dom}\,\varrho^k$. **qed**

---
[15]In this subsection, by $[s, t]$ we may understand the interval spanned by $s$ and $t$, for both cases $s \geq t$ and $s \leq t$.

**Proof Proposition 5.1**

Let us argue indirectly assuming $\varrho(t) \neq t$. By exponentiality, there are $\varrho_k \in \mathbf{P}$ extending $\varrho^k$. By Lemma 5.2, both functions coincide at least on $[t, \varrho(t)]$. Now, by positivity, $\varrho_k(t) \equiv \varrho^k(t)$ is monotonous in $I$, hence converging to some $s \in I$. Obviously, $\varrho(s) = s$, but now both $\varrho_k(t)$ and $\varrho_k(\varrho(t))$ converge to $s$. This contradicts pointwise properness.

**qed**

**Proposition 5.3** Let $\mathbf{P}$ be standard, exponential, involutive, pointwise proper. Then we have for all $\varrho \in \mathbf{P}$
$$\operatorname{dom} \varrho = [a,b] \iff \operatorname{im} \varrho = [a,b]$$

**Proof** It suffices to show the $\Longrightarrow$-direction. In fact, the other direction is then provided by involutivity and the fact that inversion of maps exchanges domain and image. Thus, let the domain of $\varrho$ equal $[a,b]$ and recall that each standard map is oriented.

- If $\varrho$ is positive, Proposition 5.1 gives $\varrho(t) = t$ for all $t \in [a,b]$. Hence, $\operatorname{im} \varrho = [a,b]$.
- If $\varrho$ is negative, observe that $\varrho \circ \varrho$ has domain $[a,b]$ again, hence $\varrho \bullet \varrho \in \mathbf{P}$ exists. It even coincides with $\varrho \circ \varrho$ on full non-discrete $[a,b]$. Moreover, obviously, it is increasing, hence positive. Hence, by the preceding point, $\operatorname{im} \varrho \supseteq \operatorname{im} \varrho \circ \varrho = \operatorname{im} \varrho \bullet \varrho$ equals $[a,b]$.

**qed**

As any reparametrization set is standard, exponential, involutive and pointwise proper, we see that indeed only[16] the topological types in the upper three lines of Table 1 on page 14 may occur when we study essential reparametrizations. This important notion is summarized in

**Definition 5.1** A standard map $\varrho$ is called
$$\textbf{perfect} \iff \big(\operatorname{dom} \varrho = [a,b] \iff \operatorname{im} \varrho = [a,b]\big)$$

Complementing Definitions 2.6 and 4.3, we set

**Definition 5.2** A set $\mathbf{P}$ of mappings between closed subsets of $[a,b]$ is called
- **perfect** $\iff$ all elements of $\mathbf{P}$ are perfect maps.

This implies immediately

**Proposition 5.4** Let $\mathbf{P}$ be exponential, involutive, pointwise proper. Then
$$\mathbf{P} \text{ standard} \iff \mathbf{P} \text{ perfect.}$$
In particular, any reparametrization set is a motion set.

Although some of the statements below will hold also for standard mappings, we will now focus on perfect mappings only. In particular, we introduce

**Definition 5.3** A set $\mathbf{P}$ is called **motion set** iff it is a

perfect, unital, analytic, involutive, multiplicative, pointwise proper

set of mappings between closed subsets of $[a,b]$. In particular, we use the notions

| | | |
|---:|:---:|:---|
| **shift** | ... | positive element in a motion set |
| **reflection** | ... | negative element in a motion set |

---

[16]Recall the comments around that table concerning the notion "type". In particular, note that all the situations can occur that arise from those sketched in the table by flipping left and right or exchanging domain and image.

## 5.2 Domains of Perfect Maps

It is clear that the domain (as well as the image) of a perfect map is the union of one or two compact intervals each containing $a$ or $b$. Such sets shall be called margins.

**Definition 5.4** A subset $K \subseteq [a, b]$ is called

$$\begin{aligned}\textbf{left margin} &\iff K \text{ is a compact interval that contains } a. \\ \textbf{right margin} &\iff K \text{ is a compact interval that contains } b.\end{aligned}$$

The only interval being both a left and a right margin is the full interval $[a, b]$. Now, the maximal left margin in the domain of $\varrho$ will be called left domain. As we will need this notion also for more general mappings (usually concatenations of perfect mappings), we will fix that in

**Definition 5.5** Let $\varrho$ be some map between subsets of $[a, b]$. Then we call

$$\begin{aligned}\mathbf{L}(\varrho) &:= \text{ connected component of } \mathbf{I}(\varrho) \text{ containing } a \\ &\qquad\qquad\qquad\qquad\qquad\qquad \ldots \textbf{ left domain } \text{of } \varrho \\ \mathbf{R}(\varrho) &:= \text{ connected component of } \mathbf{I}(\varrho) \text{ containing } b \\ &\qquad\qquad\qquad\qquad\qquad\qquad \ldots \textbf{ right domain } \text{of } \varrho\end{aligned}$$

If there is no such component, the respective left/right domain is assumed empty.

Clearly, for perfect $\varrho$, its left domain $\mathbf{L}(\varrho)$ is nothing but the maximal nontrivial interval contained in $\mathrm{dom}\,\varrho$ that includes $a$, unless there exists no such interval. As for $\mathbf{I}$, we see that $\mathbf{L}(\varrho)$ always contains $\mathbf{L}(\sigma \circ \varrho)$; the same relation applies to $\mathbf{R}$. For completeness, we list the explicit expressions for the margins of perfect maps.

**Lemma 5.5** If $\varrho$ is perfect and positive, then

$$\begin{aligned}\mathbf{L}(\varrho) &= [a, \varrho^{-1}(b)] &&(\text{if } a \in \mathrm{dom}\,\varrho, \text{ otherwise } \varnothing) \\ \mathbf{R}(\varrho) &= [\varrho^{-1}(a), b] &&(\text{if } b \in \mathrm{dom}\,\varrho, \text{ otherwise } \varnothing)\end{aligned}$$

If $\varrho$ is perfect and negative, then

$$\begin{aligned}\mathbf{L}(\varrho) &= [a, \varrho^{-1}(a)] &&(\text{if } a \in \mathrm{dom}\,\varrho, \text{ otherwise } \varnothing) \\ \mathbf{R}(\varrho) &= [\varrho^{-1}(b), b] &&(\text{if } b \in \mathrm{dom}\,\varrho, \text{ otherwise } \varnothing)\end{aligned}$$

We would like to emphasize that perfect mappings exhibit a crucial property that is not shared by just standard mappings: a perfect mapping is always defined in at least one of the boundary points $a$ and $b$. It might now be tempting to use this for decomposing any perfect $\mathbf{P}$ into two sets $\mathbf{P}_a$ and $\mathbf{P}_b$. Indeed, this is possible for the decomposition $\mathbf{P} = \mathbf{P}^+ \sqcup \mathbf{P}^-$ into the positive and the negative part. Here, however, we may have perfect mappings that are defined on both $a$ and $b$. The trivial example is the identity, but often there are more elements in the intersection of $\mathbf{P}_a$ and $\mathbf{P}_b$. This is, e.g., the case for the shifts on $S^1$, where the interval $I$ covers more than half of $S^1$. Nevertheless, $\mathbf{P}_a$ and $\mathbf{P}_b$ will span any perfect $\mathbf{P}$. Moreover, the statements on $\mathbf{P}_a$ can usually be transferred immediately to $\mathbf{P}_b$ and vice versa. Therefore, we will usually restrict ourselves to the case of $\mathbf{P}_a$ only. But first, let us define these notions in general.

**Definition 5.6** Let $\mathbf{P}$ consist of mappings between subsets of $[a, b]$. We set

$$\mathbf{P}_t := \{\varrho \in \mathbf{P} \mid t \in \mathrm{dom}\,\varrho\}$$

and

$$\mathbf{P}(t) := \{\varrho(t) \mid \varrho \in \mathbf{P}_t\} \equiv \{\varrho(t) \mid \varrho \in \mathbf{P},\ t \in \mathrm{dom}\,\varrho\}.$$

**Lemma 5.6** If $\mathbf{P}$ is perfect, then $\mathbf{P} = \mathbf{P}_a \cup \mathbf{P}_b$.

## 5.3 Operations involving Perfect Maps

The homeomorphisms on a topological space always form a group. The perfect maps we are now focussing on, are homeomorphisms, indeed. However, they are homeomorphisms that may have different domains or images. This will give us restrictions on "group" operations. The first one, namely the inversion, is simple:

**Lemma 5.7** *The inverse of any perfect map is perfect.*

The composition of perfect maps, however, is more tricky. Observe, first of all, that even the composition of a map $\varrho$ and its inverse is only the identity *on the domain of* $\varrho^{-1}$ which by no means has to be full $[a,b]$. Indeed, if the image $\varrho^{-1}$ is not of full $[a,b]$, then $\varrho \circ \varrho^{-1} \equiv \mathbf{1}_{\operatorname{im}\varrho}$ is even not standard. Nevertheless, it can be extended to a standard, even perfect map. This extendability will be the crucial point in the following. Indeed, recall from Proposition 2.8 that any two essential reparametrizations $\varrho_{g_1}$ and $\varrho_{g_2}$ fulfill $\varrho_{g_1} \circ \varrho_{g_2} = \varrho_{g_1 g_2}$ on the non-discrete part of $\varrho$. Thus, what we now have to study comprises two issues: under which conditions does the domain of the concatenation of perfect maps contain a non-discrete part and when can this composition be continued to a map being perfect again?

As the extendability problem is minor as soon as we are given multiplicativity, let us focus here on the domain issue. Considering $\varrho_i$ as a function from $\operatorname{dom}\varrho_i \subseteq [a,b]$ to $[a,b]$, we can define $\varrho_1 \circ \varrho_2$ in $t \in [a,b]$ iff $\varrho_2(t) \in \operatorname{dom}\varrho_1$, i.e., $t \in \varrho_2^{-1}(\operatorname{dom}\varrho_1)$. In other words,

$$\operatorname{dom}\varrho_1 \circ \varrho_2 \;=\; \varrho_2^{-1}(\operatorname{dom}\varrho_1) \;\equiv\; \varrho_2^{-1}(\operatorname{dom}\varrho_1 \cap \operatorname{im}\varrho_2).$$

As $\varrho_2^{-1}$ is a homeomorphism on $\operatorname{dom}\varrho_1 \cap \operatorname{im}\varrho_2$, the topology of $\operatorname{dom}\varrho_1 \circ \varrho_2$ is completely given by that of $\operatorname{dom}\varrho_1 \cap \operatorname{im}\varrho_2$. One easily checks that the latter set always consists of up to two margins plus a possibly trivial interval in the "middle". The twelve possible topological types are as follows:

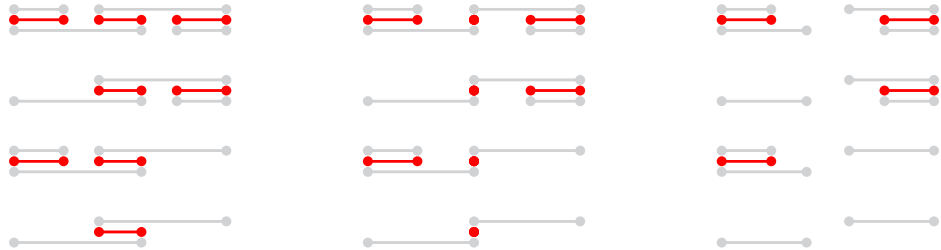

In particular, it may happen that there is a single or isolated dot. For instance, consider the right shift $\varrho$ by 1 on $\mathbb{R}$, restricted to the interval $[0,2]$. Then $\operatorname{dom}\varrho = [0,1]$ and $\operatorname{im}\varrho = [1,2]$ giving $\operatorname{dom}\varrho \cap \operatorname{im}\varrho = \{1\}$. This now implies $\operatorname{dom}\varrho \circ \varrho = \varrho^{-1}(\{1\}) = \{0\}$ and $\operatorname{im}\varrho \circ \varrho = \varrho(\{1\}) = \{2\}$. Observe that the singleton appears for the case that the boundary point 0 is mapped to the boundary point 2. Indeed, for positive perfect mappings, this is a very general feature. One shows easily that here $\operatorname{dom}\varrho_1 \circ \varrho_2$ has a discrete part iff $\varrho_1 \circ \varrho_2$ maps $a$ to $b$ or vice versa. For multiple concatenations the structure is similar: The singular part of the domain of $\varrho_k \circ \ldots \circ \varrho_1$ is given by all that $t$ that pass both $a$ and $b$ when successively mapped by the $\varrho_i$. In turn, this means if $t$ is never touching $a$ or never touching $b$, it will be in the non-discrete part of the domain. This motivates

**Definition 5.7** Let $\varrho_1, \ldots, \varrho_k$ be perfect and positive.
  Then $t \in [a,b]$ is **snaking along** $\varrho_k \circ \ldots \circ \varrho_1$ iff
  - $t$ is in the domain of $\varrho_k \circ \ldots \circ \varrho_1$.
  - $[\varrho_i \circ \ldots \circ \varrho_1](t)$ never equals $a$ or never equals $b$, for any $i$.

We often simply say "snaking", if the perfect maps $\varrho_1, \ldots, \varrho_k$ are clear.

**Proposition 5.8** Let $\varrho_1, \ldots, \varrho_k$ be perfect and positive. Then we have

$$\mathbf{I}(\varrho_k \circ \ldots \circ \varrho_1) = \{t \mid t \text{ is snaking along } \varrho_k \circ \ldots \circ \varrho_1\}$$

Since the domain of $\varrho_k \circ \ldots \circ \varrho_1$ is closed, the above proposition is an immediate consequence of

**Lemma 5.9** Let $\varrho_1, \ldots, \varrho_k$ be perfect and positive. Moreover, let $s = [\varrho_k \circ \ldots \circ \varrho_1](t)$.
Then the following statements are equivalent:
1. $\mathrm{dom}(\varrho_k \circ \ldots \circ \varrho_1)$ contains no interval $[t, t + \varepsilon]$.
2. $\mathrm{im}(\varrho_k \circ \ldots \circ \varrho_1)$ contains no interval $[s, s + \varepsilon]$.
3. $b = [\varrho_i \circ \ldots \circ \varrho_1](t)$ for some $0 \leq i \leq k$.

**Proof** Write shortly $\varrho_{i \ldots j} := \varrho_i \circ \ldots \circ \varrho_j$.
1. $\implies$ 2. $\varrho_{1 \ldots k}$ is strictly monotonous on intervals.
3. $\implies$ 1. If $\mathrm{dom}\, \varrho_{k \ldots 1}$ contains $[t, t+\varepsilon]$, then $\mathrm{dom}\, \varrho_{i \ldots 1}$ does so as well. By monotonicity, $\varrho_{i \ldots 1}$ maps $[t, t+\varepsilon]$ to $[\varrho_{i \ldots 1}(t), \varrho_{i \ldots 1}(t)+\varepsilon'] \subseteq [a, b]$. This contradicts $\varrho_{i \ldots 1}(t) = b$.
2. $\implies$ 3. The statement is trivial for $k = 1$. Therefore, assume $k > 1$ and $\varrho_{i \ldots 1}(t) < b$ for all $i$. By induction, $\mathrm{im}\, \varrho_{k-1 \ldots 1}$ contains some $[r, r + \varepsilon]$ with $r := \varrho_{k-1 \ldots 1}(t)$. Since, however, neither $r$ nor $\varrho_k(r)$ equals $b$, there is an $\varepsilon'$ with $[r, r + \varepsilon'] \subseteq \mathrm{dom}\, \varrho_k \cap \mathrm{im}\, \varrho_{k-1 \ldots 1}$. As $\varrho_k$ is an increasing homeomorphism there, this is a contradiction. **qed**

Completely analogously, we can prove the respective statement for $a$ instead of $b$. Here, we only have to replace the "right-bound" intervals $[t, t + \varepsilon]$ by the "left-bound" ones $[t - \varepsilon, t]$.

For perfect maps, we may now transfer Proposition 5.8 to the $\bullet$-product using Proposition 4.5.

**Corollary 5.10 Snaking Lemma**
Let $\mathbf{P}$ be perfect, multiplicative, positive and analytic, and let $\varrho_1, \ldots, \varrho_k \in \mathbf{P}$.
Moreover, let there be some element in $[a, b]$ that is snaking along $\varrho_k \circ \ldots \circ \varrho_1$.
Then $\varrho_k \bullet \ldots \bullet \varrho_1 \in \mathbf{P}$ is well defined and coincides with $\varrho_k \circ \ldots \circ \varrho_1$ on all $t$ snaking along $\varrho_k \circ \ldots \circ \varrho_1$.

Of course, the similar result holds for any $\varrho_k * \ldots * \varrho_1$. Moreover, in particular, $\mathbf{L}(\varrho_k \circ \ldots \circ \varrho_1)$ is non-empty and contained in $\mathbf{L}(\varrho_k \bullet \ldots \bullet \varrho_1)$, if $a$ is snaking.

We close with some useful properties.

**Lemma 5.11** Let $\mathbf{P}$ be perfect and positive. Then we have for all $\varrho, \sigma \in \mathbf{P}_a$

$$\begin{aligned}
\sigma(a) \leq \varrho(a) &\iff \varrho(a) \in [\sigma(a), b] \equiv \mathbf{R}(\sigma^{-1}) \\
&\iff \mathbf{L}(\varrho) = \mathbf{L}(\sigma^{-1} \circ \varrho) \\
&\iff \mathbf{L}(\varrho) = \mathbf{L}(\sigma \circ \sigma^{-1} \circ \varrho) \\
&\iff \varrho = \sigma \circ \sigma^{-1} \circ \varrho \quad \text{on } \mathbf{L}(\varrho) \\
&\implies \sigma^{-1}(\varrho(a)) \in \mathbf{L}(\sigma)
\end{aligned}$$

If $\mathbf{P}$ is, moreover, multiplicative, involutive and analytic, $\varrho(a) \geq \sigma(a)$ implies
1. $\sigma^{-1} \bullet \varrho$ and $\sigma \bullet \sigma^{-1} \bullet \varrho$ are a well-defined elements in $\mathbf{P}_a$.
2. $\sigma^{-1} \bullet \varrho$ coincides with $\sigma^{-1} \circ \varrho$ on $\mathbf{L}(\varrho)$.
3. $\sigma \bullet \sigma^{-1} \bullet \varrho$ coincides with $\varrho$.

**Proof** Use $\mathbf{L}(\sigma^{-1} \circ \varrho) \equiv \mathbf{L}(\sigma^{-1} \circ \sigma \circ \sigma^{-1} \circ \varrho) \subseteq \mathbf{L}(\sigma \circ \sigma^{-1} \circ \varrho) \subseteq \mathbf{L}(\sigma^{-1} \circ \varrho) \subseteq \mathbf{L}(\varrho)$ to see the third equivalence. The first and the final one are trivial. For the second one, use

$$\begin{aligned}
\varrho(a) \in [\sigma(a), b] &\iff \varrho(\mathbf{L}(\varrho)) = [\varrho(a), b] \subseteq [\sigma(a), b] = \sigma(\mathbf{L}(\sigma)) \subseteq \mathrm{dom}\, \sigma^{-1} \\
&\iff \mathbf{L}(\varrho) \subseteq \mathrm{dom}(\sigma^{-1} \circ \varrho) \iff \mathbf{L}(\varrho) \subseteq \mathbf{L}(\sigma^{-1} \circ \varrho)
\end{aligned}$$

and again $\mathbf{L}(\sigma^{-1} \circ \varrho) \subseteq \mathbf{L}(\varrho)$. To show $\sigma^{-1}(\varrho(a)) \in \mathbf{L}(\sigma)$, use $\sigma^{-1}(\mathbf{R}(\sigma^{-1})) = \mathbf{L}(\sigma)$. The final implications follow immediately from the definitions. **qed**

**Corollary 5.12** Let **P** be perfect, multiplicative, analytic and positive.
Then we have for all $\varrho, \varrho_1, \varrho_2 \in \mathbf{P}_a$ with $\varrho_1 \geq \varrho_2$

$$\varrho_2^{-1} \circ \varrho_1 \geq \varrho \text{ in } a \implies \varrho(a) \in \mathbf{L}(\varrho_2) \text{ and } \varrho_1 \geq \varrho_2 \circ \varrho \geq \varrho_2 \text{ in } a$$

**Proof** Define $\sigma := \varrho_2^{-1} \circ \varrho_1$. By Lemma 5.11, $\mathbf{L}(\sigma) = \mathbf{L}(\varrho_1)$, whence $\sigma$ maps $[a, \varrho_1^{-1}(b)]$ to

$$[\sigma(a), \sigma(\varrho_1^{-1}(b))] = [\sigma(a), \varrho_2^{-1}(b)] \subseteq [a, \varrho_2^{-1}(b)] = \mathbf{L}(\varrho_2).$$

Since, by assumption, $\sigma(a) \geq \varrho(a) \geq a$, we have $\varrho(a) \in \mathbf{L}(\varrho_2)$ and the proof. **qed**

Finally, we would like to emphasize that the snaking lemma can be generalized to concatenations of any perfect maps, i.e., the positivity condition is not necessary. However, the definition of being snaking along has to be modified. To see this, let us for simplicity assume that $[a, b] = [0, 2]$ and consider the flipping reflection $\varsigma$ defined by $\varsigma(t) = 2 - t$ on full $[0, 2]$. Of course, $\varsigma \circ \varsigma = \mathbf{1}$ is also defined on the full interval $[0, 2]$ and it is positive there. In particular, we have $\mathbf{I}(\varsigma \circ \varsigma) = [0, 2]$. Nevertheless, if we applied our definition in the positive case, 0 is not snaking along $\varsigma \circ \varsigma$; in fact, $\varsigma(0) = 2$. On the other hand, let $\varrho$ be the unit right shift, i.e., $\varrho(t) := t + 1$. It is defined on $[0, 1]$ only. It is easy to check that $\varrho^{-1} \circ \varsigma \circ \varrho$ maps 0 via 1 and 1 to 0. Thus, it should be snaking. Nevertheless, one easily sees that $\varsigma \circ \varrho$ maps $t \in [0, 1]$ to $1 - t$ which is in the domain of $\varrho^{-1}$ iff $t = 0$. Hence, the domain of $\varrho^{-1} \circ \varsigma \circ \varrho$ is just 0 having no non-discrete part at all.

Now, the main idea how to modify the snaking definition is as follows. To be not snaking means in the positive case that both $a$ and $b$ are touched by the concatenation. This is to be replaced by the condition that it is forbidden that between touching two end points (of the interval) there is an odd number of reflections if both end points coincide or there is an even number of reflections if both end points differ. In the positive-only case, the number of reflections is always zero, hence even, thus it is not allowed that both end points are passed by $t$. This is precisely the definition above. To check the non-positive case, reconsider the example $\varrho^{-1} \circ \varsigma \circ \varrho$ again. There, the endpoint 0 is reached again after two positive and one negative mapping. Hence, 0 is not snaking. On the other hand, $\varrho \circ \varsigma \circ \varrho$ maps $t$ to $2 - t$. In particular, 0 is mapped to 2. This now shows that 0 is in the non-discrete part. Even more, we see that any point in dom $\varrho$ is in the non-discrete part. Indeed, $\varrho \circ \varsigma \circ \varrho = \varsigma$ on the domain of $\varrho$, hence $\varrho \bullet \varsigma \bullet \varrho = \varsigma$.

Let us now state the precise extended definition. The snaking lemma applies then verbatim; just the positivity assumption is dropped. The proof of the generalized snaking lemma is left to the reader.

**Definition 5.8** Let $\varrho_1, \ldots, \varrho_k$ be perfect. Then $t \in [a, b]$ is **snaking along** $\varrho_k \circ \ldots \circ \varrho_1$ iff
- $t$ is in the domain of $\varrho_k \circ \ldots \circ \varrho_1$.
- If $[\varrho_i \circ \ldots \circ \varrho_1](t)$ and $[\varrho_j \circ \ldots \circ \varrho_1](t)$ are in $\{a, b\}$ with $j > i$, then $\varrho_j \circ \ldots \circ \varrho_{i+1}$ is a product of an even number of reflections if both points coincide or a product of an odd number of reflections if both points differ.

# 6 Ordering within Motion Sets

An obvious, but very remarkable feature of positive motion sets on intervals of $\mathbb{R}$ or $S^1$ is the fact that any shift is uniquely given by its action on a single point of its domain. Similarly, a reflection is fully determined by its fixed point(s). Are there now similar results for general motion sets?

Indeed, there are. We even may get further. At least on $\mathbb{R}$, we can easily compare shifts by their length. Trivially, we can even relate the length of the shift to the distance between any point and its image under the shift. This way, we can easily order the shifts by means of the ordering on $\mathbb{R}$. Only in the case of $S^1$, the situation is a bit more delicate, as there is no ordering that is compatible with shifting. Nevertheless, if we restrict ourselves to the shift of one of the boundary points, say $a$, we indeed may order the shifts by where they map $a$ to. Of course, there we should restrict the shifts to those defined on $a$.

Again, this idea can be transferred directly to general motion sets. Thus, let us start with

**Definition 6.1** Let $\mathbf{P}$ be a motion set. We define for $\varrho_i \in \mathbf{P}_a$

$$\varrho_1 < \varrho_2 \iff \varrho_1(s) < \varrho_2(s) \qquad \text{for all } s \in \mathbf{L}(\varrho_1) \cap \mathbf{L}(\varrho_2)$$

Similarly the relations $\leq$, $\geq$ and $>$ defined.

We already know from Lemma 2.16 that any negative standard map has a fixed point. For positive standard maps, the situation is less clear a priori. The identity $\mathbf{1}$, of course, has a fixed point (even very many), but also other standard maps can have one. For instance, $\varrho(t) = t + \frac{1}{2}\sin t$ with $t \in [0, 2\pi]$ is standard and has $0$, $\pi$ and $2\pi$ as fixed points. However, it will be a very important fact that such functions cannot appear in motion sets. In particular, pointwise properness prevents them to exist. In our particular example, one can easily see that $\varrho^k$ is defined everywhere, but $\varrho^k(t)$ converges to $\pi$ for all $t \in (0, 2\pi)$ contradicting pointwise properness. The general statement to be proven is now

**Proposition 6.1** Let $\mathbf{P}$ be a motion set. Then we have for all $\varrho \in \mathbf{P}$:

$$\varrho \text{ has a fixed point} \iff \varrho \text{ is a reflection or equals } \mathbf{1}$$

This non-existence of (nontrivial) fixed points for positive motion sets will turn out to be extremely relevant for the further claims. In particular, is allows to identify any positive map by its value at the boundary.

**Proposition 6.2** Let $\mathbf{P}$ be a positive motion set and $\varrho_i \in \mathbf{P}_a$ with $s \in \mathbf{L}(\varrho_1) \cap \mathbf{L}(\varrho_2)$. Then

$$\begin{aligned}
\varrho_1 = \varrho_2 &\iff \varrho_1^{-1}(b) = \varrho_2^{-1}(b) \iff \mathbf{L}(\varrho_1) = \mathbf{L}(\varrho_2) \iff \varrho_1(s) = \varrho_2(s) \\
\varrho_1 < \varrho_2 &\iff \varrho_1^{-1}(b) > \varrho_2^{-1}(b) \iff \mathbf{L}(\varrho_1) \supset \mathbf{L}(\varrho_2) \iff \varrho_1(s) < \varrho_2(s)
\end{aligned}$$

Any reflection in our prime examples fulfills $\sigma \bullet \sigma = \mathbf{1}$. This transfers to

**Proposition 6.3** Let $\mathbf{P}$ be a motion set. Then $\sigma = \sigma^{-1}$ for all $\varrho \in \mathbf{P}^-$.

As for positive elements, we can identify elements of $\mathbf{P}^-$ with their values at $a$. Moreover, there is a similar ordering, even compatible with the ordering of fixed points.

**Proposition 6.4** Let $\mathbf{P}$ be a motion set and $\sigma_i \in \mathbf{P}_a^-$ with $s \in \mathbf{L}(\sigma_1) \cap \mathbf{L}(\sigma_2)$. Then

$$\begin{aligned}
\sigma_1 = \sigma_2 &\iff \mathbf{x}_1 = \mathbf{x}_2 \iff \mathbf{L}(\sigma_1) = \mathbf{L}(\sigma_2) \iff \sigma_1(s) = \sigma_2(s) \\
\sigma_1 < \sigma_2 &\iff \mathbf{x}_1 < \mathbf{x}_2 \iff \mathbf{L}(\sigma_1) \subset \mathbf{L}(\sigma_2) \iff \sigma_1(s) < \sigma_2(s)
\end{aligned}$$

Here, $\mathbf{x}_i$ denotes the fixed point of $\sigma_i$ in its left domain.

Note that the relation between the left domains differs for negative $\varrho$ from that for positive ones.[17] Observe that Propositions 6.2 and 6.4 imply

**Proposition 6.5** If $\mathbf{P}$ is a motion set, then the evaluation maps

$$\begin{aligned}
\mathbf{\Phi}: \mathbf{P}_a^{\pm} &\longrightarrow \mathbf{P}^{\pm}(a) \\
\varrho &\longmapsto \varrho(a)
\end{aligned}$$

are order-preserving bijections.

---

[17] We should note that the Propositions 6.2 and 6.4 have analogous counterparts for $\mathbf{P}_b$ instead of $\mathbf{P}_a$. One just has to exchange the entities $a$ and $\mathbf{L}$ by their respective counterparts $b$ and $\mathbf{R}$, and adapt Definition 6.1. Be aware, however, that the inclusion relations between left and right margins do *not* get reversed. The same exchanges will apply to the other statements within this section. Although we may state them in a lemma, corollary etc., we will refrain from proving them.

This will allow us to define intervals in $\mathbf{P}_a^{\pm}$. They correspond via $\Phi$ to intersections of $\mathbf{P}^{\pm}(a)$ with intervals in $[a,b]$. For instance, in the positive case, $[\mathbf{1}, \varrho)$ corresponds to $[a, \varrho(a)) \cap \mathbf{P}(a)$.

We are now going to prove the propositions above step by step. Sometimes, we might even relax the assumptions or sharpen the claims somewhat. Ultimately, Proposition 6.1 will be a direct consequence of Corollary 6.7 (positive $\varrho$) and Lemma 2.16 (negative $\varrho$). Proposition 6.2 is basically Proposition 6.12 together with $\mathbf{L}(\varrho) = [a, \varrho^{-1}(b)]$ for positive perfect $\varrho$. Proposition 6.3 is Proposition 6.21. And Proposition 6.4 is an immediate consequence of Propositions 6.24 and 6.27 with $\mathbf{L}(\sigma) = [a, \sigma(a)]$ for negative $\sigma \in \mathbf{P}$ from Corollary 6.22. We start in Subsection 6.1 with discussing the positive motions sets only, and then include reflections in Subsection 6.2.

## 6.1 Shifts

A central consequence of pointwise properness in the positive case is Corollary 6.7 saying that the only positive element having a fixed point is the identity. This implies that each positive element in $\mathbf{P}_a$ is uniquely determined by its value at $a$. Even more, relations between two positive elements at $a$ transfer to the full left margins of their domain. Ultimately, this will be the key to induce a notion convergence on $\mathbf{P}_a$ from that on $[a,b]$.

**Lemma 6.6** Let $\mathbf{P}$ be homeomorphic, exponential, positive, pointwise proper.
Then we have for all $\varrho \in \mathbf{P}$:
- If $t \in \mathbf{L}(\varrho)$, then $\varrho(t) = t$ implies $\varrho(s) = s$ for all $s < t$.
- If $t \in \mathbf{R}(\varrho)$, then $\varrho(t) = t$ implies $\varrho(s) = s$ for all $s > t$.

**Proof** In the first case, choose $I := [a, t] \subseteq \mathrm{dom}\,\varrho$ and observe that positivity implies $\varrho(I) = [\varrho(a), \varrho(t)] = [\varrho(a), t] \subseteq I$. Inductively, $\varrho^k(I) \subseteq I$, whence any $s \in [a,t]$ fulfills $\varrho(s) = s$ by Proposition 5.1. The second case is analogous. **qed**

**Corollary 6.7** Let $\mathbf{P}$ be perfect, unital, analytic, exponential, positive, pointwise proper.
Then we have for all $\varrho \in \mathbf{P}$
$$\varrho \text{ has a fixed point} \iff \varrho = \mathbf{1}$$

**Proof** ($\implies$ only) If $t \in \mathbf{R}(\varrho)$ is a fixed point, then $t = \varrho(t) \in \mathbf{L}(\varrho^{-1})$ is a fixed point for $\varrho^{-1}$. As $\varrho$ is the identity iff $\varrho^{-1}$ is the identity, we may assume w.l.o.g. that $t \in \mathbf{L}(\varrho)$ be a fixed point. Assume first $t = a$ and $\varrho(s) \neq s$ for all other $s \in \mathbf{L}(\varrho)$. Then, by positivity, we see that $[a, \varrho(s)] = [\varrho(a), \varrho(s)] = \varrho[a, s] \subseteq \mathrm{im}\,\varrho$. As $\varrho$ is perfect, $[a, \varrho(s)]$ (being the image of a left margin) is contained in the right margin of $\mathrm{im}\,\varrho$. Consequently, $\mathrm{im}\,\varrho$ must equal $[a, b]$ as well as (by perfectness) $\mathrm{dom}\,\varrho$. By positivity, we get $\varrho(b) = b$. This means we may in any case assume that $t$ is a fixed point greater than $a$. But, now Lemma 6.6 implies that $\varrho$ is the identity on $[a, t]$. By unitality and analyticity, we get the proof. **qed**

**Corollary 6.8** Let $\mathbf{P}$ be perfect, unital, analytic, exponential, positive, pointwise proper.
Let $\varrho \in \mathbf{P}_t$. Then
$$t \in \mathbf{L}(\varrho) \iff \varrho(t) \geq t \implies \varrho \in \mathbf{P}_a$$
$$t \in \mathbf{R}(\varrho) \iff \varrho(t) \leq t \implies \varrho \in \mathbf{P}_b$$

**Proof** If $\varrho(t) < t$ and $t \in \mathbf{L}(\varrho)$, then $\varrho(s) = s$ for some $s \in [a, t]$ by $\varrho(a) \geq a$. Hence, $\varrho = \mathbf{1}$, giving a contradiction. If $\varrho(t) > t$ and $t \in \mathbf{R}(\varrho)$, we get a similar contradiction. This proves the other equivalences. The implications are trivial. **qed**

**Corollary 6.9** Let $\mathbf{P}$ be a positive motion set.
Then, for all $\varrho_1, \varrho_2 \in \mathbf{P}$ with nonempty $A := \mathrm{dom}\,\varrho_1 \cap \mathrm{dom}\,\varrho_2$, we have
$$\varrho_1 = \varrho_2 \text{ somewhere on } A \iff \varrho_1 = \varrho_2 \text{ everywhere}$$

As always $\varrho$ or $\varrho^{-1}$ have a left margin, the corollary above simply says that $\varrho$ is uniquely determined by its values $\varrho(a)$ and $\varrho^{-1}(a)$, whichever exists.

**Proof** Let $t \in A$ with $\varrho_1(t) = \varrho_2(t)$. By Corollary 6.8, $t$ is in $\mathbf{L}(\varrho_1) \cap \mathbf{L}(\varrho_2)$ or in $\mathbf{R}(\varrho_1) \cap \mathbf{R}(\varrho_2)$. In particular, there is an interval $I \subseteq A$ containing $t$. Hence, $\mathrm{dom}(\varrho_1 \circ \varrho_2^{-1})$ contains $\varrho_2(I)$ being again an interval. Therefore, $\varrho_1 \bullet \varrho_2^{-1} \in \mathbf{P}$, by multiplicativity, coincides with $\varrho_1 \circ \varrho_2^{-1}$ on $\varrho_2(I)$. Since now
$$[\varrho_1 \bullet \varrho_2^{-1}](\varrho_2(t)) \;=\; [\varrho_1 \circ \varrho_2^{-1}](\varrho_2(t)) \;=\; \varrho_1(t) \;=\; \varrho_2(t),$$
Corollary 6.7 gives $\varrho_1 \bullet \varrho_2^{-1} = \mathbf{1}$, hence $\varrho_1 \equiv \varrho_2$ by Corollary 4.9. **qed**

**Proposition 6.10** Let $\mathbf{P}$ be a positive motion set. Then
$$\begin{aligned}\Phi_t: \mathbf{P}_t &\longrightarrow \mathbf{P}(t)\\ \varrho &\longmapsto \varrho(t)\end{aligned}$$
is a bijection for each $t \in [a,b]$.

**Proof** Injectivity follows from Corollary 6.9, surjectivity from the definition of $\mathbf{P}(t)$. **qed**

**Corollary 6.11** Let $\mathbf{P}$ be a positive motion set. Then we have for all $\varrho_1, \varrho_2 \in \mathbf{P}_a$
$$\varrho_1(a) > \varrho_2(a) \iff \varrho_1^{-1}(b) < \varrho_2^{-1}(b) \iff \mathbf{L}(\varrho_1) \subset \mathbf{L}(\varrho_2)$$
and
$$\varrho_1(a) = \varrho_2(a) \iff \varrho_1^{-1}(b) = \varrho_2^{-1}(b) \iff \mathbf{L}(\varrho_1) = \mathbf{L}(\varrho_2)$$

**Proof** For the second equivalence in the upper line use $\mathbf{L}(\varrho) = [a, \varrho^{-1}(b)]$ for $\varrho \in \mathbf{P}_a$. For the first equivalence in the upper line, we may restrict ourselves to the $\Longrightarrow$ direction. So assume $\varrho_1^{-1}(b) \geq \varrho_2^{-1}(b)$. This implies $\varrho_2^{-1}(b) \in \mathbf{L}(\varrho_1)$, hence $\varrho_1(\varrho_2^{-1}(b)) \leq b = \varrho_2(\varrho_2^{-1}(b))$. As $\varrho_1(a) > \varrho_2(a)$, there must be a $t \in [a, \varrho_2^{-1}(b)] \subseteq \mathbf{L}(\varrho_2)$ with $\varrho_1(t) = \varrho_2(t)$. But, now Corollary 6.9 gives $\varrho_1(a) = \varrho_2(a)$.
The lower line follows now immediately from Corollary 6.9 as well. **qed**

**Proposition 6.12** Let $\mathbf{P}$ be a positive motion set.
Then we have for all $\varrho_1, \varrho_2 \in \mathbf{P}_a$ exactly one of the following cases:
$$\varrho_1 > \varrho_2 \quad \text{or} \quad \varrho_1 = \varrho_2 \quad \text{or} \quad \varrho_1 < \varrho_2$$
on $\mathbf{L}(\varrho_1) \cap \mathbf{L}(\varrho_2)$. They correspond to the cases
$$\mathbf{L}(\varrho_1) \subset \mathbf{L}(\varrho_2) \quad \text{or} \quad \mathbf{L}(\varrho_1) = \mathbf{L}(\varrho_2) \quad \text{or} \quad \mathbf{L}(\varrho_1) \supset \mathbf{L}(\varrho_2).$$

Analogous statements hold for $\varrho_1, \varrho_2 \in \mathbf{P}_b$; then the left domains are replaced by the right domains as well as the inclusion relations inverted.

**Proof** If the relation between values of $\varrho_1$ and $\varrho_2$ is changing on $\mathbf{L}(\varrho_1) \cap \mathbf{L}(\varrho_2)$ being connected, then they have to coincide somewhere thereon, hence $\varrho_1 \equiv \varrho_2$ by Corollary 6.9, contradicting the change of relation. – The correspondence between the values and the left domains is due to Corollary 6.11. **qed**

So far, we have only studied the relation between elements of $\mathbf{P}_a$ on their left domains. Sometimes, we have remarked that similar relations are true for elements of $\mathbf{P}_b$. However, are these findings compatible? In fact, there may be a nontrivial intersection of $\mathbf{P}_a$ and $\mathbf{P}_b$. But, indeed, the relations are compatible. Unless $\varrho$ is the identity, we will see in a moment that $\varrho_1 > \varrho_2$ is not only given on the intersection of the left domains, but on that of the right domains as well. Similarly, one can see that $\varrho_1 > \varrho_2$ is valid as soon as it holds somewhere, unless it is at the intersection of $\mathbf{R}(\varrho_1)$ with $\mathbf{L}(\varrho_2)$. Note that the latter condition already shows that the identity shall be excluded here.

**Lemma 6.13** Let **P** be a positive motion set and $\varrho_1, \varrho_2 \in \mathbf{P}$. Then
$$\mathbf{L}(\varrho_1) \subseteq \mathbf{L}(\varrho_2) \quad \text{and} \quad \mathbf{R}(\varrho_1) \subseteq \mathbf{R}(\varrho_2) \implies \varrho_2 \text{ equals } \varrho_1 \text{ or } \mathbf{1}.$$

**Proof**
- Since $\varrho_1^{-1}(b) \in \mathbf{L}(\varrho_1) \subseteq \mathbf{L}(\varrho_2)$, we have $\varrho_2(\varrho_1^{-1}(b)) \in \mathbf{R}(\varrho_2^{-1})$.
- Since $\varrho_1^{-1}(a) \in \mathbf{R}(\varrho_1) \subseteq \mathbf{R}(\varrho_2)$, we have $\varrho_2(\varrho_1^{-1}(a)) \in \mathbf{L}(\varrho_2^{-1})$.
  Obviously, both $a$ and $b$ are snaking along $\varrho_2 \circ \varrho_1^{-1}$, whence $\varrho := \varrho_2 \bullet \varrho_1^{-1}$ is well defined.
- If $\varrho$ is the identity, Corollary 4.9 gives the proof.
- If $\varrho$ is not the identity, then $\varrho(a) > \varrho(b)$. At the same time, as shown above, we have $\varrho(b) \in \mathbf{R}(\varrho_2^{-1})$ and $\varrho(a) \in \mathbf{L}(\varrho_2^{-1})$, whence the left and the right domains of $\varrho_2^{-1}$ intersect. This, of course, implies that $\varrho_2$ is the identity. **qed**

The following corollary is now obvious.

**Corollary 6.14** Let **P** be a positive motion set and let $\varrho_1, \varrho_2 \in \mathbf{P}$ with $\varrho_2 \neq \mathbf{1}$. Then
$$\mathbf{L}(\varrho_1) \subset \mathbf{L}(\varrho_2) \iff \mathbf{R}(\varrho_1) \supset \mathbf{R}(\varrho_2)$$

**Proposition 6.15** Let **P** be a positive motion set.
Then we have for all $\varrho_1, \varrho_2 \in \mathbf{P}_a$ with $\varrho_2 \neq \mathbf{1}$
$$\begin{aligned}\varrho_1 > \varrho_2 &\iff \varrho_1(t) > \varrho_2(t) \quad \text{for all } t \in \mathbf{R}(\varrho_1) \cap \mathbf{R}(\varrho_2) \\ &\iff \varrho_1(t) > \varrho_2(t) \quad \text{for some } t \in \mathbf{R}(\varrho_1) \cap \mathbf{R}(\varrho_2)\end{aligned}$$

**Proof** By Proposition 6.12, $\varrho_1 > \varrho_2$ is equivalent to $\mathbf{L}(\varrho_1) \subset \mathbf{L}(\varrho_2)$. As $\varrho_2$ does not equal $\mathbf{1}$, this is equivalent to $\mathbf{R}(\varrho_1) \supset \mathbf{R}(\varrho_2)$ by Corollary 6.14. Now, we get the equivalence to the right hand side by the **R**-analogue of Proposition 6.12. **qed**

**Proposition 6.16** Let **P** be a positive motion set.
Then we have for all $\varrho_1, \varrho_2 \in \mathbf{P}_a$
$$\varrho_1 > \varrho_2 \quad \text{on } \mathbf{L}(\varrho_1) \cap \mathbf{L}(\varrho_2) \iff \varrho_1^{-1} < \varrho_2^{-1} \quad \text{on } \mathbf{R}(\varrho_1^{-1}) \cap \mathbf{R}(\varrho_2^{-1})$$

**Proof** If $\varrho_1 > \varrho_2$, then $[a, \varrho_1^{-1}(b)] = \mathbf{L}(\varrho_1) \subset \mathbf{L}(\varrho_2) = [a, \varrho_2^{-1}(b)]$, hence $\varrho_1^{-1}(b) < \varrho_2^{-1}(b)$. Now, the implication follows since the relation between two elements in $\mathbf{P}_a$ and $\mathbf{P}_b$ is determined by the relations between their values at $a$ and $b$, respectively. – The reverse implication is analogous. **qed**

## 6.2 Reflections

After having identified $\mathbf{P}_a^+$, i.e., the positive part of $\mathbf{P}_a$, with $\mathbf{P}^+(a) \subseteq [a, b]$, we are now going to discuss the negative part. As already mentioned, fixed points will play a crucial rôle for this. Recall that any standard map has a unique fixed point in the interior of each connected domain component. We are now going to show that this fixed point is characteristic for elements in motion sets. Indeed, fixed points correspond to a unique reflection in **P**. This way, fixed points will play the same rôle as $\varrho(a)$ or $\varrho(b)$ did for the characterization of $\varrho \in \mathbf{P}^+$.

Thus, let us start with

**Proposition 6.17** Let **P** be perfect, multiplicative, analytic, involutive.
Let $\sigma \in \mathbf{P}^-$ and $\tau \in \mathbf{P}$. Assume that $\mathbf{x} \in \text{dom}\, \tau$ with $\tau(\mathbf{x}) \neq a, b$. Then
$$\mathbf{x} \text{ fixed point of } \sigma \implies \tau(\mathbf{x}) \text{ fixed point of } \tau \bullet \sigma \bullet \tau^{-1}.$$
In particular, then $\tau \bullet \sigma \bullet \tau^{-1}$ is a well-defined element of $\mathbf{P}^-$.

**Corollary 6.18** Let **P** be perfect, multiplicative, analytic, involutive. Then
$$\mathbf{P}(\mathbf{x}) \subseteq \mathbf{F} \sqcup \{a, b\} \quad \text{for all } \mathbf{x} \in \mathbf{F}.$$

**Proof Proposition 6.17**
By assumption, neither $a$ nor $b$ equal $\mathbf{x}$ or $\tau(\mathbf{x})$. As $\tau \circ \sigma \circ \tau^{-1}$ maps $\tau(\mathbf{x})$ via $\mathbf{x}$ and $\mathbf{x}$ to $\tau(\mathbf{x})$, we see that $\tau(\mathbf{x})$ is snaking. Now, the generalized version of the snaking lemma (see Corollary 5.10 and the discussion towards the end of Section 5) gives that $\tau \bullet \sigma \bullet \tau^{-1}$ is well defined. Moreover, we have
$$[\tau \bullet \sigma \bullet \tau^{-1}](\tau(\mathbf{x})) = [\tau \circ \sigma \circ \tau^{-1}](\tau(\mathbf{x})) = \tau(\mathbf{x}).$$
It is obvious that $\tau \bullet \sigma \bullet \tau^{-1}$ is in $\mathbf{P}^-$ again. **qed**

The proposition above crucially depends on the assumption that the fixed point $\mathbf{x}$ is in the domain of $\tau$, but does not get mapped to the boundary. Indeed, otherwise, $\tau \bullet \sigma \bullet \tau^{-1}$ need not be well defined. For products of just two factors, we may recall that any perfect $\sigma \in \mathbf{P}_a^-$ has not only nontrivial left domain $\mathbf{L}(\sigma)$, but this left domain is also mapped to some left margin. This implies that its image has always nontrivial intersection with the domain of any $\tau \in \mathbf{P}_a$, in particular, with itself. This proves already the second part of

**Lemma 6.19** Let $\mathbf{P}$ be perfect, multiplicative, analytic, and let $\sigma \in \mathbf{P}_a^-$.
1. $\sigma \bullet \sigma$ is well defined and the fixed point $\mathbf{x}$ of $\sigma$ is fixed by $\sigma \bullet \sigma$ as well.[18]
2. $\tau \bullet \sigma$ is well defined for all $\tau \in \mathbf{P}_a$.

**Proof** If $\mathbf{x}$ is a fixed point of $\sigma \in \mathbf{P}^-$, then $\mathbf{x}$ lies in the interior of $\mathrm{dom}\,\sigma$ and of $\mathrm{im}\,\sigma$. Therefore, $\mathrm{dom}\,\sigma \circ \sigma$ contains a nontrivial interval around $\mathbf{x}$, giving the claim. **qed**

In particular, $\sigma \bullet \sigma \in \mathbf{P}^+$ has a fixed point. Now, Corollary 6.7 implies

**Lemma 6.20** Let $\mathbf{P}$ be perfect, multiplicative, analytic, and let $\mathbf{P}^+$ be a motion set. Then $\sigma \bullet \sigma = \mathbf{1}$ for any $\sigma \in \mathbf{P}^-$.

Now, Corollary 4.9 gives

**Proposition 6.21** Let $\mathbf{P}$ be a motion set. Then $\sigma = \sigma^{-1}$ for any $\sigma \in \mathbf{P}^-$.

**Corollary 6.22** Let $\mathbf{P}$ be a motion set and $\sigma \in \mathbf{P}^-$. Then:
1. $\sigma$ has left domain $\mathbf{L}(\sigma) = [a, \sigma(a)]$, provided $\sigma \in \mathbf{P}_a$.
2. $\sigma$ has right domain $\mathbf{R}(\sigma) = [\sigma(b), b]$, provided $\sigma \in \mathbf{P}_b$.
3. $\sigma$ maps $\mathbf{L}(\sigma)$ onto itself, as well as it keeps $\mathbf{R}(\sigma)$ invariant.
4. $\sigma$ maps $[a, \mathbf{x}]$ onto $[\mathbf{x}, \sigma(a)]$ and vice versa, provided $\sigma$ fixes $\mathbf{x} \in \mathbf{L}(\sigma)$.

**Corollary 6.23** Let $\mathbf{P}$ be a motion set and let $\sigma, \tau \in \mathbf{P}_a$ with $\sigma \in \mathbf{P}^-$. Then
$$\mathbf{L}(\sigma) \subseteq \mathbf{L}(\tau) \implies \mathbf{L}(\sigma) = \mathbf{L}(\tau \circ \sigma) \subseteq \mathbf{L}(\tau \bullet \sigma).$$

**Proof** $\sigma(\mathbf{L}(\sigma)) = \mathbf{L}(\sigma) \subseteq \mathbf{L}(\tau)$ implies $\mathbf{L}(\sigma) \subseteq \mathrm{dom}(\tau \circ \sigma)$, hence $\mathbf{L}(\sigma) \subseteq \mathbf{L}(\tau \circ \sigma)$. The other direction is trivial. **qed**

Let us now attack the proof of Proposition 6.4. Given $\sigma_1, \sigma_2 \in \mathbf{P}_a^-$, we surely have $\mathbf{L}(\sigma_1) \subseteq \mathbf{L}(\sigma_2)$ iff $\sigma_1(a) \leq \sigma_2(a)$, as shown above. Similarly, the other relations are proved. Let us now show

**Proposition 6.24** Let $\mathbf{P}$ be a motion set, $\sigma_i \in \mathbf{P}_a^-$ with fixed points $\mathbf{x}_i \in \mathbf{L}(\sigma_i)$. Then
$$\begin{aligned} \mathbf{x}_1 = \mathbf{x}_2 &\iff \mathbf{L}(\sigma_1) = \mathbf{L}(\sigma_2) \iff \sigma_1 = \sigma_2 \\ \mathbf{x}_1 < \mathbf{x}_2 &\iff \mathbf{L}(\sigma_1) \subset \mathbf{L}(\sigma_2) \end{aligned}$$

All these cases can easily be deduced from

---

[18] One can even show inductively that $\sigma^{2n} \in \mathbf{P}^+$ and $\sigma^{2n+1} \in \mathbf{P}^-$.

**Lemma 6.25** Let $\mathbf{P}$ be a motion set, $\sigma_i \in \mathbf{P}_a^-$ with fixed points $\mathbf{x}_i \in \mathbf{L}(\sigma_i)$. Then
$$\mathbf{x}_1 \geq \mathbf{x}_2 \quad \text{and} \quad \mathbf{L}(\sigma_1) \subseteq \mathbf{L}(\sigma_2) \implies \sigma_1 = \sigma_2$$

**Proof** As fixed points are interior points of the respective domain, we have $a < \mathbf{x}_i < \sigma_i(a)$. In particular, $a < \mathbf{x}_2 \leq \mathbf{x}_1 < \sigma_1(a) \leq \sigma_2(a)$; for the last inequality we have used that $\mathbf{L}(\sigma_1) \subseteq \mathbf{L}(\sigma_2)$. Now, Corollary 6.23 gives us $[\mathbf{x}_1, \sigma_1(a)] \subseteq \mathbf{L}(\sigma_1) \subseteq \mathbf{L}(\sigma_2 \circ \sigma_1)$. On the other hand, $[\sigma_2 \circ \sigma_1](\sigma_1(a)) = \sigma_2(a) \geq \sigma_1(a)$ and $[\sigma_2 \circ \sigma_1](\mathbf{x}_1) = \sigma_2(\mathbf{x}_1) \leq \sigma_2(\mathbf{x}_2) = \mathbf{x}_2 \leq \mathbf{x}_1$ imply that $\sigma_2 \circ \sigma_1$ has a fixed point in $\mathbf{L}(\sigma_2 \circ \sigma_1)$ by the mean value theorem, hence $\sigma_2 \bullet \sigma_1$ thereon as well. Corollary 6.7 implies that $\sigma_2 \bullet \sigma_1 = \mathbf{1}$, hence $\sigma_1 = \sigma_2^{-1} = \sigma_2$ by Corollary 4.9 and Proposition 6.21. **qed**

The proposition above comprises most of the equivalences to be proven in Proposition 6.4. The remaining ones are easily obtained from the two following statements.

**Lemma 6.26** Let $\mathbf{P}$ be a motion set and $\sigma_i \in \mathbf{P}^-$. Then we have
$$\sigma_1(s) = \sigma_2(s) \quad \text{for some } s \in \mathbf{L}(\sigma_1) \cap \mathbf{L}(\sigma_2) \implies \sigma_1 = \sigma_2.$$

**Proof** Assume $s \in \mathbf{L}(\sigma_1) \subseteq \mathbf{L}(\sigma_2)$. Then $\sigma_2(s) = \sigma_1(s) \in \mathbf{L}(\sigma_1)$, hence $s$ is snaking along $\sigma_1 \circ \sigma_2$. Hence, $\sigma_1 \bullet \sigma_2$ is a well-defined element in $\mathbf{P}^+$ with $[\sigma_1 \bullet \sigma_2](s) = \sigma_1(\sigma_2(s)) = \sigma_1(\sigma_1(s)) = s$. Hence, $\sigma_1 \bullet \sigma_2 = \mathbf{1}$ by Corollary 6.7. This gives $\sigma_1 = \sigma_2$ and the proof as above. **qed**

**Proposition 6.27** Let $\mathbf{P}$ be a motion set and $\sigma_i \in \mathbf{P}^-$ with fixed points $\mathbf{x}_i \in \mathbf{L}(\sigma_i)$. Then
$$\mathbf{x}_1 < \mathbf{x}_2 \implies \sigma_1(s) < \sigma_2(s) \qquad \text{for all } s \in \mathbf{L}(\sigma_1)$$

**Proof** Recall from Proposition 6.24 and Corollary 6.23 that
$$\mathbf{x}_1 < \mathbf{x}_2 \iff \mathbf{L}(\sigma_1) \subset \mathbf{L}(\sigma_2) \implies \mathbf{L}(\sigma_1) \subseteq \mathbf{L}(\sigma_2 \circ \sigma_1) \subseteq \mathbf{L}(\sigma_2 \bullet \sigma_1)$$
Now we have
$$\begin{aligned}
\mathbf{x}_1 < \mathbf{x}_2 &\implies \sigma_1(\mathbf{x}_1) = \mathbf{x}_1 < \mathbf{x}_2 \in \mathbf{L}(\sigma_2) \\
&\implies [\sigma_2 \bullet \sigma_1](\mathbf{x}_1) = \sigma_2(\sigma_1(\mathbf{x}_1)) > \sigma_2(\mathbf{x}_2) = \mathbf{x}_2 > \mathbf{x}_1 \in \mathbf{L}(\sigma_2 \bullet \sigma_1) \\
&\implies [\sigma_2 \bullet \sigma_1](s) > s \qquad \text{for all } s \in \mathbf{L}(\sigma_1) \subseteq \mathbf{L}(\sigma_2 \bullet \sigma_1)
\end{aligned}$$
by Proposition 6.2. From $\mathbf{L}(\sigma_2 \circ \sigma_2) = \mathbf{L}(\sigma_2)$, we see that $\mathbf{L}(\sigma_2 \circ \sigma_2 \circ \sigma_1) = \mathbf{L}(\sigma_1)$, hence $\sigma_1(s) = [\sigma_2 \bullet \sigma_2 \bullet \sigma_1](s) = \sigma_2([\sigma_2 \bullet \sigma_1](s)) < \sigma_2(s)$ for all $s \in \mathbf{L}(\sigma_1)$. **qed**

The last statements referred to fixed points in left margins, hence reflections in $\mathbf{P}_a$ only. Of course, similar results can be derived also for $\mathbf{P}_b$. One might now ask whether there are any orders between fixed points in left margins and those in right margins. Can one even say whether a fixed point is a left one or a right one? Well, in general, we cannot. Indeed, if there is a reflection $\varsigma$ defined on whole $[a, b]$, then its single fixed point corresponds to both the left and the right ones. However, as we will see in a minute, this is, if any, the only element in $\mathbf{P}^-$ that has a fixed point contained in both the left and in the right margin. The simple reason for that is again that $\varsigma \bullet \varsigma$ is positive with fixed point, hence $\mathbf{1}$; and any other reflection must equal $\varsigma$ for similar reasons. Any other fixed point will clearly be assignable to the left ones or the right ones. At the same time, of course, it may happen that a reflection $\sigma$ has both a left and a right margin, hence both a left fixed point and a right fixed point. Thus, only the leftness or the rightness of the particular fixed point will be a good notion. For $\mathbf{P}^-$, this is not possible, in general.

To sum up, let us introduce the notions
$$\begin{aligned}
\mathbf{F}_a &:= \{\mathbf{x} \in [a,b] \mid \exists \sigma \in \mathbf{P}^- : \mathbf{x} \in \mathbf{L}(\sigma) \text{ is a fixed point of } \sigma\} \\
\mathbf{F}_b &:= \{\mathbf{x} \in [a,b] \mid \exists \sigma \in \mathbf{P}^- : \mathbf{x} \in \mathbf{R}(\sigma) \text{ is a fixed point of } \sigma\} \\
\mathbf{F} &:= \{\mathbf{x} \in [a,b] \mid \exists \sigma \in \mathbf{P}^- : \mathbf{x} \in \operatorname{dom} \sigma \text{ is a fixed point of } \sigma\}
\end{aligned}$$
We will call the elements of $\mathbf{F}_a$ **left fixed points**, those of $\mathbf{F}_b$ **right fixed points**.

**Lemma 6.28** Let $\mathbf{P}$ be a motion set. Moreover, let $\mathbf{x}_i$ be fixed points of some $\sigma_i \in \mathbf{P}^-$. Then

$$\mathbf{x}_a \geq \mathbf{x}_b \text{ with } \mathbf{x}_a \in \mathbf{L}(\sigma_a) \text{ and } \mathbf{x}_b \in \mathbf{R}(\sigma_b) \implies \mathbf{x}_a = \mathbf{x}_b \text{ and } \sigma_a = \sigma_b.$$

**Proof** By $\mathbf{L}(\sigma_a) = [a, \sigma_a(a)]$ and $\mathbf{R}(\sigma_b) = [\sigma_b(b), b]$, we have $a \leq \sigma_b(b) < \mathbf{x}_b \leq \mathbf{x}_a < \sigma_a(a) \leq b$, by assumption. In particular, $[a, \mathbf{x}_a] \subseteq \operatorname{dom} \sigma_b \circ \sigma_a$, as $\sigma_a[a, \mathbf{x}_a] = [\mathbf{x}_a, \sigma_a(a)] \subseteq \operatorname{dom} \sigma_b$. Therefore, $\sigma_b \bullet \sigma_a \in \mathbf{P}_a^+$ and $\mathbf{x}_a \in \mathbf{L}(\sigma_b \bullet \sigma_a)$, whence

$$\mathbf{x}_a \leq [\sigma_b \bullet \sigma_a](\mathbf{x}_a) = [\sigma_b \circ \sigma_a](\mathbf{x}_a) = \sigma_b(\mathbf{x}_a) \leq \sigma_b(\mathbf{x}_b) = \mathbf{x}_b \leq \mathbf{x}_a$$

using $\mathbf{x}_a \geq \mathbf{x}_b$ twice. Thus, $\mathbf{x}_a$ is a fixed point of $\sigma_b \bullet \sigma_a \in \mathbf{P}_a^+$, whence $\sigma_b \bullet \sigma_a = \mathbf{1}$ by Corollary 6.7. This gives, $\sigma_b = \sigma_a$. **qed**

This immediately provides us with a separation statement.

**Proposition 6.29** Let $\mathbf{P}$ be a motion set. Then $\mathbf{x}_a \leq \mathbf{x}_b$ for all $\mathbf{x}_a \in \mathbf{F}_a$ and $\mathbf{x}_b \in \mathbf{F}_b$.

Recall that reflections can have one or two fixed points. Therefore, a reflection does not determine a single fixed point. However, any fixed point uniquely determines its respective reflection.

**Proposition 6.30** Let $\mathbf{P}$ be a motion set, and let $\mathbf{x}_1, \mathbf{x}_2$ be fixed points of $\sigma_1, \sigma_2 \in \mathbf{P}^-$. Then

$$\mathbf{x}_1 = \mathbf{x}_2 \implies \sigma_1 = \sigma_2.$$

**Proof** If both $\mathbf{x}_i$ are in $\mathbf{L}(\sigma_i)$, the claim comes from Proposition 6.24. Similarly, one argues for $\mathbf{x}_i$ in $\mathbf{R}(\sigma_i)$. If, finally, $\mathbf{x}_1$ is in $\mathbf{L}(\sigma_1)$ and $\mathbf{x}_2$ is in $\mathbf{R}(\sigma_2)$, then Lemma 6.28 gives the proof. **qed**

**Corollary 6.31** Let $\mathbf{P}$ be a motion set and let $\mathbf{x} \in \mathbf{F}_a \cap \mathbf{F}_b$. Then there is a unique $\sigma \in \mathbf{P}^-$ having $\mathbf{x}$ as a fixed point. This $\sigma$ has even full domain $[a, b]$.

**Proof** Uniqueness comes from the preceding proposition. As $\mathbf{x}$ is contained in both the left domain and the right domain of $\sigma$, these domains need to have nontrivial intersection. As $\sigma$ is perfect, the domain must be full $[a, b]$. **qed**

## 7 Right-Moving Products

In the subsequent section, we will be going to derive our first classification result. Indeed, we will explicitly determine (up to isomorphism) all finite motion sets $\mathbf{P}$. In the (nontrivial) positive case, it will turn out, that there is always a generator $\mu \in \mathbf{P}_a$ in the sense that any element of $\mathbf{P}$ can be written as $\mu^k$ for some $k$. In the general case, we may need an additional reflection to generate $\mathbf{P}$.

This already shows that we shall now focus on the product of elements in motion sets. When we calculate the values of a product of functions at specific points, it is usually easier to do this for the standard concatenation than for the $\bullet$-product; indeed, in the latter case we first have to check whether the argument is snaking. In constrast, when we need to use general arguments that are generally true for perfect maps or elements of a motion set only, we prefer to study the $\bullet$-products instead of the simple concatenation. Therefore, it would be nice to have criteria when both products lead to the same results. In particular, as we shall focus on $\mathbf{P}_a$, when do the standard concatenation and the $\bullet$-product have the same left domain? Let us discuss this in the positive case for two situations.

1. If $\varrho$ is in $\mathbf{P}_a$, then $\varrho^{-1} \circ \varrho$ is the identity on $\operatorname{dom} \varrho$. In particular, its left domain equals that of $\varrho$. On the other hand, $\varrho^{-1} \bullet \varrho$ equals the identity, whence its left domain is full $[a, b]$. So, unless $\varrho = \mathbf{1}$, the left domains differ.

2. Going now back to the prime example on $S^1$, we see that the critical moment is when the right-shift is that "large" that it moves the interval in a way such that it "re-enters" from the left-hand side. In other words, even if $a$ is getting shifted to the right, $b$ is getting effectively shifted to the left.

The main question is to find criteria that exclude such a behaviour. This will lead to the notion of right-moving products $\varrho_k \circ \ldots \circ \varrho_1$. It works specifically for elements of $\mathbf{P}_a^+$ which is sufficient for positive motion sets as any element or its inverse is in $\mathbf{P}_a$. More precisely, there are two equivalent definitions: either the (finite) sequence $([\varrho_k \circ \ldots \circ \varrho_1](a))$ is monotonous, without reaching $b$, or $[\varrho_i \circ \ldots \circ \varrho_1](a) \in \text{int } \mathbf{L}(\varrho_{i+1})$ for all $i < k$. Then, the left domains of the normal concatenation $\varrho_k \circ \ldots \circ \varrho_1$ and the product $\varrho_k \bullet \ldots \bullet \varrho_1$ have the same nontrivial domain, hence coincide thereon.

The right-moving property will be particularly important for the case of $\varrho^n$. Indeed, if $\varrho \circ \cdots \circ \varrho$ is right-moving, then $\varrho \bullet \cdots \bullet \varrho$ is not only well defined, but coincides with the usual concatenation on its left domain. This will simplify many arguments. Remarkably, for $\varrho \neq \mathbf{1}$, we always find a maximal $n$ with $\varrho^n$ being right-moving; $n$ will be called multiplicity of $\varrho$. This structure will turn out crucial later. Indeed it will be used to prove, first, that any finite positive motion set is given by powers of some generating element $\mu \in \mathbf{P}_a$. This will be done in Section 8. Later, in the infinite case, we will be able to define square roots and even powers of $\varrho \in \mathbf{P}_a$ with real exponents; this will show that then $\mathbf{P}$ is given by all fractional powers of some $\varrho \in \mathbf{P}_a$.

## 7.1 Definition and Elementary Properties

**Definition 7.1** A product $\varrho_k \circ \ldots \circ \varrho_1$ of maps is called **right-moving** iff $\varrho_i \in \mathbf{P}_a$ for all $i$ and
$$a \ \leq \ \varrho_1(a) \ \leq \ [\varrho_2 \circ \varrho_1](a) \ \leq \ \ldots \ \leq \ [\varrho_k \circ \ldots \circ \varrho_1](a) \ < \ b$$
with all expressions being well defined.

Actually, we should better denote the $k$-tuple $(\varrho_1, \ldots, \varrho_k)$ of functions right-moving instead of the product. However, we do not expect misconceptions writing the product of the functions. Note that we always decompose such a product w.r.t. the normal concatenation operation $\circ$, ignoring, in particular, $\bullet$-subproducts. This means that, for instance, $\varrho_4 \circ [\varrho_3 \bullet \varrho_2] \circ \varrho_1$ is right-moving iff
$$a \ \leq \ \varrho_1(a) \ \leq \ [\varrho_3 \bullet \varrho_2](\varrho_1(a)) \ \leq \ \varrho_4([\varrho_3 \bullet \varrho_2](\varrho_1(a))) \ < \ b \,.$$

Obviously, we have

**Lemma 7.1** If $\varrho_k \circ \ldots \circ \varrho_1$ is right-moving, then $\varrho_l \circ \ldots \circ \varrho_1$ is right-moving for all $l \leq k$.

Moreover, Corollary 6.8 implies

**Lemma 7.2** Let $\mathbf{P}$ be a positive motion set. Then, given $b \neq [\varrho_k \circ \ldots \circ \varrho_1](a)$, we have
$$\varrho_k \circ \ldots \circ \varrho_1 \text{ right-moving} \quad \Longleftrightarrow \quad [\varrho_i \circ \ldots \circ \varrho_1](a) \in \mathbf{L}(\varrho_{i+1}) \text{ for all } i \,.$$

**Proposition 7.3** Let $\mathbf{P}$ be a positive motion set, and let $\varrho_k \circ \ldots \circ \varrho_1$ be right-moving. Then
1. $\mathbf{L}(\varrho_k \bullet \ldots \bullet \varrho_1)$ equals $\mathbf{L}(\varrho_k \circ \ldots \circ \varrho_1)$ and is non-empty.
2. $\varrho_k \bullet \ldots \bullet \varrho_1$ equals $\varrho_k \circ \ldots \circ \varrho_1$ thereon.

In particular, $\varrho_k \bullet \ldots \bullet \varrho_1 \in \mathbf{P}_a$ again.

**Proof** As $a$ is snaking along $\varrho_k \circ \ldots \circ \varrho_1$, we get almost all statements from Corollary 5.10 and the lines following it. Only the equality of the left domains is still to be proven. For this, let $[a, t] := \mathbf{L}(\varrho_k \circ \ldots \circ \varrho_1) \subseteq \mathbf{L}(\varrho_k \bullet \ldots \bullet \varrho_1)$ with $t > a$. By perfectness, we will be done as soon as $b = [\varrho_k \bullet \ldots \bullet \varrho_1](t) = [\varrho_k \circ \ldots \circ \varrho_1](t)$. Lemma 5.9 shows $b = [\varrho_i \circ \ldots \circ \varrho_1](t)$ at least for some $i$. Since, by assumption, $\varrho_i \circ \ldots \circ \varrho_1$ maps $[a, t]$ fully into the domain of $\varrho_{i+1}$ and maps $a$ into $\mathbf{L}(\varrho_{i+1})$, it also maps $t$ into $\mathbf{L}(\varrho_{i+1})$. Hence,
$$b \ \geq \ \varrho_{i+1}\bigl([\varrho_i \circ \ldots \circ \varrho_1](t)\bigr) \ \geq \ [\varrho_i \circ \ldots \circ \varrho_1](t) \ = \ b \,.$$
Inductively, we get $b = [\varrho_k \circ \ldots \circ \varrho_1](t) = [\varrho_k \bullet \ldots \bullet \varrho_1](t)$. **qed**

**Lemma 7.4** Let **P** be a positive motion set and let $\varrho_k \circ \ldots \circ \varrho_1$ be right-moving. Then
$$[\varrho_{i-1} \circ \ldots \circ \varrho_1](t) \in \mathbf{L}(\varrho_i) \qquad \text{for all } t \in \mathbf{L}(\varrho_i \circ \ldots \circ \varrho_1) \text{ and } i \leq k.$$
Moreover, we have $\mathbf{L}(\varrho_k \circ \ldots \circ \varrho_1) \subseteq \mathbf{L}(\varrho_i \circ \ldots \circ \varrho_1)$ for all $i \leq k$, as well as for $\bullet$.

**Proof** Observe that $\varrho_i \circ \ldots \circ \varrho_1$ is right-moving by Lemma 7.1, and define $\varrho := \varrho_{i-1} \circ \ldots \circ \varrho_1$.
- Since $[a, t] \subseteq \mathbf{L}(\varrho_i \circ \varrho)$ by assumption, we get $[\varrho(a), \varrho(t)] \equiv \varrho[a, t] \subseteq \mathbf{I}(\varrho_i)$, by Lemma 4.2. Since, on the other hand, $\varrho_i(\varrho(a)) \geq \varrho(a)$ by assumption, hence $\varrho(a) \in \mathbf{L}(\varrho_i)$ by Corollary 6.8, we get $\varrho(t) \in \mathbf{L}(\varrho_i)$.
- Using Proposition 7.3, we get
$$\mathbf{L}(\varrho_i \bullet \ldots \bullet \varrho_1) = \mathbf{L}(\varrho_i \circ \ldots \circ \varrho_1) \supseteq \mathbf{L}(\varrho_k \circ \ldots \circ \varrho_1) = \mathbf{L}(\varrho_k \bullet \ldots \bullet \varrho_1).$$
**qed**

**Lemma 7.5** Let **P** be a positive motion set and $\varrho_1, \varrho_2, \sigma \in \mathbf{P}_a$ with $\varrho_1 < \varrho_2$. Then we have
$$\begin{array}{lll}
\varrho_2 \circ \sigma \text{ right-moving} & \Longrightarrow & \varrho_1 \circ \sigma \text{ right-moving and } \varrho_1 \bullet \sigma < \varrho_2 \bullet \sigma \\
\sigma \circ \varrho_2 \text{ right-moving} & \Longrightarrow & \sigma \circ \varrho_1 \text{ right-moving and } \sigma \bullet \varrho_1 < \sigma \bullet \varrho_2
\end{array}$$

**Proof** For the upper line, observe that $\varrho_2 \bullet \sigma$ is well defined by the right-movement assumption. Moreover, $\sigma(a) \in \mathbf{L}(\varrho_2) \subset \mathbf{L}(\varrho_1)$. Hence, $\varrho_1(\sigma(a)) < \varrho_2(\sigma(a)) < b$. This gives the claim. For the lower line, observe that $\sigma \bullet \varrho_2$ is well defined with $\varrho_1(a) < \varrho_2(a) \in \mathbf{L}(\sigma)$, hence $\sigma(\varrho_1(a)) < \sigma(\varrho_2(a))$ giving the claim. **qed**

## 7.2 Multiplicity

**Definition 7.2** The **multiplicity** $\mathbf{m}(\varrho)$ of an element $\varrho \in \mathbf{P}_a$ is the supremum of all $n$ for which
$$\varrho^n \text{ is right-moving.}$$

More explicitly, this means that $\mathbf{m}(\varrho)$ is the supremum of all $n$ for which
$$a \equiv \varrho^0(a), \quad \varrho^1(a), \quad \varrho^2(a), \quad \ldots, \quad \varrho^n(a) < b$$
is defined and monotonous. Of course, the multiplicity of $\mathbf{1}$ is always $\infty$. But, this is the only case due to pointwise properness:

**Lemma 7.6** Let **P** be a positive motion set.
Then $\mathbf{m}(\varrho)$ is positive and finite for all $\mathbf{1} \neq \varrho \in \mathbf{P}_a$.

**Proof** Positivity is trivial, since $\varrho \in \mathbf{P}_a$. Assume that $\mathbf{m}(\varrho)$ is infinite. By Lemma 7.4, we have $\varrho^k(a) \in \mathbf{L}(\varrho)$ for all $k$. Now, Proposition 5.1 gives $\varrho(a) = a$, hence $\varrho = \mathbf{1}$. **qed**

**Remark** In the following, unless stated otherwise, we will always denote the $k$-fold concatenation of $\varrho$ by $\varrho_\circ^k$, whereas $\varrho^k$ is now reserved for the $k$-fold $\bullet$-product. Moreover, recall that $t \in \mathbf{L}(\varrho)$ is mapped by any positive $\varrho$ to $b$ iff $t$ is the right boundary point of $\mathbf{L}(\varrho)$. We will therefore often need the interval $\mathbf{L}(\varrho) \setminus \{\varrho^{-1}(b)\}$. Shortly, we will now define int $\mathbf{L}(\varrho) := [a, t)$ for $\mathbf{L}(\varrho) = [a, t]$. Unless $\varrho = \mathbf{1}$, this notion indeed coincides with the topological interior of $[a, t]$ within $[a, b]$.

The following lemma is a direct consequence of Lemma 7.2 and Proposition 7.3.

**Lemma 7.7** Let **P** be a positive motion set. Then we have for $\mathbf{1} \neq \varrho \in \mathbf{P}_a$:
1. $\mathbf{m}(\varrho)$ is the largest number with $\varrho_\circ^k(a) \in \text{int } \mathbf{L}(\varrho)$ for all smaller $k$.
2. $\varrho_\circ^k$ extends to a well-defined element $\varrho^k$ of $\mathbf{P}_a$ for all $k \leq \mathbf{m}(\varrho)$.
3. $\varrho_\circ^k$ and $\varrho^k$ have the same left domain $[a, (\varrho^{-1})_\circ^k(b)]$ and coincide thereon.

**Corollary 7.8** Let **P** be a positive motion set. Then **m** is non-increasing, i.e., for all $\varrho_1, \varrho_2 \in \mathbf{P}_a$

$$\varrho_1 \leq \varrho_2 \quad\Longrightarrow\quad \mathbf{m}(\varrho_1) \geq \mathbf{m}(\varrho_2)$$

Even more, $\varrho_1 \leq \varrho_2$ implies $\varrho_1^k(a) \leq \varrho_2^k(a)$ for all $k \leq \mathbf{m}(\varrho_2)$.

**Proof** It is sufficient to show $\varrho_1^k(a) \leq \varrho_2^k(a)$ for all $k \leq \mathbf{m}(\varrho_2)$. In fact, then for all these $k$, we have $\varrho_1^k(a) \leq \varrho_2^k(a) \in \mathrm{int}\, \mathbf{L}(\varrho_2) \subseteq \mathrm{int}\, \mathbf{L}(\varrho_1)$, hence $\mathbf{m}(\varrho_1)$ is not smaller than $\mathbf{m}(\varrho_2)$. The inequality, on the other hand, follows inductively from

$$\varrho_1^k(a) \;=\; \varrho_1(\varrho_1^{k-1}(a)) \;\leq\; \varrho_2(\varrho_1^{k-1}(a)) \;\leq\; \varrho_2(\varrho_2^{k-1}(a)) \;=\; \varrho_2^k(a)$$

**qed**

**Lemma 7.9** Let **P** be a positive motion set and let $\varrho \in \mathbf{P}_a$. Then we have

$$\varrho^{k_1} \bullet \varrho^{k_2} \;=\; \varrho^{k_1+k_2} \qquad \text{for all } |k_1|, |k_2|, |k_1+k_2| \leq \mathbf{m}(\varrho).$$

**Proof** Unless $k_1$ and $k_2$ have opposite sign, the statement is trivial by the properties of multiplicity. Thus, let us assume $k_1 < 0 < k_2$ with $k_1 + k_2 \geq 0$, hence $k_1 + k_2 \leq \mathbf{m}(\varrho)$. As $a$ is snaking along $(\varrho^{-1})_\circ^{-k_1} \circ \varrho_\circ^{k_2}$, and this function equals $\varrho_\circ^{k_1+k_2}$ in $a$, we get

$$\varrho^{k_1} \bullet \varrho^{k_2} \;=\; (\varrho^{-1})^{-k_1} \bullet \varrho^{k_2} \;=\; \varrho^{k_1+k_2}.$$

The proof for $k_1 + k_2 < 0$ is completely analogous. **qed**

**Corollary 7.10** Let **P** be a positive motion set.
Then we have for all $\mathbf{1} \neq \varrho \in \mathbf{P}_a$ and all $|k_i| \leq \mathbf{m}(\varrho)$ with $k_1 k_2 \geq 0$:

$$\varrho^{k_1} = \varrho^{k_2} \quad\Longleftrightarrow\quad k_1 = k_2.$$

**Proof** Let $k_1 > k_2$. By assumption, $|k_1 - k_2| \leq \mathbf{m}(\varrho)$. Therefore, $\mathbf{1} = \varrho^{k_1} \bullet \varrho^{-k_2} = \varrho^{k_1-k_2}$, giving $\varrho^{k_1-k_2}(a) = a$, hence $\varrho(a) = a$ by the monotonicity condition in the definition of multiplicity, hence $\varrho = \mathbf{1}$. **qed**

## 7.3 Squeezing Lemma

The multiplicity of some element $\varrho$ tells us, how often we can apply $\varrho$ until either $a$ is kicked off the interval or re-enters. Looking at the particular example of the unit right-shift at some interval in $\mathbb{R}$ having length $x$, we can surely apply $\varrho^\mathbf{m}$ as long as $\mathbf{m}$ is natural and smaller than $x$. If we take the half-unit right shift, then we can do it $2\mathbf{m}$ times, but possibly even once more, if $\mathbf{m} + \frac{1}{2} < x$. Similarly, we can do it for right-shifts by $\frac{1}{n}$. This gives a more "accurate" information how "long" $\varrho$ in the general setting jumps. To put this on solid grounds, we should study how the multiplicities of $\varrho$ and $\varrho^n$ are related. Note, however, that the multiplicity is defined only referring to the usual concatenation of mappings, while $\varrho^n$ refers explcitly to the $\bullet$-product. Hence, we shall investigate how the right-moving property of the $\circ$-product is related to case where some $\circ$ are replaced by $\bullet$.

**Lemma 7.11** Let **P** be a positive motion set and $\varrho_i, \sigma_j \in \mathbf{P}$.
Then the following statements are equivalent:
1. $\sigma_l \circ \ldots \circ \sigma_1 \circ \varrho_k \circ \ldots \circ \varrho_1$ is right-moving.
2. $\sigma_l \bullet \ldots \bullet \sigma_1 \circ \varrho_k \circ \ldots \circ \varrho_1$ and $\sigma_l \circ \ldots \circ \sigma_1$ are right-moving.

For brevity, let us denote

$$\begin{aligned}\varrho_\circ &:= \varrho_k \circ \ldots \circ \varrho_1 \\ \varrho &:= \varrho_k \bullet \ldots \bullet \varrho_1\end{aligned} \qquad \text{and} \qquad \begin{aligned}\sigma_\circ &:= \sigma_l \circ \ldots \circ \sigma_1 \\ \sigma &:= \sigma_l \bullet \ldots \bullet \sigma_1\end{aligned}$$

We now have to show that $\sigma_\circ \circ \varrho_\circ$ is right-moving iff $\sigma_\circ$ and $\sigma \circ \varrho_\circ$ are right-moving.

**Proof** First observe that all $\varrho_i, \sigma_j$ are in $\mathbf{P}_a$ and, by Lemma 7.1, $\varrho_\circ$ is right-moving in any case. Hence $\varrho \in \mathbf{P}_a$ is well defined, extends $\varrho_\circ$ and fulfills $\mathbf{L}(\varrho) = \mathbf{L}(\varrho_\circ)$, by Proposition 7.3.

$\implies$
- $\sigma$ is well defined and extends $\sigma_\circ$, since $\varrho_\circ(a)$ is snaking along $\sigma_\circ$.
- Now, $\sigma(\varrho_\circ(a)) = \sigma_\circ(\varrho_\circ(a)) \geq \varrho_\circ(a)$. Consequently, $\sigma \circ \varrho_\circ$ is right-moving.
- Since $a \leq \varrho_\circ(a) \in \mathbf{L}(\sigma_1)$ and $\sigma_1(\varrho_\circ(a)) \in \mathbf{L}(\sigma_2)$, we have $\sigma_1(a) \leq \sigma_1(\varrho_\circ(a))$ and $\sigma_1(a) \in \mathbf{L}(\sigma_2)$ again. Inductively, $[\sigma_i \circ \ldots \circ \sigma_1](a) \in \mathbf{L}(\sigma_{i+1})$. Lemma 7.2 shows that $\sigma_\circ$ is right-moving.

$\impliedby$
- By Proposition 7.3, $\sigma \in \mathbf{P}_a$ is well defined and extends $\sigma_\circ$.
- By Lemma 7.2, $\varrho(a) = \varrho_\circ(a) \in \mathbf{L}(\sigma) = \mathbf{L}(\sigma_\circ)$. Hence, for all $i$
$$(\sigma_i \circ \ldots \circ \sigma_1)[a, \varrho(a)] \subseteq \mathbf{I}(\sigma_{i+1}).$$
- As $\sigma_\circ$ is right-moving, we have $(\sigma_i \circ \ldots \circ \sigma_1)(a) \in \mathbf{L}(\sigma_{i+1})$, hence even
$$(\sigma_i \circ \ldots \circ \sigma_1)[a, \varrho(a)] \subseteq \mathbf{L}(\sigma_{i+1}).$$
- Thus, $(\sigma_i \circ \ldots \circ \sigma_1)(\varrho(a)) \in \mathbf{L}(\sigma_{i+1})$. As $\sigma_\circ(\varrho(a)) = \sigma(\varrho(a)) = \sigma(\varrho_\circ(a)) \neq b$ by assumption, we get the proof from Lemma 7.2. **qed**

Inductively, we now can show that we may freely replace subproducts by the $\bullet$-product of the respective factors; then the original product is right-moving iff these $\circ$-subproducts are all right-moving as well as the "squeezed" product. For example, $\varrho_6 \circ \varrho_5 \circ \varrho_4 \circ \varrho_3 \circ \varrho_2 \circ \varrho_1$ is right-moving iff $\varrho_5 \circ \varrho_4 \circ \varrho_3$ and $\varrho_2 \circ \varrho_1$ as well as the "squeezed" product $\varrho_6 \circ [\varrho_5 \bullet \varrho_4 \bullet \varrho_3] \circ [\varrho_2 \bullet \varrho_1]$ are right-moving. A special case will be very relevant in a moment: $\varrho \circ \varrho \circ \varrho \circ \varrho \circ \varrho \circ \varrho$ is right-moving iff $\varrho \circ \varrho$ and $\varrho^2 \circ \varrho^2 \circ \varrho^2$ are right-moving. Here, $\varrho^2 \equiv \varrho \bullet \varrho$.

Using that $\varrho^k$ is right-moving for $k \leq \mathbf{m}(\varrho)$, we get immediately from the lemma above

**Corollary 7.12** Let $\mathbf{P}$ be a positive motion set, $\varrho \in \mathbf{P}_a$ and $k \leq \mathbf{m}(\varrho)$. Then
$$\varrho_\circ^{km} \text{ right-moving} \iff (\varrho^k)_\circ^m \text{ right-moving}.$$

Altogether, we have

**Proposition 7.13** Let $\mathbf{P}$ be a positive motion set, $\varrho \in \mathbf{P}_a$ and $k \leq \mathbf{m}(\varrho)$. Then we have
$$\mathbf{m}(\varrho^k) \leq \frac{\mathbf{m}(\varrho)}{k} < \mathbf{m}(\varrho^k) + 1.$$

### 7.4 Order of Elements

How often may we minimally shift an element in $[a, b]$ to the right by some $\varrho \in \mathbf{P}_a$ without throwing it out of $[a, b]$? The answer for $a$ appears to be already given by the multiplicity giving the number. But, this need not be correct. Indeed, first, we may enter from the left again, since $\mathbf{P}$ might be cyclic. And, second, as for the multiplicity of some $\varrho$, the boundary $b$ is critical; again, we will let the shift ending in $b$ not count here. To make these ideas more precise, we re-use the idea that led to the notion of multiplicity.

**Definition 7.3** Let $\mathbf{P}$ be a nontrivial positive motion set and let $\mathbf{1} \neq \varrho \in \mathbf{P}_a$.
Then the **order** $\mathbf{o}_\varrho(t)$ of $t \in [a, b]$ w.r.t. $\varrho$ is the maximal $k \in \mathbb{N}$, such that $t \in \mathrm{dom}\, \varrho^k$ with $t < \varrho(t) < \ldots < \varrho_\circ^k(t) < b$.

For convenience, we will assume any $t$ to be of order 0 if $\mathbf{P}$ is trivial. It is clear that the order of $a$ w.r.t. $\varrho$ is now precisely the multiplicity of $\varrho$.

**Lemma 7.14** Let $\mathbf{P}$ be a nontrivial positive motion set, let $\mathbf{1} \neq \varrho \in \mathbf{P}_a$ and let $t \in [a, b)$. Then
1. $\mathbf{o}_\varrho(t)$ is well defined with
$$\mathbf{o}_\varrho(t) = \max\{k \mid \varrho_\circ^i(t) \in \mathrm{int}\, \mathbf{L}(\varrho) \text{ for all } 0 \leq i < k\}.$$

2. $\mathbf{o}_\varrho(a)$ is the multiplicity $\mathbf{m}(\varrho)$.
3. $\mathbf{o}_\varrho$ is non-increasing with $\mathbf{o}_\varrho \circ \varrho = 1 + \mathbf{o}_\varrho$ on $\mathbf{L}(\varrho)$.
4. $\varrho_\circ^i$ equals $\varrho^i$ on $[a,t]$ for all $i \leq \mathbf{o}_\varrho(t)$.

**Proof** First observe that clearly the following two statements are equivalent:
- $t \in \mathrm{dom}\, \varrho_\circ^k$ and $\varrho_\circ^i(t) < \varrho_\circ^{i+1}(t) \equiv \varrho(\varrho_\circ^i(t)) < b$ for all $i < k$;
- $t \in \mathrm{dom}\, \varrho_\circ^i$ and $\varrho_\circ^i(t) \in \mathrm{int}\, \mathbf{L}(\varrho)$ for all $i < k$.

The equivalence remains if we change "$< b$" into "$\leq b$" in the upper line and the drop the condition to be in the interior in the lower line. Now, the proof follows easily:

1. If $\varrho_\circ^i(t) < \varrho_\circ^{i+1}(t)$ for all $i$, then $\varrho_\circ^i(t) \in \mathbf{L}(\varrho)$ for all $i$, hence $\varrho(t) = t$ by Proposition 5.1 and now even $\varrho = \mathbf{1}$. Thus, the maximum is attained and the formula above holds.
2. Trivial.
3. As $s \leq t \in \mathbf{L}(\varrho)$ implies $s \in \mathbf{L}(\varrho)$ and $\varrho(s) \leq \varrho(t)$, we get $\mathbf{o}_\varrho(s) \geq \mathbf{o}_\varrho(t)$ inductively. The formula is obvious.
4. Use that $\varrho_\circ^i(t) < b$ for all $i \leq \mathbf{o}_\varrho(t)$, whence $t$ is snaking along $\varrho_\circ^k$. The claim for the remaining elements of $[a,t]$ follows as in 3. **qed**

**Corollary 7.15** Let $\mathbf{P}$ be a nontrivial positive finite motion set, $\mathbf{1} \neq \varrho \in \mathbf{P}_a$ and $t \in (a,b)$. Then there is a unique $0 \leq k \leq \mathbf{m}(\varrho)$ with $\varrho^{-k}(t) \in (a, \varrho(a)]$.

**Proof** Let $\mathbf{o} := \mathbf{o}_\varrho(t)$ and $\mathbf{m} := \mathbf{m}(\varrho)$. By Lemma 7.14, $\varrho^\mathbf{o}(t) \notin \mathrm{int}\, \mathbf{L}(\varrho)$, but $\varrho^{\mathbf{o}-1}(t) \in \mathrm{int}\, \mathbf{L}(\varrho)$.

- Assume that $\varrho^\mathbf{m}(a) < \varrho^\mathbf{o}(t)$. Then $\varrho^\mathbf{o}(t) \in \mathbf{R}(\varrho^{-\mathbf{m}})$, hence $t$ is snaking along $\varrho^{-\mathbf{m}} \circ \varrho^\mathbf{o}$. In particular, $\varrho^{-\mathbf{m}+\mathbf{o}}(t)$ is well defined and equals $\varrho^{-\mathbf{m}}(\varrho^\mathbf{o}(t)) \in \mathbf{L}(\varrho^\mathbf{m})$. Assume that $\varrho(a) < \varrho^{-\mathbf{m}+\mathbf{o}}(t)$, then also $\varrho(a)$ is in $\mathrm{int}\, \mathbf{L}(\varrho^\mathbf{m})$, where $\varrho^\mathbf{m}$ coincides with $\varrho_\circ^\mathbf{m}$. Thus,

$$\varrho(a) \leq \varrho^\mathbf{m}(\varrho(a)) = \varrho_\circ^\mathbf{m}(\varrho(a)) \equiv \varrho(\varrho_\circ^\mathbf{m}(a)) = \varrho(\varrho^\mathbf{m}(a)),$$

giving $\varrho^\mathbf{m}(a) \in \mathbf{L}(\varrho)$. By definition of multiplicity, $\varrho^\mathbf{m}(\varrho(a)) = \varrho(\varrho^\mathbf{m}(a)) = b$. This, however, contradicts $\varrho(a) \in \mathrm{int}\, \mathbf{L}(\varrho^\mathbf{m})$.
- Assume that $\varrho^\mathbf{m}(a) \geq \varrho^\mathbf{o}(t)$. If we had $\varrho^{\mathbf{m}-1}(a) \geq \varrho^\mathbf{o}(t)$, then $\varrho^\mathbf{o}(t) \in \mathrm{int}\, \mathbf{L}(\varrho)$ which contradicts the lemma above. Hence $\varrho^{\mathbf{m}-1}(a) < \varrho^\mathbf{o}(t) \leq \varrho^\mathbf{m}(a)$. As the left-hand side is in $\mathbf{R}(\varrho^{1-\mathbf{m}})$, we get immediately $a < \varrho^{1-\mathbf{m}+\mathbf{o}}(t) \leq \varrho(a)$.

Uniqueness is trivial. **qed**

# 8 Finite Motion Sets

In this section we are going to prove our first classification theorems for finite motion sets $\mathbf{P}$. They rest on a few observations:

1. Recall that $\mathbf{P}_a^+$ can be identified with $\mathbf{P}^+(a)$, whence (in the nontrivial case) there is some $\mu \in \mathbf{P}_a^+$ with minimal $\mu(a) > a$. This can be understood as a minimal jump a right-shift can make. We will then prove that any further element in $\mathbf{P}_a^+$ is simply given by some nonnegative power of $\mu$. As $\mathbf{P}_b^+$ contains the inverses of the elements of $\mathbf{P}_a^+$, we get a complete description of $\mathbf{P}^+$. We only have still to distinguish between the two cases where $\mathbf{P}_a^+$ and $\mathbf{P}_b^+$ have trivial or nontrivial intersection. From the point of view of our prime examples this refers to the case where the interval $I \subseteq S^1$ is shorter or longer than a half-circle.

2. If $\mathbf{P}$ contains also negative elements, there are sort-of generic situations and some exceptional ones. In the first one, we can assume that the fixed point set of $\mathbf{P}^-$ contains at least two right or two left fixed points. In fact, then $\mathbf{P}^+$ is nontrivial. Exploiting now the properties of $\mu$, we see that the fixed point set is given by the $\mathbf{P}_a^+$-images of the two smallest fixed points (as long as these points are within $(a,b)$). Correspondingly, we see that $\mathbf{P}^-$ is generated by some reflection together with $\mu$. The only exceptions arise when there is at most a single left and at most a single right fixed point. These cases have to be studied in case-by-case analysis.

## 8.1 Shifts

Assume now that **P** is positive, finite and nontrivial. As $\mathbf{P} = \mathbf{P}_a \cup \mathbf{P}_b$, at least one of both sets is nontrivial. Since $\mathbf{P}_a = \mathbf{P}_b^{-1}$, this is true even for both sets. This now means that $\mathbf{P}_a$ contains at least one element that is larger than **1**. As $\mathbf{P}_a$ is finite, there is even a minimal element of that kind. Let us fix this in a definition which already anticipates the following theorem.

**Definition 8.1** Let **P** be a nontrivial positive motion set.
Then the **generator** of **P** is the minimizer $\mu$ of $\mathbf{P}_a \setminus \{\mathbf{1}\}$.

Recall that the ordering on $\mathbf{P}_a$ is the same as that on $\mathbf{P}(a) \subseteq [a,b]$ via the evaluation map $\varrho \longmapsto \varrho(a)$. Therefore, $\mu(a)$ is also the minimizer of $\mathbf{P}(a) \setminus \{a\}$. Finally, observe that the left domain of any $\varrho \neq \mathbf{1}$ is contained in that of $\mu$. In fact, assume $\mathbf{L}(\varrho)$ is non-empty, i.e., $\varrho \in \mathbf{P}_a$. If $\mathbf{L}(\varrho) \supset \mathbf{L}(\mu)$, then $\varrho(a) < \mu(a)$ by Proposition 6.2, hence $\varrho = \mathbf{1}$.

**Theorem 8.1** Let **P** be a nontrivial finite positive motion set and $\mu \in \mathbf{P}_a$ its generator. Then
$$\begin{aligned} \mathbf{P}_a &= \{\mu^k \mid 0 \leq k \leq \mathbf{m}(\varrho)\} \\ \mathbf{P}_b &= \{\mu^k \mid 0 \leq -k \leq \mathbf{m}(\varrho)\} \end{aligned}$$

**Proof** As $\mathbf{P}_b = \mathbf{P}_a^{-1}$, we only have to show that any $\varrho \in \mathbf{P}_a$ equals $\mu^k$ for some $0 \leq k \leq \mathbf{m}(\mu)$. We will show this for $k \leq \mathbf{m}(\mu)$ being maximal with $\mu^k \leq \varrho$.
- $k$ exists as $\mu^0 \equiv \mathbf{1} \leq \varrho$ and as $\mu^i \in \mathbf{P}_a$ for all $i \leq \mathbf{m}(\mu)$ by the definition of $\mathbf{m}(\mu)$.
- $\mu^{-k} \bullet \varrho \in \mathbf{P}_a$, as $a$ is snaking along $\mu^{-k} \circ \varrho$ by $\varrho(a) \in [\mu^k(a), b) \subseteq \mathbf{R}(\mu^{-k})$.
- $\mu^{-k} \bullet \varrho = \mathbf{1}$: If, otherwise, $\mu^{-k} \bullet \varrho > \mathbf{1}$, then even $\mu^{-k} \bullet \varrho \geq \mu$ by maximality of $k$.. Now, Corollary 5.12 implies that $\varrho(a) \geq \mu^{k+1}(a) \geq \mu^k(a)$ with $\mu(a) \in \mathbf{L}(\mu^k)$, hence $k+1 \leq \mathbf{m}(\varrho)$. This contradicts the maximality of $k$.
- Consequently, $\varrho = \mu^k$.  **qed**

The proposition above justifies to denote $\mu$ as generator. Note, however, that the proposition above does not exclude the case that $\mu^{k_1} = \mu^{k_2}$ for different $k_i$ with $|k_i| \leq \mathbf{m}(\mu)$. Of course, by Corollary 7.10, $k_1$ and $k_2$ cannot be both non-negative or both non-positive at the same time. Thus $k_1$ and $k_2$ have to be of opposite sign. We even have that $|k_1 - k_2|$ is of a fixed value.

**Proposition 8.2** Let **P** be a finite positive motion set with nontrivial $\mathbf{P}_a \cap \mathbf{P}_b$. Moreover, let $n$ be the minimal positive number with $\mu^n \in \mathbf{P}_b$, and let $\mathbf{u} := \mathbf{m}(\mu) + n$.
Then we have for all $|k_i| \leq \mathbf{m}(\mu)$
$$\mu^{k_1} = \mu^{k_2} \iff k_1 - k_2 \in \{-\mathbf{u}, 0, \mathbf{u}\}$$

**Proof**
- First observe that $\mu^{-n}$ equals $\mu^{\mathbf{m}}$ for $\mathbf{m} := \mathbf{m}(\mu)$. In fact, $\mu^n \in \mathbf{P}_b$ implies $\mu^{-n} \in \mathbf{P}_a$, hence $\mu^{-n} = \mu^k$ for some $0 \leq k \leq \mathbf{m}$. If we had $k < \mathbf{m}$, then, by Lemma 7.9, $\mu^{-n+1} = \mu^{-n} \bullet \mu = \mu^k \bullet \mu = \mu^{k+1} \in \mathbf{P}_a$ contradicting minimality of $n$. Hence, $k = \mathbf{m}$.
- Next, again by Lemma 7.9, we see that $\mu^{-n-i} = \mu^{\mathbf{m}-i}$ for all $i = 0, \ldots, \mathbf{m} - n$. In particular, this shows that $\mu^{k_1} = \mu^{k_2}$ for all $k_2 = k_1 \pm \mathbf{u}$ with $|k_i| \leq \mathbf{m}$.
- It remains to show the "$\Longrightarrow$" direction. For this, as we have seen above, we may assume that $k_1 < 0 < k_2$. Moreover, from $\mu^{-k_1} = \mu^{-k_2} \in \mathbf{P}_a^{-1} = \mathbf{P}_b$, we get $k_1 \leq -n$. Thus, $\mu^{k_2} = \mu^{k_1} = \mu^{k_1 + \mathbf{u}}$, hence $k_2 = k_1 + \mathbf{u}$ as $n = -\mathbf{m} + \mathbf{u} \leq k_1 + \mathbf{u} = k_1 + \mathbf{m} + n \leq \mathbf{m}$. Here we have used Corollary 7.10 and $\mu \neq \mathbf{1}$.  **qed**

**Corollary 8.3** Let **P** be a nontrivial finite positive motion set. Then we have
- If $\mathbf{P}_a \cap \mathbf{P}_b$ is nontrivial, then **P** has exactly **u** elements.
- If $\mathbf{P}_a \cap \mathbf{P}_b$ is trivial, then **P** has exactly $2\mathbf{m}(\mu) + 1$ elements.

**Theorem 8.4  Classification of Finite Positive Motion Sets**

Any finite nontrivial positive motion set is isomorphic to one prime integer-shift sets

| circumference of $S^1$ | length of interval $I$ | characterization |
|---|---|---|
| $\mathbf{u}$ | $\mathbf{m}(\mu)+\mathbf{r}$ | $\mathbf{P}_a \cap \mathbf{P}_b$ nontrivial |
| $2(\mathbf{m}(\mu)+1)$ | $\mathbf{m}(\mu)+\mathbf{r}$ | $\mathbf{P}_a \cap \mathbf{P}_b$ trivial |

Here $0 < \mathbf{r} \leq 1$ is some number with $\mathbf{r} = 1$ iff $\mu^{\mathbf{m}(\mu)+1}(a)$ is well-defined and equals $b$.

**Proof**  Denote again $\mathbf{m} := \mathbf{m}(\mu)$, and let $\mathbf{u} := 2(\mathbf{m}+1)$ if $\mathbf{P}_a \cap \mathbf{P}_b$ is trivial. Define the desired homeomorphism $\varphi : [a, b] \longrightarrow [0, \mathbf{m}+\mathbf{r}]$ as follows.
- Choose any positive homeomorphism $\varphi : [a, \mu(a)] \longrightarrow [0, 1]$.
- Define $\varphi : [\mu^i(a), \mu^{i+1}(a)] \longrightarrow [i, i+1]$ recursively by $\varphi := i + \varphi \circ \mu^{-i}$.
- Define $\mathbf{r} := \varphi(\mu^{-\mathbf{m}}(b))$ and let $\varphi := \mathbf{m} + \varphi \circ \mu^{-\mathbf{m}}$ to give $\varphi : [\mu^{\mathbf{m}}(a), b] \longrightarrow [\mathbf{m}, \mathbf{m}+\mathbf{r}]$.

One immediately checks that $\varphi$ is the desired homeomorphism that intertwines the given motion set with the prototypical one. **qed**

Note that $\mathbf{r}$ is arbitrary, unless 1. In fact, given any two points in the interior of $[0, 1]$, we can always find a homeomorphism of $[0, 1]$ that intertwines these two points.

## 8.2  Reflections – General Case

After having considered so far positive finite motion sets, we will now drop the positivity assumption. Nevertheless, the classifying Theorem 8.1 will again play an important rôle. In fact, we already know that the product of two reflections in a motion set $\mathbf{P}$ lies, if defined, in the positive motion set $\mathbf{P}^+$. This is finite again, hence contains only powers of the generator $\mu \in \mathbf{P}_a$. On the other hand, we will show that $\mathbf{P}_a$, i.e., the nonnegative powers of $\mu$, generate all possible fixed points (of reflections) out of the two minimal ones $\mathbf{x}_0$ and $\mathbf{x}_1$. Therefore, any reflection is given as a conjugation of the corresponding minimal reflections $\sigma_0$ and $\sigma_1$ by elements of $\mathbf{P}_a$. Even more, we will show that $\sigma_1 = \sigma_0 \bullet \mu$. Therefore, we will just need a single reflection together with the generator of the positive part to generate the full motion set.

But, this is not the full truth yet. Note that we have made an important assumption: there is indeed an nontrivial element in $\mathbf{P}^+$. This, however, is only guaranteed if there are at least two different reflections in $\mathbf{P}_a$ or in $\mathbf{P}_b$. In fact, let us have a look at $S^1$ and an interval $I$ comprising more than half the circle. Then the reflection at any interior point of $I$ forms a well-defined motion set together with the identity. The similar is true for the reflection at any diameter that hits $I$ twice in its interior. Here, we even have two fixed points, namely, the two intersection points with the diameter. However, one point is a left fixed point, one a right one. Finally, assume that $I$ is a half-circle, such that it contains two of the three cubic roots of unity in its interior. One easily checks that the two corresponding reflections have no common domain, whence they cannot be multiplied. Also their reflection points fall in the different categories as one being left and the other one right. Nevertheless, the reflections form a motion set together with $\mathbf{1}$. One can easily check that (up to isomorphism) the three[19] cases comprise all possibilities for having trivial positive part.

In the following, it will usually suffice for our purposes to assume that there are at least two left (or two right) fixed points. So let us start with the formalization of the initial argument.

**Lemma 8.5**  Let $\mathbf{P}$ be a motion set.
Then $\mathbf{P}^+$ is nontrivial as soon as $\mathbf{F}_a$ or $\mathbf{F}_b$ contain at least two elements.

---
[19]We will subdivide the first case into whether the domain is full $[a, b]$ or not.

**Proof** Assume that $|\mathbf{F}_a| \geq 2$. Denote the minimal elements of $\mathbf{F}_a$ by $\mathbf{x}_0 < \mathbf{x}_1$ and the corresponding reflections by $\sigma_0$ and $\sigma_1$, respectively. By Proposition 6.4, we have $\mathbf{L}(\sigma_0) \subset \mathbf{L}(\sigma_1)$, hence $\sigma_1 \bullet \sigma_0 \in \mathbf{P}_a^+$. Observe that $\sigma_1 \bullet \sigma_0 \neq \mathbf{1}$ as otherwise $\sigma_1 = \sigma_0$, hence $\mathbf{x}_0 = \mathbf{x}_1$. In particular, we see that $\mathbf{P}^+$ is not trivial, i.e., contains more elements than just $\mathbf{1}$. – The proof for $|\mathbf{F}_b| \geq 2$ is analogous. **qed**

**Lemma 8.6** Let $\mathbf{P}$ be a motion set with nontrivial finite $\mathbf{P}^-$, and let $\sigma_0$ be the reflection having fixed point $\mathbf{x}_0 := \min \mathbf{F}_a$. Then at least one of the following statements is true:
1. $\mathbf{F}$ consists of a single element, namely $\mathbf{x}_0$.
2. $\sigma_0(a) \leq \mathbf{x}$ for all $\mathbf{x} \in \mathbf{F}$ with $\mathbf{x} \neq \mathbf{x}_0$.

**Proof** Assume $\mathbf{x}_0 \leq \mathbf{x} < \sigma_0(a) \in \mathbf{L}(\sigma_0)$ for some $\mathbf{x} \in \mathbf{F}$. Then $\mathbf{x}_0 = \sigma_0(\mathbf{x}_0) \geq \sigma_0(\mathbf{x}) > a$, hence $\sigma_0(\mathbf{x}) \neq a, b$. Consequently, $\sigma_0(\mathbf{x})$ is again a fixed point. As $\mathbf{F}_a \leq \mathbf{F}_b$, $\mathbf{x}_0$ even minimizes full $\mathbf{F}$, whence $\sigma_0(\mathbf{x})$ must be $\mathbf{x}_0$, giving $\mathbf{x}_0 = \mathbf{x}$. **qed**

Until the end of this subsection, let us be given a $\mathbf{P}$ is a finite motion set for which $\mathbf{F}_a$ contains two fixed points $\mathbf{x}_0 < \mathbf{x}_1$. We may choose them minimally. As usual, we denote the corresponding reflections by $\sigma_0$ and $\sigma_1$, respectively. Moreover, since $\mathbf{P}^+$ is nontrivial, it has a generator, say, $\mu$.

**Lemma 8.7**
- $\mu \circ \sigma_0$ and $\sigma_1$ coincide on $\mathbf{L}(\sigma_0)$, hence $\mu \bullet \sigma_0 = \sigma_1$.
- $\sigma_1 \circ \sigma_0$ and $\mu$ coincide on $\mathbf{L}(\sigma_0)$, hence $\sigma_1 \bullet \sigma_0 = \mu$.

**Proof**
- As in the proof of Lemma 8.5, we see that $\mathbf{1} \neq \sigma_1 \bullet \sigma_0 \in \mathbf{P}_a^+$, whence $\mu \leq \sigma_1 \bullet \sigma_0$ by minimality of $\mu$. Thus, $\mathbf{L}(\mu) \supseteq \mathbf{L}(\sigma_1 \bullet \sigma_0) \supseteq \mathbf{L}(\sigma_0) \ni \sigma_0(a)$ by Corrolary 6.23, hence

$$[\mu \bullet \sigma_0](a) \;=\; \mu(\sigma_0(a)) \;\leq\; [\sigma_1 \bullet \sigma_0](\sigma_0(a)) \;=\; [\sigma_1 \bullet \sigma_0 \bullet \sigma_0](a) \;=\; \sigma_1(a).$$

  As $\mu \bullet \sigma_0 \in \mathbf{P}_a^-$, we know from Proposition 6.4 that the (left) fixed point $\mathbf{x}$ of $\mu \bullet \sigma_0$ is smaller or equal to that of $\sigma_1$ being $\mathbf{x}_1$. By assumption, $\mathbf{x}$ must be $\mathbf{x}_0$ or $\mathbf{x}_1$. If it is $\mathbf{x}_1$, both functions coincide by the proposition just cited and we are done. If it is $\mathbf{x}_0$, then $\mu \bullet \sigma_0$ and $\sigma_0$ would coincide, implying $\mu = \mathbf{1}$, giving a contradiction.
- The second assertion follows, since $\mu \circ \sigma_0 \circ \sigma_0$ equals $\mu$ and, as just proven, also $\sigma_1 \circ \sigma_0$ on $\mathbf{L}(\sigma_0) \subseteq \mathbf{L}(\mu)$. **qed**

**Lemma 8.8** We have $a < \mathbf{x}_0 < \mathbf{x}_1 \leq \mu(a) \in \mathbf{L}(\sigma_1)$.
Moreover, $\mathbf{x}_0$ and $\mathbf{x}_1$ are the only fixed points within $[a, \mu(a)]$.

**Proof**
- If $\mu(a) < \mathbf{x}_1$, then $\mathbf{x}_1 \in \mathbf{R}(\mu^{-1})$ with $\mu^{-1}(\mathbf{x}_1) \neq a, b$. Because, by Proposition 6.17, $\mu^{-1}(\mathbf{x}_1) < \mathbf{x}_1$ is a fixed point, it coincides with $\mathbf{x}_0$ by assumption. This, however, is impossible, since now $[\mu \bullet \sigma_0](\mathbf{x}_0) = \mu(\mathbf{x}_0) = \mathbf{x}_1 = [\mu \bullet \sigma_0](\mathbf{x}_1)$ implies $\mathbf{x}_0 = \mathbf{x}_1$.
- Observe next that $\mathbf{x}_0 \in \mathbf{L}(\sigma_1) \cap \mathbf{L}(\mu)$, whence

$$\mu(a) \;<\; \mu(\mathbf{x}_0) \;=\; \mu(\sigma_0(\mathbf{x}_0)) \;=\; \sigma_1(\mathbf{x}_0) \;\in\; \mathbf{L}(\sigma_1).$$

- Assume that $\mathbf{x} \in [a, \mu(a)]$ is a fixed point of some $\sigma \in \mathbf{P}^-$. We may assume $\mathbf{x} > \mathbf{x}_1$. Since, as just seen, $\mathbf{x} \in \mathbf{L}(\sigma_1)$, we have $[\sigma_1 \bullet \sigma](\mathbf{x}) = \sigma_1(\mathbf{x}) < \sigma_1(\mathbf{x}_1) = \mathbf{x}_1 < \mathbf{x}$, hence $\varrho := \sigma_1 \bullet \sigma \in \mathbf{P}_b$. Therefore, $\mathbf{1} \neq \varrho^{-1} \in \mathbf{P}_a$ and $\varrho(\mathbf{x}) \in \mathbf{L}(\varrho^{-1}) \subseteq \mathbf{L}(\mu)$. Therefore, $\mu(a) \leq \mu(\varrho(\mathbf{x})) \leq \varrho^{-1}(\varrho(\mathbf{x})) = \mathbf{x}$, hence $\mathbf{x} = \mu(a)$ and $\mu = \varrho^{-1}$. Now, $\sigma_1(\mathbf{x}) = \varrho(\mathbf{x}) = \mu^{-1}(\mathbf{x}) = a$, hence $\mu(a) = \mathbf{x} = \sigma_1(a) = \mu(\sigma_0(a))$. This gives, finally, $\sigma_0(a) = a$. Contradiction. **qed**

Having now investigated the behaviour of the two smallest fixed points, we will construct the other ones. For this, define

$$\begin{aligned}\mathbf{x}_{2i} &:= \mu^i(\mathbf{x}_0) &&\text{for } i \leq \mathbf{o}(\mathbf{x}_0) \\ \mathbf{x}_{2i+1} &:= \mu^i(\mathbf{x}_1) &&\text{for } i \leq \mathbf{o}(\mathbf{x}_1)\end{aligned}$$

and

$$\begin{aligned}\tau_{2i} &:= \mu^i &&\text{for } i \leq \mathbf{o}(\mathbf{x}_0)\\ \tau_{2i+1} &:= \mu^i \bullet \sigma_1 &&\text{for } i \leq \mathbf{o}(\mathbf{x}_1)\end{aligned}$$

with $\sigma_1 \in \mathbf{P}^-$ being the reflection that corresponds to the fixed point $\mathbf{x}_1$. Here, we shortly wrote $\mathbf{o}$ instead of $\mathbf{o}_\mu$; we keep this notion until the end of this section.

**Proposition 8.9** Let $\mathbf{P}$ be a finite motion set with $|\mathbf{F}_a| \geq 2$ and let $n := \mathbf{o}(\mathbf{x}_0) + \mathbf{o}(\mathbf{x}_1) + 1$. Then the points $\mathbf{x}_0 < \mathbf{x}_1 < \ldots < \mathbf{x}_n$ comprise all fixed points in $\mathbf{F}$.

**Proof** $\mathbf{x}_k$ is well-defined by the definition of the order function $\mathbf{o}$. Starting with Lemma 8.8, we see inductively that $\mu^i(a) < \mu^i(\mathbf{x}_0) < \mu^i(\mathbf{x}_1) \leq \mu^{i+1}(a)$ as long as the exponent does not surpass the order of the argument. Here, we have used monotonicity on $\mathbf{L}(\mu)$. Next, we deduce from Proposition 6.17 that each $\mathbf{x}_k$ is indeed a fixed point. Thus, we are left with the proof that any fixed point is given by some of the $\mathbf{x}_k$. Indeed, let $\mathbf{x}$ be any fixed point. By Corollary 7.15, there is a $j$ with $\mu^{-j}(\mathbf{x}) \in (a, \mu(a)]$. By Proposition 6.17, it is again a fixed point; by Lemma 8.8, it must be $\mathbf{x}_0$ or $\mathbf{x}_1$. The claim is now obvious from the definition of $\mathbf{x}_k$. **qed**

**Proposition 8.10** Let $\mathbf{P}$ be a finite motion set with $|\mathbf{F}_a| \geq 2$, and let $n := \mathbf{o}(\mathbf{x}_0) + \mathbf{o}(\mathbf{x}_1) + 1$. Then we have $n \geq 2$ with $\mathbf{x}_2 = \sigma_1(\mathbf{x}_0)$, and the following maps

$$\begin{aligned}\sigma_0 \equiv \tau_k : &\quad [\mathbf{x}_0, \sigma_0(a)] &&\longrightarrow [a, \mathbf{x}_0] &&\text{for } k = -1\\ \tau_k : &\quad [\mathbf{x}_0, \mathbf{x}_1] &&\longrightarrow [\mathbf{x}_k, \mathbf{x}_{k+1}] &&\text{for } 0 \leq k < n\\ \tau_k : &\quad [\mathbf{x}_0, \tau_n^{-1}(b)] &&\longrightarrow [\mathbf{x}_n, b] &&\text{for even } k = n\\ \tau_k : &\quad [\tau_n^{-1}(b), \mathbf{x}_1] &&\longrightarrow [\mathbf{x}_n, b] &&\text{for odd } k = n\end{aligned}$$

are homeomorphisms.

**Proof** From above, we have $\sigma_1(\mathbf{x}_0) = \mu(\sigma_0(\mathbf{x}_0)) = \mu(\mathbf{x}_0) \equiv \mathbf{x}_2$, hence $\mathbf{o}(\mathbf{x}_0) \geq 1$. Next, observe that the homeomorphy claims are obvious for the cases of $\sigma_0$ and of $\tau_k$ for even $k$.

- For odd $k = 2i + 1 < n$, we have $2i < n - 1 = \mathbf{o}(\mathbf{x}_0) + \mathbf{o}(\mathbf{x}_1)$, hence $\mathbf{o}(\mathbf{x}_0) \geq i + 1$ and $\mathbf{o}(\mathbf{x}_1) \geq i$ as both orders differ by at most 1 (join $\mathbf{x}_0, \mathbf{x}_1 \in [a, \mu(a)]$ with Lemma 7.14.) Since now $\sigma_1$ maps $[\mathbf{x}_0, \mathbf{x}_1]$ to $[\mathbf{x}_1, \mathbf{x}_2] \subseteq \mathbf{L}(\mu^i)$, the interval $[\mathbf{x}_0, \mathbf{x}_1]$ is contained in $\mathbf{I}(\mu^i \circ \sigma_1)$, whence $\tau_{2i+1} \equiv \tau_k$ is well defined. Now, the statement is clear from the definition of $\mathbf{x}_k$ and $\mathbf{x}_{k+1}$.
- For odd $k = 2i + 1 = n$, we have $\mathbf{o}(\mathbf{x}_0) = \mathbf{o}(\mathbf{x}_1) = i \geq 1$, as above. Define $t := \mu^{-i}(b)$. Then $\mathbf{o} \geq i$ on $[a, t]$ and $\mathbf{o} < i$ on $[t, b]$ by Lemma 7.14. From $\mathbf{o}(\mathbf{x}_2) \equiv \mathbf{o}(\mu(\mathbf{x}_0)) = \mathbf{o}(\mathbf{x}_0) - 1 = i - 1$, we get $\mathbf{x}_1 < t \leq \sigma_1(\mathbf{x}_0) \in \mathbf{L}(\sigma_1)$. Now, $\sigma_1$ maps $[\sigma_1(t), \mathbf{x}_1]$ to $[\mathbf{x}_1, t] \subseteq \mathbf{L}(\mu^i)$, whence $\tau_n$ is well defined with the desired property. **qed**

**Proposition 8.11** Let $\mathbf{P}$ be a finite motion set with $|\mathbf{F}_a| \geq 2$, and let $n := \mathbf{o}(\mathbf{x}_0) + \mathbf{o}(\mathbf{x}_1) + 1$. Then $\mathbf{x}_k$ is a fixed point of $\sigma_k = \tau_k \bullet \tau_{k-1}^{-1}$ for all $0 \leq k \leq n$.

**Proof** The previous proposition shows that $\tau_{k-1}^{-1}$ maps $[\mathbf{x}_{k-1}, \mathbf{x}_k]$ to $[\mathbf{x}_0, \mathbf{x}_1]$ for $k = 1, \ldots, n$, whereas $\tau_k$ is defined at least on some nontrivial subinterval thereof. Thus, $\tau_k \bullet \tau_{k-1}^{-1} \in \mathbf{P}^-$. Argue similarly for $k = 0$. One easily checks that $\mathbf{x}_k$ is the respective fixed point. **qed**

As the powers of $\mu$ span $\mathbf{P}^+$ and as any element in $\mathbf{P}^-$ is given by its fixed points, we have

**Corollary 8.12** Any finite motion set $\mathbf{P}$ with $|\mathbf{F}_a| \geq 2$ is generated by $\mu$ and $\sigma_1$.

## 8.3 Reflections – Special Cases

Now, only the cases remain where $\mathbf{F}_a$ and $\mathbf{F}_b$ contain at most a single element. Moreover, we may assume that $\mathbf{F}$ contains at least one element, as otherwise $\mathbf{P}$ is positive. Finally, we may assume that $\mathbf{F}_a$ contains exactly one element, namely $\mathbf{x}_0$, corresponding to the reflection $\sigma_0$; otherwise conjugate $\mathbf{P}$ with a flip exchanging the left and the right fixed points.

For further specification, consider first the case that $\mathbf{P}^+$ is nontrivial. Let for the moment $\varrho$ be any element in $\mathbf{P}_a^+$ that is not the identity. Then, $\varrho \bullet \sigma_0$ is a well-defined element in $\mathbf{P}^-$. Moreover, $\varrho \circ \sigma_0 \circ \sigma_0$ is defined at least on the nontrivial intersection $\mathbf{L}(\varrho) \cap \mathbf{L}(\sigma_0)$ and coincides with $\varrho$ there. Consequently, $\varrho \bullet \sigma_0 \bullet \sigma_0 = \varrho$. Thus, $\varrho \bullet \sigma_0 = \sigma_0$ would imply $\varrho = \mathbf{1}$, which is excluded. Therefore, $\varrho \bullet \sigma_0$ must equal another reflection $\sigma_1 \in \mathbf{P}^-$. Recall that $\mathbf{P}^-$ was assumed to contain at most two elements, whence $\sigma_1$ cannot depend on the shift $\varrho$ chosen in the beginning. As, moreover, $\sigma_1 \bullet \sigma_0 = (\varrho \bullet \sigma_0) \bullet \sigma_0 = \varrho$, for all $\varrho \neq \mathbf{1}$, we see that $\mathbf{P}_a^+$ can contain just a single element, namely $\mu$, beyond the identity. The only remaining options comprise $\mu = \mu^{-1}$ or not.

If, on the other hand, $\sigma_1$ and $\sigma_0$ are two reflections that can be multiplied, then their product is a positive element in $\mathbf{P}$. If it were $\mathbf{1}$, then $\sigma_1 = (\sigma_1 \bullet \sigma_0) \bullet \sigma_0 = \sigma_0$. Otherwise, the positive part of $\mathbf{P}$ is nontrivial, whence we are in the situation above.

It is now an easy exercise that exactly the situations of Table 2 on page 50 remain. Indeed any of them occurs as a motion set on $S^1$. To see this, let $S^1$ have circumference $\mathbf{u}$ that allows to identify $S^1$ with $\mathbb{R}/\mathbf{u}\mathbb{Z}$. Moreover, let $\sigma_0$ be the reflection at the horizontal diameter, having fixed points at $\frac{1}{2}\mathbf{u}\mathbb{Z}$. Also, we let $\sigma_1$ always be a reflection on a diameter; here, on one that passes $2 \in S^1$. This corresponds to the vertical diameter for $\mathbf{u} = 8$ and to the increasing diagonal diameter for $\mathbf{u} = 16$. Then, we can reproduce all the cases above for an appropriate choice of the interval $[a,b] \subseteq S^1$, see Table 3 on page 50. Of course, the restrictions to $[a,b]$ of the reflections above will again be denoted by $\sigma_0$ and $\sigma_1$, respectively.

| # | $|\mathbf{F}_a|$ | $|\mathbf{F}_b|$ | $|\mathbf{F}|$ | $|\mathbf{P}|$ | $\mathbf{P}^+$ | $\mathbf{P}^-$ | with |
|---|---|---|---|---|---|---|---|
| 1 | 1 | 0 | 1 | 2 | $\{\mathbf{1}\}$ | $\{\sigma_0\}$ | $\sigma_0$ not defined on full $[a,b]$ |
| 2 | 1 | 1 | 1 | 2 | $\{\mathbf{1}\}$ | $\{\sigma_0\}$ | $\sigma_0$ flipping $[a,b]$ |
| 3 | 1 | 1 | 2 | 2 | $\{\mathbf{1}\}$ | $\{\sigma_0\}$ | $\sigma_0$ has two fixed points |
| 4 | 1 | 1 | 2 | 3 | $\{\mathbf{1}\}$ | $\{\sigma_0, \sigma_1\}$ | $\sigma_1 \bullet \sigma_0$ not defined |
| 5 | 1 | 1 | 2 | 4 | $\{\mathbf{1}, \mu\}$ | $\{\sigma_0, \sigma_1\}$ | $\sigma_1 \bullet \sigma_0 = \mu = \mu^{-1}$ |
| 6 | 1 | 1 | 2 | 5 | $\{\mu^{-1}, \mathbf{1}, \mu\}$ | $\{\sigma_0, \sigma_1\}$ | $\sigma_1 \bullet \sigma_0 = \mu \neq \mu^{-1}$ |

Table 2: Motion Sets with Few Fixed Points (Abstract)

| # | circumference $\mathbf{u}$ | interval $[a,b]$ | fixed points 0 | 2 | 4 | shift length $\mu$ |
|---|---|---|---|---|---|---|
| 1 | 8 | $[-1, 2]$ | $\sigma_0$ | | | |
| 2 | 8 | $[-2, 2]$ | $\sigma_0$ | | | |
| 3 | 8 | $[-1, 5]$ | $\sigma_0$ | | $\sigma_0$ | |
| 4 | 8 | $[-1, 3]$ | $\sigma_0$ | $\sigma_1$ | | |
| 5 | 8 | $[-2, 4]$ | $\sigma_0$ | $\sigma_1$ | | 4 |
| 6 | 16 | $[-2, 4]$ | $\sigma_0$ | $\sigma_1$ | | 4 |

Table 3: Motion Sets with Few Fixed Points (Concrete)

## 8.4 Classification

**Theorem 8.13  Classification of Finite Motion Sets**

Any finite motion set is isomorphic to one of the finite prime examples on $S^1$. In particular, if it is nontrivial, we can assume that the shifts all have integer step size and we can make the following choices[20]:

| if ... is | nontrivial, ... | trivial, ... | can be chosen as ... |
|---|---|---|---|
| $\mathbf{P}_a^+ \cap \mathbf{P}_b^+$ | $|\mathbf{P}^+|$ | $2|\mathbf{P}^+| + 2$ or more | circumference of $S^1$ |
| $\mathbf{P}^-$ | $\frac{1}{2}\mathbb{Z} \cap \operatorname{int} I$ | — | reflection points |
| $\mathbf{P}^+$ | $\mathbf{m} + \mathbf{r}$ | $\mathbf{r}$ | length of interval $I$ |

Here $\mathbf{m}$ denotes the multiplicity of the generator $\mu$ of $\mathbf{P}^+$. Moreover, $0 < \mathbf{r} \leq 1$ is some number with $\mathbf{r} = 1$ iff $\mu^{\mathbf{m}+1}(a)$ is well-defined and equals $b$. Moreover, if $\mathbf{P}^-$ is nontrivial, then $\mathbf{r}$ has to be chosen, such that the number of half-integers in the interior of $I$ equals the number $|\mathbf{F}|$ of reflecting fixed points. Finally, the interval has to be starting or ending at a half-integer iff $\mathbf{x}_1 = \mu(a)$ or $\mathbf{x}_{|\mathbf{F}|} = \mu^{-1}(b)$, respectively.

The idea of the proof is the same as that for finite positive motion sets. Basically, the proof is a simple, but toilsome case-by-case analysis. As this theorem will not be relevant for our ultimate goal, namely the symmetry behaviour of paths, we refrain from giving this proof here and leave it as an exercise.

## 9  Free Intervals

The basic constituent of a brick path is a free segment. This is a path that equals its $g$-translate as soon as it non-trivially overlaps it. In the particular situaton of $\gamma$ being a path in $\mathbb{R}$, where $G = \mathbb{Z}$ is acting as integer shifts, we see that $\gamma$ overlaps $\varphi_1 \circ \gamma$ nontrivially iff the length of $\gamma$ is larger than 1. Transferred to the reparametrization function, this means that $\varrho_1^{-1}(I \cap \operatorname{dom} \varrho_1) \cap I$ with $I = \operatorname{dom} \gamma$ is non-trivial. Indeed, this is a general property as we have seen in Proposition 2.13 that $\gamma$ is a free segment iff $\varrho^{-1}(I) \cap I$ implies $\varrho = \mathbf{1}$. Similarly, we have already mentioned that a path is a concatenation of translates of free segments (possibly cut at the ends) iff the sets $\varrho_g^{-1}(I)$ with $g$ running over $G$ cover $I := \operatorname{dom} \gamma$. This transfers the two crucial notions for brick paths to the level of motion sets. Let us fix these observations in

**Definition 9.1**  An interval $I$ is called
- **free** iff it is nontrivial and for all $\varrho \in \mathbf{P}$
$$\varrho^{-1}(I) \cap I \text{ nontrivial} \iff \varrho = \mathbf{1};$$
- **generating** iff it is closed, free and $\{\varrho^{-1}(I) \mid \varrho \in \mathbf{P}\}$ covers $[a, b]$.

Obviously, any nontrivial subinterval of a free interval is free again.

We are now going to identify explicitly the free and the generating intervals of finite motion sets. We start with some useful criteria, continue with positive motion sets and end up with the general case. In the positive case the generating intervals are given by one "jump" of $\mu$, i.e., by the intervals $[t, \mu(t)]$ with $t \in \mathbf{L}(\mu)$. In the non-positive case the intervals between any two neighbouring fixed points are generating. There might be further intervals containing $a$ or $b$, but this depends on the special situation. In any case, there is a generating interval for any finite motion set.

---

[20] Reading guide: For each of the four entities choose the column according to whether the expression in the right column is trivial or not in the respective situations. Note that in some lines, the left middle column may apply and in some other lines, the right middle column. Finally, note that in some cases the given choice is the only possible.

## 9.1 Equivalent Criteria

**Lemma 9.1** The following statements are equivalent for any homeomorphic $\mathbf{P}$:
1. The interval $I$ is free.
2. $\varrho = \mathbf{1}$ provided $\varrho^{-1}(I) \cap I$ is nontrivial.
3. $\varrho = \mathbf{1}$ provided $\varrho(I \cap \operatorname{dom} \varrho) \cap I$ is nontrivial.

If $\mathbf{P}$ is a positive motion set, they are also equivalent to any of the following:

4. $\varrho = \mathbf{1}$ provided $\varrho(I \cap \mathbf{L}(\varrho)) \cap I$ is nontrivial.
5. $\varrho = \mathbf{1}$ provided $\varrho(I \cap \mathbf{R}(\varrho)) \cap I$ is nontrivial.

**Proof**  1. $\iff$ 2. Trivial.

2. $\iff$ 3. Use $\varrho^{-1}(I) \cap I = \varrho^{-1}(I \cap \varrho(I \cap \operatorname{dom} \varrho))$ and the fact that $\varrho^{-1}$ is homeomorphic on $I \cap \varrho(I \cap \operatorname{dom} \varrho) \subseteq \operatorname{im} \varrho \equiv \operatorname{dom} \varrho^{-1}$.

4. $\iff$ 5. Use $\varrho(\mathbf{L}(\varrho)) = \mathbf{R}(\varrho^{-1})$ to obtain $\varrho(I \cap \mathbf{L}(\varrho)) \cap I = \varrho[I \cap \varrho^{-1}(I \cap \mathbf{R}(\varrho^{-1}))]$. Hence, $\varrho(I \cap \mathbf{L}(\varrho)) \cap I$ is nontrivial iff $\varrho^{-1}(I \cap \mathbf{R}(\varrho^{-1})) \cap I$ is nontrivial. Since $\varrho = \mathbf{1}$ iff $\varrho^{-1} = \mathbf{1}$, we get the desired equivalence.

5. $\implies$ 3. Since $\operatorname{dom} \varrho = \mathbf{L}(\varrho) \cup \mathbf{R}(\varrho)$, the nontriviality of $\varrho(I \cap \operatorname{dom} \varrho) \cap I$ implies that of $\varrho(I \cap \mathbf{L}(\varrho)) \cap I$ or $\varrho(I \cap \mathbf{R}(\varrho)) \cap I$, hence $\varrho = \mathbf{1}$. Here we used that 4. and 5. are equivalent.

3. $\implies$ 4. As $\mathbf{L}(\varrho) \subseteq \operatorname{dom} \varrho$, this is obvious.   **qed**

**Lemma 9.2** Let $I$ be a non-trivial interval, $M \subseteq [a,b]$ be arbitrary and let $\varrho$ extend $\sigma$. Then
$$I \subseteq \sigma^{-1}(M) \implies I \subseteq \varrho^{-1}(M).$$

**Proof** If $I \subseteq \sigma^{-1}(M)$, then $I \subseteq \operatorname{dom} \sigma$, hence $I \subseteq \mathbf{I}(\sigma)$. Consequently, $\varrho$ and $\sigma$ coincide on $I$; in particular, $I \subseteq \mathbf{I}(\varrho)$. Now, any $t \in I$ fulfills $\varrho(t) = \sigma(t) \in M$, hence $t \in \varrho^{-1}(M)$.   **qed**

**Proposition 9.3** Let $\mathbf{P}$ be a motion set. Then for all $\sigma \in \mathbf{P}$ and all nontrivial intervals $I$
$$I \text{ free} \implies \sigma^{-1}(I) \text{ free or trivial}.$$

**Proof** Assume $J$ is a nontrivial interval in $\varrho^{-1}(\sigma^{-1}(I)) \cap \sigma^{-1}(I)$. Now, $\sigma(J)$ is a well-defined nontrivial interval, since $\sigma^{-1}(I) \subseteq \operatorname{dom} \sigma$. It is, moreover, contained in $[\sigma \circ \varrho \circ \sigma^{-1}]^{-1}(I)$, hence in $\mathbf{I}(\sigma \circ \varrho \circ \sigma^{-1})$. Thus, $\sigma \bullet \varrho \bullet \sigma^{-1}$ is well defined and extends $\sigma \circ \varrho \circ \sigma^{-1}$. By Lemma 9.2, $\sigma(J)$ is contained in $[\sigma \bullet \varrho \bullet \sigma^{-1}]^{-1}(I)$. At the same time, it is contained in $I$. Now, by freeness of $I$, we get $\sigma \bullet \varrho \bullet \sigma^{-1} = \mathbf{1}$, hence $\varrho = \mathbf{1}$ by Lemma 4.11. Thus, $\sigma^{-1}(I)$ is free by Lemma 9.1 above.   **qed**

## 9.2 Positive Finite Motion Sets

For positive finite motion sets, the classification of free intervals is rather easy. In fact, an interval is free iff it fits into an interval $[s, \mu(s)]$. Here and throughout the whole section, $\mu$ will be the generator of $\mathbf{P}^+$, provided $\mathbf{P}^+$ is non-trivial.

**Proposition 9.4** Let $\mathbf{P}$ be a nontrivial finite positive motion set. Then
$$[s,t] \subseteq [a,b] \text{ is free} \iff t \leq \mu(s) \text{ or } s \notin \mathbf{L}(\mu)$$

**Proof** $\implies$ Assume $s \in \mathbf{L}(\mu)$ and $\mu(s) < t$. Then

$t \in \mathbf{L}(\mu) \implies [s,t] \cap \mathbf{L}(\mu) = [s,t] \implies \mu([s,t] \cap \mathbf{L}(\mu)) = [\mu(s), \mu(t)]$
$t \notin \mathbf{L}(\mu) \implies [s,t] \cap \mathbf{L}(\mu) = [s, \mu^{-1}(b)] \implies \mu([s,t] \cap \mathbf{L}(\mu)) = [\mu(s), b]$

In both cases, the resulting interval intersects $[s,t]$ in the nontrivial interval $[\mu(s), t]$. As $\mu \neq \mathbf{1}$, the interval $[s,t]$ not free by Lemma 9.1.

⇐ Let $\varrho \in \mathbf{P}$ with $\varrho \neq \mathbf{1}$. One easily shows

$$\varrho\bigl([s,t] \cap \mathbf{L}(\varrho)\bigr) = \begin{cases} [\varrho(s), \varrho(t)] & \text{for } s \in \mathbf{L}(\varrho),\, t \in \mathbf{L}(\varrho) \\ [\varrho(s), b] & \text{for } s \in \mathbf{L}(\varrho),\, t \notin \mathbf{L}(\varrho) \\ \varnothing & \text{for } s \notin \mathbf{L}(\varrho),\, t \notin \mathbf{L}(\varrho) \end{cases}$$

In the upper two cases, we have $s \in \mathbf{L}(\varrho) \subseteq \mathbf{L}(\mu)$ by choice of $\mu$ (see the lines preceding Theorem 8.1), hence $\varrho(s) \geq \mu(s) \geq t$. Thus, $\varrho\bigl([s,t] \cap \mathbf{L}(\varrho)\bigr)$ intersects $[s,t]$ at most in $t$. Lemma 9.1 gives the proof.

**qed**

**Corollary 9.5** Let $\mathbf{P}$ be a nontrivial finite positive motion set.
Then the maximal free intervals are given by $[s, \mu(s)]$ with $s$ running over $\mathbf{L}(\mu)$.

For the trivial motion set, of course, $[a,b]$ is free and maximal.

## 9.3 General Finite Motion Sets

We are now going to identify the free intervals for non-positive finite motion sets. First observe that no free interval can contain a fixed point in its interior as the reflection w.r.t. the fixed point overlaps the left with the right hand side. On the other hand, we will see that any interval between neighbouring fixed points is free. Up to the boundary points, this completely classifies the free intervals.

Before we start, let us denote by $\overline{\mathbf{F}} := \mathbf{F} \cup \{a,b\}$ the set all fixed points of reflections and all boundary points.

**Lemma 9.6** Let $\mathbf{P}$ be a motion set with finite $\mathbf{P}^+$, and let $\mathbf{x}, \mathbf{y}$ be neighbouring points in $\overline{\mathbf{F}}$.
Then $[\mathbf{x}, \mathbf{y}]$ is a free interval w.r.t. $\mathbf{P}^+$.

The main idea behind the proof is to show that the fixed points cut $[a,b]$ into intervals that are at most as long as the generator $\mu$ of $\mathbf{P}^+$ jumps.

**Proof** We may assume that $\mathbf{P}^+$ is nontrivial, i.e., has a generator $\mu$. By Corollary 9.5, we only have to show that $[\mathbf{x}, \mathbf{y}]$ with $\mathbf{x} < \mathbf{y}$ is contained in a set of the type $[t, \mu(t)]$.
- If $\mathbf{x} = a$, then $[\mathbf{x}, \mathbf{y}] \subseteq [a, \mu(a)]$. Otherwise, we would have $\mathbf{y} > \mu(a)$, whence $\mu^{-1}(\mathbf{y})$ would be a fixed point by Proposition 6.17 that is smaller than $\mathbf{y}$.
- If $\mathbf{x}$ is a fixed point in $\mathbf{L}(\mu)$, then $\mu(\mathbf{x}) > \mathbf{x}$ equals $b$ or is a fixed point by Proposition 6.17. In any case, $[\mathbf{x}, \mathbf{y}] \subseteq [\mathbf{x}, \mu(\mathbf{x})]$.
- If $\mathbf{x} \notin \mathbf{L}(\mu)$, then $\mu^{-1}(b) < \mathbf{x}$, giving $[\mathbf{x}, \mathbf{y}] \subseteq [\mu^{-1}(b), b]$. **qed**

Lemma 9.1 implies the following freeness criterion.

**Corollary 9.7** Let $\mathbf{P}$ be a motion set with finite $\mathbf{P}^+$, and let $\mathbf{x}, \mathbf{y}$ be neighbouring points in $\overline{\mathbf{F}}$.
Then $I := [\mathbf{x}, \mathbf{y}]$ is a free interval w.r.t. $\mathbf{P}$ provided $\sigma(I \cap \mathbf{L}(\sigma)) \cap I$ is trivial for all reflections $\sigma \in \mathbf{P}_a^-$.

**Lemma 9.8** Let $\mathbf{P}$ be a motion set.
If an interval is free, then its interior does not contain a reflection fixed point.

**Proof** Assume $\mathbf{x} \in \text{int}\, I$ is fixed point of $\sigma \in \mathbf{P}^-$. Then there is a connected neighbourhood $U \subseteq \text{dom}\, \sigma$ of $\mathbf{x}$ contained in $\text{int}\, I$. As $\sigma$ fixes $\mathbf{x}$, also $\sigma(U)$ is a connected neighbourhood of $\mathbf{x}$. Shrinking $U$, if necessary, we may assume that $\sigma(U) \subseteq \text{int}\, I$ as well. Now, $\sigma(U) \cap U$ is nontrivial with $\sigma \neq \mathbf{1}$. Therefore, $I$ cannot be free. **qed**

**Theorem 9.9** Let $\mathbf{P}$ be a finite motion set with nontrivial $\mathbf{P}^-$.
Let $\mathbf{x}_0 < \mathbf{x}_1 < \ldots < \mathbf{x}_n$ be the fixed points of $\mathbf{P}^-$, and let $\mathbf{x}_{-1} := a$, $\mathbf{x}_{n+1} := b$.
Then the intervals $[\mathbf{x}_k, \mathbf{x}_{k+1}]$ with $k = -1, \ldots, n$ are maximally free.

**Proof** We may assume that $|\mathbf{F}_a| \geq |\mathbf{F}_b|$; otherwise conjugate by the flip $\varsigma$. Let now $I = [\mathbf{x}_k, \mathbf{x}_{k+1}]$ for some $k = -1, \ldots, n$. According to the freeness criterion of Corollary 9.7, we only have to check whether $\sigma(I \cap \mathbf{L}(\sigma))$ and $I$ intersect trivially for $\sigma \in \mathbf{P}_a^-$.

- Let $\mathbf{F}_a$ contain at least two fixed points.
  By Proposition 8.10, $I$ is the preimage of a nontrivial subinterval of $[\mathbf{x}_0, \mathbf{x}_1]$ under $\tau_k^{-1}$. Hence, the claim follows from Proposition 9.3, as soon as we have proven freeness of $I := [\mathbf{x}_0, \mathbf{x}_1]$. Using $\sigma_0(a) \leq \mathbf{x}_1$ by Lemma 8.6, as well as $\sigma_k(\mathbf{x}_1) \geq \sigma_k(\mathbf{x}_k) = \mathbf{x}_k \geq \mathbf{x}_1$ for $\sigma_k \in \mathbf{P}_a^-$, we see that

  |  | $I \cap \mathbf{L}(\sigma_k)$ | $\sigma_k(I \cap \mathbf{L}(\sigma_k))$ |
  |---|---|---|
  | $k = 0$ | $[\mathbf{x}_0, \sigma_0(a)]$ | $[a, \mathbf{x}_0]$ |
  | $k > 0$ | $[\mathbf{x}_0, \mathbf{x}_1]$ | $[\sigma_k(\mathbf{x}_1), \sigma_k(\mathbf{x}_0)]$ |

  By inspection, $\sigma_k(I \cap \mathbf{L}(\sigma_k)) \cap I$ is nontrivial.

- Let $\mathbf{F}_a$ contain a single point and $\mathbf{x}_1 < b$.
  Then $\mathbf{x}_1$ is a second fixed point in $\mathbf{F}$. Again by Lemma 8.6, we have $\sigma_0(a) \leq \mathbf{x}_1$, whence $\sigma_0$ exchanges $[a, \mathbf{x}_0]$ and $[\mathbf{x}_0, \sigma_0(a)] = [\mathbf{x}_0, \mathbf{x}_1] \cap \mathbf{L}(\sigma_0)$. It is now clear that $\sigma_0(I \cap \mathbf{L}(\sigma_0)) \cap I$ is trivial for all three cases $I = [a, \mathbf{x}_0], [\mathbf{x}_0, \mathbf{x}_1], [\mathbf{x}_1, b]$.

- Let $\mathbf{F}_a$ contain a single point and $\mathbf{x}_1 = b$.
  Then $\mathbf{P}^-$ consists of the single element $\sigma_0$. As $\sigma_0$ maps $[a, \mathbf{x}_0]$ to $[\mathbf{x}_0, \sigma_0(a)] \subseteq [\mathbf{x}_0, \mathbf{x}_1]$, the statement is trivial.

The maximality of the intervals follows from Lemma 9.8. **qed**

Finally, we are study the generating intervals. We begin with

**Lemma 9.10** If $\mathbf{P}$ is a finite motion set, then any generating interval is a maximal free interval.

**Proof** Let $I$ be a generating interval that is non-maximal, hence there is some free interval $J \supset I$. Let $t$ be in the interior or $J \setminus I$. Then there is some $\varrho \in \mathbf{P}$ with $\varrho(t) \in I$. Assume that $\varrho(t)$ neither equals $a$ nor $b$. As also $t \neq a, b$ by assumption, there is some nontrivial interval $K \subseteq (J \setminus I) \cap \mathrm{dom}\, \varrho$ containing $t$ with $\varrho(K) \subseteq I$. Hence $K \subseteq \varrho^{-1}(I) \cap (J \setminus I) \subseteq \varrho^{-1}(J) \cap J$. This gives $\varrho^{-1} = \mathbf{1}$ and $\varrho = \mathbf{1}$ by freeness of $J$. Consequently $t = \varrho(t) \in I$ contradicting the assumption. Hence, $\varrho(t)$ equals $a$ or $b$. In other words, any $t \in J \setminus I$ is in some preimage $\varrho^{-1}(\{a, b\})$. As there are only finitely many $\varrho$, this is a contradiction. **qed**

Now, Corollary 9.5 and Theorem 9.9 immediately give[21]

**Proposition 9.11** Let $\mathbf{P}$ be a motion set. Then the generating intervals for $\mathbf{P}$ are given by

| $\mathbf{P}$ | generating intervals |
|---|---|
| trivial | $[a, b]$ |
| non-trivial, positive | $[s, \mu(s)]$ with $s \in \mathbf{L}(\mu)$ |
| non-positive | all that connect neighbouring elements in $\mathbf{F}$ |

Here, $\mathbf{F}$ contains the reflection fixed points $\mathbf{x}_0, \ldots, \mathbf{x}_n$. Additionally,
- $[a, \mathbf{x}_0]$ is generating iff $\mathbf{x}_1 = \mu(a)$ or ($|\mathbf{F}| = 1$ and not $\sigma_0(a) < b$);
- $[\mathbf{x}_n, b]$ is generating iff $\mu(\mathbf{x}_{n-1}) = b$ or ($|\mathbf{F}| = 1$ and not $a < \sigma_0(b)$),
provided the respective expressions exist.

This implies

**Theorem 9.12** Any finite motion set has a generating interval.

---
[21] For the exceptional cases with $|\mathbf{F}| = 1$ check also Table 2 on page 50.

Recall that this statement has turned out crucial for the proof that any path with finite $\mathbf{P}_\gamma$ is a brick path. In the proposition below, we will even give an explicit and complete description for which images of a generating interval cover $[a, b]$. Only note that we have skipped there the case of a non-positive $\mathbf{P}$ with non-trivial $\mathbf{P}^+$ and $|\mathbf{P}_b^-| \geq 2$ instead of $|\mathbf{P}_a^-| \geq 2$. However, there the statement remains valid if $\mathbf{x}_0 > \mathbf{x}_1$ are now the two largest fixed points and $\mu$ has to be replaced by $\mu^{-1}$. Accordingly, the order and the multiplicity functions have to be adjusted. We refrain from stating this here more explicitly.

**Proposition 9.13** Let $\mathbf{P}$ be a non-trivial finite motion set.
1. If $\mathbf{P}$ is positive, then $[a, b]$ is covered by the images of $[a, \mu(a)]$ under the mappings
$$\mathbf{1},\, \mu,\, \mu^2,\, \ldots,\, \mu^{\mathbf{m}(\mu)}\,.$$
2. If $\mathbf{P}$ is non-positive with non-trivial $\mathbf{P}^+$ and $|\mathbf{P}_a^-| \geq 2$, then $[a, b]$ is covered by the images of $[\mathbf{x}_0, \mathbf{x}_1]$ under the mappings
$$\mu^{-1} \bullet \sigma,\, \mathbf{1},\, \sigma,\, \mu,\, \mu \bullet \sigma,\, \mu^2,\, \mu^2 \bullet \sigma,\, \ldots,\, \mu^{\mathbf{m}(\mu)},\, \mu^{\mathbf{m}(\mu)} \bullet \sigma\,.$$
Here, $\mathbf{x}_0 < \mathbf{x}_1$ are the smallest fixed points for reflections in $\mathbf{P}$, and $\sigma$ is the reflection w.r.t. $\mathbf{x}_1$. The ultimate term appears iff $\mathbf{o}(\mathbf{x}_1) = \mathbf{m}(\mu)$, the penultimate iff $\mathbf{o}(\mathbf{x}_0) = \mathbf{m}(\mu)$.
3. If $\mathbf{P}$ is non-positive with trivial $\mathbf{P}^+$ and $|\mathbf{P}^-| = 2$, then $[a, b]$ is covered by the images of $[\mathbf{x}_0, \mathbf{x}_1]$ under the mappings
$$\sigma_0,\, \mathbf{1},\, \sigma_1\,.$$
Here, $\mathbf{x}_0 < \mathbf{x}_1$ are the two fixed points for the reflections $\sigma_0$ and $\sigma_1$ in $\mathbf{P}$.
4. If $\mathbf{P}$ is non-positive with trivial $\mathbf{P}^+$ and $|\mathbf{P}^-| = 1$, then $[a, b]$ is covered by the images of $[a, \mathbf{x}]$ or $[\mathbf{x}, b]$ under the mappings
$$\mathbf{1},\, \sigma \quad \text{or} \quad \sigma,\, \mathbf{1}\,.$$
respectively. Here $\mathbf{x}$ is the fixed point of the reflection $\sigma \in \mathbf{P}^-$.

In any of the preceding cases, subsequent images share exactly a common boundary point. Moreover, the first and the final image might be cut, i.e., the image of some subinterval of $[a, b]$, etc.

## 10 Topology on Motion Sets

In the last two sections, we have concentrated on finite motion sets and completed their classification including the existence of free and generating intervals. Now, we will bring the infinite motion sets into focus. As we already know that motion sets can be characterized by the values of their elements on boundary points, infinite motion sets correspond to infinite $\mathbf{P}(a)$. As $[a, b]$ is compact, there will be accumulation points suggesting to introduce some notion of limit or topology into the game. Indeed, we will use the order-preserving bijection between $\mathbf{P}_a^+$ and $\mathbf{P}^+(a) \subseteq [a, b)$ to transfer topological properties from $[a, b]$ to $\mathbf{P}$.

Let us motivate the constructions below by our prime example. For brevity, we restrict ourselves to a positive motion set $\mathbf{P}$ of shift operators on an interval $[0, 1]$ in $\mathbb{R}$. There, we can identify $\mathbf{P}_0$ with $\mathbf{P}(0)$, i.e., the points reached by shifts starting in $0$. Assume now that $\mathbf{P}(0)$ has an accumulation point in $0$, i.e., there are right shifts $\varrho_{\lambda_i}$ with $\lambda_i = \varrho_{\lambda_i}(0) \to 0$. Now, obviously, any $t \in [0, 1]$ is an accumulation point in $\mathbf{P}(t)$ as well. But, even more, then each $s \in \mathbf{P}(t)$ is an accumulation point. In fact, $\sigma_i := \varrho_{\lambda_i} \circ \varrho$ with $\varrho(t) = s$ fulfills $\sigma_i(t) = \lambda_i + s \to s = \varrho(t)$. On can easily show that even each $s$ in the closure of $\mathbf{P}(t)$ is indeed an accumulation point. This, however, implies that $\mathbf{P}(t)$ is already dense in $[a, b]$. In fact, if there is some "hole" in $[a, b]$, then we may cut $\mathbf{P}(t)$ into two pieces to the left and to the right of the missing interval and take the supremum of, say, the left part. Assuming, for simplicity, that the supremum is a maximum $t = \varrho(s)$, we may add the slight right shifts $\varrho_{\lambda_i}$ in order to get a contradiction as then $[\varrho_{\lambda_i} \circ \varrho](s)$ hits the hole. Ultimately, one easily sees that $\mathbf{P}(t)$ is dense in $[a,b]$ for all $t$ iff just one of them has just one accumulation point. This shows that $\mathbf{P}(t)$ is either dense or finite, independently of $t$. This will turn out crucial for the application to the original problem of intersections between group translates of analytic paths. Note, however, that we cannot deduce that $\mathbf{P}(t)$ covers at least[22] $(a,b)$. Indeed, consider the subgroup $U$ of $\mathbb{R}$ generated by 1 and $\sqrt{2}$. The corresponding shifts induce a positive motion set $\mathbf{P}$, for which $\mathbf{P}(0) = U \cap [0,1)$ is dense in $[0,1)$ though, but surely not full $[0,1)$.

After having presented the basic ideas, we are now going to implement them in the general case of positive motion sets. We will first transfer the notion of accumulation points which allows us to show that motion sets are either dense or finite. Afterwards we define and study the notion of convergence that turns $\mathbf{P}_a$ into a metric space isometric to $\mathbf{P}(a)$, being a subspace of $[a,b]$. In the subsequent section, we will investigate the interplay of convergence and multiplication in motion sets. This will ultimately lead us to the desired classification theorems for infinite motion sets.

Note that from now on, we will frequently write expressions like "$\varrho_i(s) \downarrow t$". This shall serve as a shorthand notation for the assumption that $\varrho_i \in \mathbf{P}_s$ for sufficiently large $i$ and that the sequence $(\varrho_i(s))$ is strictly decreasing at least for sufficiently large $i$. Similarly, we use "$\uparrow$" for strictly increasing and "$\to$" for converging. In the monotonous, but not necessarily strict cases we will use "$\nearrow$" and "$\searrow$".

## 10.1 Accumulation Points

**Lemma 10.1** Let $\mathbf{P}$ be a positive motion set. Then:
$$\varrho_i(a) \downarrow a \implies \varrho_i(\varrho_i(a)) \downarrow a.$$
In particular, the right hand size comprises $\varrho_i \bullet \varrho_i \in \mathbf{P}_a$ for large $i$.

**Proof** Since $(\varrho_i(a))$ is strictly decreasing, $(\varrho_i^{-1}(b))$ is, by Corollary 6.11, strictly increasing, hence converging to some limit larger than $a$. Consequently, $\varrho_i(a) < \varrho_i^{-1}(b)$, hence $\varrho_i(a) \in \mathbf{L}(\varrho_i)$ for large $i$. Since, using Proposition 6.2, $\varrho_i(\varrho_i(a)) > \varrho_i(\varrho_j(a)) > \varrho_j(\varrho_j(a))$ for large $i < j$, the sequence $([\varrho_i \circ \varrho_i](a))$ is strictly decreasing, hence converging to some $s$. This, however, implies $s < \varrho_i(\varrho_j(a))$ for $i < j$ and $s \leq \lim_j \varrho_i(\varrho_j(a)) = \varrho_i(\lim_j \varrho_j(a)) = \varrho_i(a)$ for large $i$. Hence, $s \leq \lim_i \varrho_i(a) = a$. Moreover, the snaking lemma gives $\varrho_i \bullet \varrho_i \in \mathbf{P}_a$. **qed**

**Proposition 10.2** Let $\mathbf{P}$ be a positive motion set. Then
$$\varrho_i(a) \downarrow a \iff \varrho_i^{-1}(b) \uparrow b.$$

**Proof** Since $(\varrho_i(a))$ is strictly decreasing, $(\varrho_i^{-1}(b))$ is strictly increasing, hence converging to some $t > a$. Choosing, if necessary, a subsequence, we may assume that $\varrho_{i-1}(a) > \varrho_i(\varrho_i(a))$ and $\varrho_i^{-1}(b) > \varrho_i(a)$, for all $i$. Consequently, $a$ is snaking along $\varrho_i \circ \varrho_i$ and $b$ is snaking along $\varrho_i^{-1} \circ \varrho_i^{-1}$. Now,

$$\begin{aligned}
\varrho_{i-1}^{-1}(b) &< [\varrho_i \bullet \varrho_i]^{-1}(b) & \text{(Proposition 6.2 with } \varrho_i \bullet \varrho_i < \varrho_{i-1}) \\
&= [\varrho_i^{-1} \bullet \varrho_i^{-1}](b) & \text{(Lemma 4.10)} \\
&= \varrho_i^{-1}(\varrho_i^{-1}(b)) & \text{(snaking lemma)} \\
&< \varrho_i^{-1}(t) & \text{(monotonicity on } [\varrho_i^{-1}(b), t] \subseteq [\varrho_i(a), b] \equiv \mathbf{R}(\varrho^{-1})) \\
&\leq \varrho_i^{-1}(b) & \text{(monotonicity on } [t, b] \subseteq [\varrho_i(a), b] \equiv \mathbf{R}(\varrho^{-1}))
\end{aligned}$$

Thus, $\lim \varrho_i^{-1}(t) = \lim \varrho_i^{-1}(b)$, giving $t = b$ by pointwise properness. The opposite direction is completely analogous. **qed**

---
[22]Note that $\varrho(a) \neq b$ and $\varrho(b) \neq a$ for all positive perfect $\varrho$, whence neither $\mathbf{P}(a)$ nor $\mathbf{P}(b)$ can be full $[a,b]$.

From $\mathbf{L}(\varrho_i) = [a, \varrho_i^{-1}(b)]$ and the previous two statements, we get

**Corollary 10.3** Let $\mathbf{P}$ be a positive motion set. Then
$$\varrho_i(a) \downarrow a \iff \mathbf{L}(\varrho_i) \uparrow [a,b] \implies \mathbf{L}(\varrho_i \bullet \varrho_i) \uparrow [a,b]$$

This, in particular, means that each $t < b$ is in $\mathbf{L}(\varrho_i)$ for large $i$.

**Lemma 10.4** Let $\mathbf{P}$ be a positive motion set. Then for all $a < s < b$
$$\varrho_i(a) \downarrow a \implies \varrho_i(s) \downarrow s$$

**Proof** By Corollary 10.3, we have $s \in \mathbf{L}(\varrho_i) \cap \mathbf{L}(\varrho_i \bullet \varrho_i)$ for large $i$. Thus, by Proposition 6.2, $\varrho_i(s)$ is strictly descreasing, hence converging to some $t \geq s$. Choosing, if necessary, a subsequence, we may assume that $\varrho_{i-1} > \varrho_i \bullet \varrho_i$ and $t < \varrho_i(s) \in \mathrm{int}\,\mathbf{L}(\varrho_i)$ for all $i$. Now, as above
$$\varrho_{i-1}(s) > [\varrho_i \bullet \varrho_i](s) = \varrho_i(\varrho_i(s)) > \varrho_i(t) \geq \varrho_i(s).$$
This implies $\lim \varrho_i(s) = \lim \varrho_i(t)$, giving $s = t$ by pointwise properness. **qed**

**Definition 10.1** $s \in [a,b]$ is called **accumulation point** of $\mathbf{P}(t)$ …

$$\begin{aligned}\ldots \text{ from above} &\iff \sigma_j(t) \downarrow s \text{ from some } (\sigma_j) \subseteq \mathbf{P}_t \\ \ldots \text{ from below} &\iff \sigma_j(t) \uparrow s \text{ from some } (\sigma_j) \subseteq \mathbf{P}_t\end{aligned}$$

**Proposition 10.5** Let $\mathbf{P}$ be a positive motion set. Then
$$\varrho_i(a) \downarrow a \text{ for some } \varrho_i \implies \mathbf{P}(t) \text{ dense in } [a,b] \text{ for all } t$$

**Proof** We are now going to show that each $s > a$ is an accumulation point from below and each $s < b$ is one from above. This will give the proof.
- First assume that $s < b$ is in the closure of $\mathbf{P}(t)$, but not an accumulation point from above. This means that there is an $r \in (s, b)$, such that $s < \sigma(t)$ even implies $r < \sigma(t)$ for $\sigma \in \mathbf{P}_t$. Nevertheless, as $s \in \overline{\mathbf{P}(t)}$, there is an increasing sequence $\sigma_j(t) \nearrow s$. By $\varrho_i(a) \downarrow a$, we have $s \in \mathbf{L}(\varrho_i)$ and $a < \varrho_i(a) \leq \varrho_i(\sigma_j(t)) \leq \varrho_i(s) < b$ for large $i$. Hence, $t$ is snaking along $\varrho_i \circ \sigma_j$, giving $\varrho_i \bullet \sigma_j \in \mathbf{P}_t$ and
$$[\varrho_i \bullet \sigma_j](t) = \varrho_i(\sigma_j(t)) \nearrow \varrho_i(s) > s$$
for large $i$. Therefore, by assumption, even $[\varrho_i \bullet \sigma_j](t) > r$ for large $i$ (and $j$). Thus, $\varrho_i(s) = \lim_j \varrho_i(\sigma_j(t)) \geq r$. Now, $s = \lim \varrho_i(s) \geq r$, by Lemma 10.4. Contradiction.
- Next, assume that $s > a$ is in the closure of $\mathbf{P}(t)$, but not an accumulation point from below. The argumentation is completely analogous. Just exchange $a$, $\downarrow$, $<$ and $\mathbf{L}$ with $b$, $\uparrow$, $>$ and $\mathbf{R}$, respectively, and observe that $\varrho_i(a) \downarrow a$ is equivalent to $\varrho_i^{-1}(b) \uparrow b$.
- Finally, assume that $s$ is not contained in the closure of $\mathbf{P}(t)$. Since $\mathbf{P}(t)$ contains at least $t \equiv \mathbf{1}(t)$, we find $\mathbf{P}(t) \cap [a, s)$ or $\mathbf{P}(t) \cap (s, b]$ not empty. If the first one is not empty, it has a supremum in $r \in \overline{\mathbf{P}(t)}$ with $r < s$. By construction, this $r$ cannot be approximated from above. This, however, contradicts our findings above. Similarly argue in the second case. **qed**

**Proposition 10.6** Let $\mathbf{P}$ be a positive motion set. Then
$$\mathbf{P}(t) \text{ has an accumulation point for some } t \implies \varrho_i(a) \downarrow a \text{ for some } \varrho_i$$

**Proof** Let $s$ be an accumulation point of $\mathbf{P}(t)$.

- Assume first $t = a$ and $\sigma_i(a) \downarrow s$ for $\sigma_i \in \mathbf{P}_a$. Then, by Lemma 5.11, $\varrho_i := \sigma_{i+1}^{-1} \bullet \sigma_i \in \mathbf{P}_a$ with $\varrho_i(a) > a$. Assume now $\varrho_i(a) \geq c > a$ for all $i$. Since, again by Lemma 5.11, we have $\varrho_i(a) \in \mathbf{L}(\sigma_{i+1})$, we get

$$\sigma_i(a) = \sigma_{i+1}(\varrho_i(a)) \geq \sigma_{i+1}(c) > \sigma_{i+1}(a),$$

  implying $\lim \sigma_i(c) = \lim \sigma_i(a)$, giving $c = a$, by pointwise properness.

- Assume next $t = a$ and $\sigma_i(a) \uparrow s$ for $\sigma_i \in \mathbf{P}_a$. Exchanging $\sigma_i$ with $\sigma_{i+1}$ above, we get the assertion again.

- Let now $t$ be arbitrary and let $\varrho_i \in \mathbf{P}_t$ with strictly monotonous $(\varrho_i(t))$ and $\varrho_i(t) \geq t$ for all (or at least large) $i$. Then $\varrho_i \in \mathbf{P}_a$, by Corollary 6.8. By Proposition 6.2, $\varrho_i(a)$ is strictly monotonous in $\mathbf{P}(a)$ as well. The statement follows now from those above.

- Let finally $t$ be arbitrary and let $\varrho_i \in \mathbf{P}_t$ with strictly monotonous $(\varrho_i(t))$ and $\varrho_i(t) \leq t$ for all (or at least large) $i$. Then $\varrho_i \in \mathbf{P}_b$, by Corollary 6.8. By the right-domain version of Proposition 6.2, $\varrho_i(b)$ is strictly monotonous as well. The same applies to $\varrho_i^{-1}(a)$. And again, we have reduced the problem to a case we have already done. **qed**

**Corollary 10.7** Let $\mathbf{P}$ be a positive motion set. Then we have for all $a < s < b$

$$\varrho_i(s) \downarrow s \implies \varrho_i(a) \downarrow a$$

**Proof** As $\varrho_i(s) \downarrow s$, there are $\sigma_i \in \mathbf{P}_a$ with $\sigma_i(a) \downarrow a$ by Proposition 10.6, hence $\sigma_i(s) \downarrow s$ by Lemma 10.4. Taking, if necessary, subsequences, we may assume that $\sigma_i(s) \geq \varrho_i(s) \geq \sigma_{i+1}(s)$. This transfers to $\sigma_i(a) \geq \varrho_i(a) \geq \sigma_{i+1}(a)$, whence $\varrho_i(a) \downarrow a$. **qed**

## 10.2 Dense or Finite?

In this subsection, we are going to show that any infinite motion set is even dense. We will first prove this for the positive part of the motion set. Before, however, we should define the corresponding notions:

**Definition 10.2** A set $\mathbf{P}$ of mappings between closed subsets of $[a,b]$ is called
- **dense** $\iff \mathbf{P}(t)$ is dense in $[a,b]$ for all $t$;
- **full** $\iff \mathbf{P}(a) = [a,b)$, $\mathbf{P}(b) = (a,b]$, and $\mathbf{P}(t) = [a,b]$ for all $t \neq a,b$;
- **closed** $\iff$ for any $\varrho_i \in \mathbf{P}_t$ with converging $(\varrho_i(t))$, there is some $\varrho \in \mathbf{P}_t$ with $\lim \varrho_i(t) = \varrho(t)$ unless $\{t, \lim \varrho_i(t)\} = \{a,b\}$.

**Lemma 10.8** Let $\mathbf{P}$ be a positive motion set. Then we have:

$$\mathbf{P} \text{ full} \iff \mathbf{P}(a) = [a,b), \mathbf{P}(b) = (a,b], \text{ or } \mathbf{P}(t) = [a,b] \text{ for some } t \neq a,b.$$

Thus, we need to check only that $\mathbf{P}(a)$ equals $[a,b)$ to provide us with fullness.

**Proof**
- Let $t \neq a,b$ be given with $\mathbf{P}(t) = [a,b]$. Thus, for any $r, s \in [a,b]$, we have $\varrho, \sigma \in \mathbf{P}_t$ with $r = \varrho(t)$ and $s = \sigma(t)$, hence $s = \sigma^{-1}(\varrho(r))$. As $r$ is snaking unless $\{r,s\} = \{a,b\}$, the product $\sigma^{-1} \bullet \varrho$ is a well defined element in $\mathbf{P}_r$. This shows that $\mathbf{P}$ is full.
- Let $t = a$ having $\mathbf{P}(t) = [a,b)$. The same argument as above shows that $\mathbf{P}(r)$ contains at least $[a,b)$ for $r \neq b$. Let now $\tau \in \mathbf{P}_b$ be nontrivial; such an element exists, as $\mathbf{P}_a$ is nontrivial and equals $\mathbf{P}_b^{-1}$. Now $\mathbf{P}(\tau^{-1}(b))$ contains $\tau(\tau^{-1}(b)) = b$, whence it even comprises full $[a,b]$. Now, the problem has been reduced to the first case.
- Let $t = b$ having $\mathbf{P}(b) = (a,b]$. This case is completely analogous. **qed**

**Theorem 10.9** Let $\mathbf{P}$ be a positive motion set. Then $\mathbf{P}$ is either dense or finite.

**Proof** If $\mathbf{P} = \mathbf{P}_a \cup \mathbf{P}_b$ is not finite, at least one of the sets $\mathbf{P}_a$ and $\mathbf{P}_b$ is not finite, say $\mathbf{P}_a$. As there is a bijection between $\mathbf{P}_a$ and $\mathbf{P}(a)$, we see that $\mathbf{P}(a) \subseteq [a,b]$ is not finite, hence has an accumulation point. Now, Propositions 10.6 and 10.5 give the proof. **qed**

**Theorem 10.10** Let $\mathbf{P}$ be a positive and closed motion set. Then $\mathbf{P}$ is either full or finite.

**Proof** If $\mathbf{P}$ is not finite, then each $\mathbf{P}(t)$ is dense in $[a,b]$. So, fix $t$ and let $s \in [a,b]$. Then there are $\varrho_i \in \mathbf{P}_t$ with $\varrho_i(t) \to s$. As $\mathbf{P}$ is closed, there is some $\varrho \in \mathbf{P}_t$ with $\varrho_i(t) \to \varrho(t)$, hence $s \in \mathbf{P}(t)$, unless $\{s,t\} = \{a,b\}$. **qed**

Now we are removing the positivity restriction. Nevertheless, we will reduce the general problem to the positive case.

**Proposition 10.11** Let $\mathbf{P}$ be a motion set. Then we have

$$\mathbf{P}^- \text{ infinite} \implies \mathbf{P}^+ \text{ infinite}.$$

**Proof** Choose infinitely many mutually different elements $\sigma_i$ of $\mathbf{P}^-$. Assume first $\sigma_i \in \mathbf{P}_a^-$ for all $i$. If necessary, taking a subsequence, we have to consider two cases:
- Assume that $(\sigma_i(a))$ is decreasing.
  Then $\mathbf{L}(\sigma_i) \subset \mathbf{L}(\sigma_j)$ and $\sigma_i(a) < \sigma_j(a)$ for all $i > j$. As $\sigma_1(a) \in \mathbf{L}(\sigma_1) \supset \mathbf{L}(\sigma_i)$ for all $i > 1$, we have $\sigma_1(\sigma_i(a)) > \sigma_1(\sigma_j(a))$ for $i > j$. Therefore, $([\sigma_1 \bullet \sigma_i](a))$ is a strictly increasing sequence in $\mathbf{P}^+(a)$, whence $\mathbf{P}^+$ is not finite.
- Assume that $(\sigma_i(a))$ is increasing.
  Then $\mathbf{L}(\sigma_i) \supset \mathbf{L}(\sigma_j)$ and $\sigma_i(a) > \sigma_j(a)$ for all $i > j$. As $\sigma_i(a) \in \mathbf{L}(\sigma_i) \supset \mathbf{L}(\sigma_1)$ for all $i > 1$, we have $\sigma_i(\sigma_1(a)) < \sigma_j(\sigma_1(a))$ for all $i > j$. Therefore, $([\sigma_i \bullet \sigma_1](a))$ is a strictly decreasing sequence in $\mathbf{P}^+(a)$, whence $\mathbf{P}^+$ is not finite.

If we can find only finitely many elements in $\mathbf{P}_a^-$, then there are infinitely many in $\mathbf{P}_b^-$. Denoting by $\varsigma : [a,b] \longrightarrow [a,b]$ with $\varsigma(t) := b + a - t$ the flip of the interval $[a,b]$, we see immediately that $\varsigma \circ \mathbf{P} \circ \varsigma$ has the same attributes as $\mathbf{P}$, whereas $\varsigma \circ \mathbf{P}^\pm \circ \varsigma = [\varsigma \circ \mathbf{P} \circ \varsigma]^\pm$. However, now $\varsigma \circ \sigma_i \circ \varsigma$ is contained in $[\varsigma \circ \mathbf{P} \circ \varsigma]_a^-$, whence the latter set infinite. As shown above, $\varsigma \circ \mathbf{P}^+ \circ \varsigma$ is infinite, giving the same claim for $\mathbf{P}^+$ immediately. **qed**

**Proposition 10.12** Let $\mathbf{P}$ be a motion set. Then we have for non-empty $\mathbf{P}^-$

$$\mathbf{P}^+ \text{ dense} \implies \mathbf{P}^- \text{ dense}.$$

**Proof** Assume that $s$ is not in the closure of $\mathbf{P}^-(t)$. Then the intersection of $\mathbf{P}^-(t)$ with $[a,s)$ or with $(s,b]$ is non-empty. We may assume the first case, the latter one is completely analogous. Thus, let $r := \sup[\mathbf{P}^-(t) \cap [a,s)] < s \leq b$ and $\sigma_i \in \mathbf{P}_t^-$ with $\sigma_i(t) \nearrow r$. As $\mathbf{P}^+$ is dense, there are $\varrho_j \in \mathbf{P}_r^+$ with $b > \varrho_j(r) \downarrow r$. Now, $r < \varrho_j(r)$ implies $r \in \text{int}\,\mathbf{L}(\varrho_j)$, hence $a < \varrho_j(a) \leq \varrho_j(\sigma_i(t)) \leq \varrho_j(r) < b$, whence $t$ is snaking along $\varrho_j \circ \sigma_i$. Thus, $\varrho_j \bullet \sigma_i$ is always a well defined element in $\mathbf{P}_t^-$. Observe now that $\lim_i[\varrho_j \bullet \sigma_i](t) = \lim_i \varrho_j(\sigma_i(t)) = \varrho_j(r) > r$, hence even $[\varrho_j \bullet \sigma_i](t) > s$ for large $i$. Consequently, $s \leq \lim_i[\varrho_j \bullet \sigma_i](t) = \varrho_j(r)$, hence $s \leq \lim_j \varrho_j(r) = r$. Contradiction. **qed**

**Theorem 10.13** Let $\mathbf{P}$ be a motion set with non-empty $\mathbf{P}^-$ having fixed point set $\mathbf{F}$. Then

$$\begin{aligned}\mathbf{P} \text{ infinite} &\iff \mathbf{P} \text{ dense} \iff \mathbf{P}^- \text{ infinite} \iff \mathbf{P}^- \text{ dense}\\ &\iff \mathbf{F} \text{ infinite} \iff \mathbf{F} \text{ dense} \iff \mathbf{P}^+ \text{ infinite} \iff \mathbf{P}^+ \text{ dense}\end{aligned}$$

Note that the assumption that $\mathbf{P}^-$ is nontrivial, cannot be removed. Indeed, any positive motion set $\mathbf{P}$ has trivial reflection set $\mathbf{P}^-$, although $\mathbf{P} \equiv \mathbf{P}^+$ easily can be infinite.

**Proof** 1. If $\mathbf{P}^-$ is dense, it is obviously infinite. This, in turn, implies that $\mathbf{P}^+$ is infinite, by Proposition 10.11. If the latter one is given, Theorem 10.9 implies that $\mathbf{P}^+$ is dense, which now implies that $\mathbf{P}^-$ is dense by Proposition 10.12.

2. If $\mathbf{F}$ is dense, it is obviously infinite. Next, observe that $\mathbf{F}$ is infinite iff $\mathbf{P}^-$ is infinite. In fact, each reflection $\sigma \in \mathbf{P}^-$ has exactly one or two fixed points and different reflections have different fixed points (Proposition 6.30), giving the claim. Therefore, $\mathbf{P}^+$ is infinite as soon as $\mathbf{F}$ is so, using the first part of the proof. Finally, Corollary 6.18 gives us $\mathbf{P}^+(\mathbf{x}) \subseteq \mathbf{F} \sqcup \{a,b\}$ for any fixed point $\mathbf{x}$. This shows that the denseness of $\mathbf{P}^+$ implies that of $\mathbf{F}$.

3. If $\mathbf{P}$ is infinite, then $\mathbf{P}^+$ or $\mathbf{P}^-$ is infinite, hence even both are infinite as shown above. The remaining implications are trivial. **qed**

## 10.3 Convergence for Motion Sets

Let us close this section by transferring the notion of convergence from $[a,b]$ to $\mathbf{P}$. We will restrict ourselves to the case of positive motion sets. The negative case can be dealt with analogously, but we will not need it in the sequel. Moreover, again foreseeing the applications below, we will study convergence on $\mathbf{P}_a$ only.

Now, Lemma 10.4 and Corollary 10.7 motivate

**Definition 10.3** We write for $\varrho, \varrho_i \in \mathbf{P}_a$

$$\begin{aligned} \varrho_i \downarrow \varrho &\iff \varrho_i(t) \downarrow \varrho(t) \quad \text{for all } t \in \mathbf{L}(\varrho) \text{ with } \varrho(t) < b \\ \varrho_i \uparrow \varrho &\iff \varrho_i(t) \uparrow \varrho(t) \quad \text{for all } t \in \mathbf{L}(\varrho) \text{ with } \varrho(t) > a \end{aligned}$$

This definition deserves a comment concerning the domains. A priori it is not clear that $t \in \mathbf{L}(\varrho)$ and $\varrho_i \in \mathbf{P}_a$ with $\varrho_i(t) \to \varrho(t)$ implies $t \in \mathbf{L}(\varrho_i)$ for at least large $i$. Of course, if $\varrho \neq \mathbf{1}$, then $\varrho(t) > t$, whence also $\varrho_i(t) > t$ for large $i$. But, if $\varrho$ is the identity, then we may might be faced with $\varrho_i(t) < t$ for infinitely many $i$. Possibly taking a subsequence, we may even assume $\varrho_i(t) \uparrow \mathbf{1}(t) \equiv t$. Indeed, Proposition 6.2 now implies that $t$ is in the right domain of any $\varrho_i$. In particular, $\varrho_i(b) \uparrow b$ by the right-domain version of Corollary 10.7. On the other hand, since $\varrho_i$ is not the identity, we have $\varrho_i(b) < \varrho_i(a)$ for all $i$, whence $\lim \varrho_i(b) = b = \lim \varrho_i(a)$ in contradiction to pointwise properness. Thus, only finitely many $\varrho_i(t)$ can be smaller than $t$, hence in the right domain. This proves

**Lemma 10.14** Let $\mathbf{P}$ be a positive motion set, $\varrho, \varrho_i \in \mathbf{P}_a$ and $s \in \mathbf{L}(\varrho)$. Then

$$\varrho_i(s) \to \varrho(s) \implies s \in \mathbf{L}(\varrho_i) \text{ for large } i.$$

We can even extend this to other points and sharpen the claim:

**Lemma 10.15** Let $\mathbf{P}$ be a positive motion set, $\varrho_i, \varrho \in \mathbf{P}_s$ with $\varrho_i(s) \to \varrho(s)$ and $s \in \mathbf{L}(\varrho)$. Then

$$t \in \text{int } \mathbf{L}(\varrho) \implies t \in \text{int } \mathbf{L}(\varrho_i) \text{ for large } i$$

**Proof** We may assume that $\mathbf{P}$ is dense. Otherwise the statement is trivial. Note that $s \in \mathbf{L}(\varrho_i)$ for large $i$ by the lemma above.
- If $\varrho(s) = b$, then $\varrho_i(s) \leq \varrho(s)$. This gives $\mathbf{L}(\varrho) \subseteq \mathbf{L}(\varrho_i)$ by Proposition 6.2, for large $i$.
- If $\varrho(s) < b$, let $r := \max\{s, t\}$. By denseness, there is an $\sigma \in \mathbf{P}_r$ with $\varrho(r) < \sigma(r) < b$. Since obviously $r < \sigma(r)$, even $t \leq r \in \mathbf{L}(\sigma) \subset \mathbf{L}(\varrho)$ by Corollary 6.8. As $\varrho_i(s) \to \varrho(s)$ and $\varrho(s) < \sigma(s)$, we have $\varrho_i(s) < \sigma(s)$ for large $i$, hence $\mathbf{L}(\sigma) \subset \mathbf{L}(\varrho_i)$. **qed**

**Lemma 10.16** Let $\mathbf{P}$ be a positive motion set. Then we have for all $\varrho_i, \varrho \in \mathbf{P}_a$

$$\begin{aligned} \varrho_i \downarrow \varrho &\iff \varrho_i(s) \downarrow \varrho(s) \quad \text{for some } s \in \mathbf{L}(\varrho) \\ \varrho_i \uparrow \varrho &\iff \varrho_i(s) \uparrow \varrho(s) \quad \text{for some } s \in \mathbf{L}(\varrho) \end{aligned}$$

**Proof** First observe that we may assume that **P** is dense; otherwise, strict monotonicity is not possible. Of course, the $\Longrightarrow$ direction is trivial. Hence, let us consider the $\Longleftarrow$ direction.

- For the upper line, let $t \in \mathbf{L}(\varrho)$ with $\varrho(t) < b$ be given. By Lemma 10.15, we have $t \in \mathbf{L}(\varrho_i)$ for large $i$. As $\varrho_i(s)$ is strictly decreasing, $\varrho_i(t)$ is strictly decreasing as well, hence converging to some $u$. Obviously, $u \geq \varrho(t)$. If $u > \varrho(t)$, then, by denseness, there is some $\tau \in \mathbf{P}_t$ with
$$\varrho_i(t) \;>\; u \;>\; \tau(t) \;>\; \varrho(t) \;\geq\; t$$
for large $i$. As above we see that $\mathbf{L}(\tau) \supset \mathbf{L}(\varrho_i)$ containing $s$, whence $\varrho_i(s) \geq \tau(s) > \varrho(s)$ for large $i$. Contradiction.
- For the lower line, the proof is completely analogous, as soon as we can guarantee for $t \in \mathbf{L}(\varrho_i)$ for large $i$. The claim is obvious as long as $\varrho(t)$ is not $b$. If it is $b$ and $\varrho \neq \mathbf{1}$, then $t < \varrho(t) = b$, hence $t < \varrho_i(t)$ for large $i$, hence $t \in \mathbf{L}(\varrho_i)$. If it is $b$ and $\varrho = \mathbf{1}$, then $\varrho_i(s) \uparrow s$, hence $\varrho_i(s) < s$ and $s \notin \mathbf{L}(\varrho_i)$ for all $i$ contradicting Lemma 10.14.

**qed**

**Corollary 10.17** Let **P** be a positive motion set. Then we have for all $\varrho_i, \varrho \in \mathbf{P}_a$
$$\varrho_i \downarrow \varrho \iff \varrho_i^{-1}(b) \uparrow \varrho^{-1}(b) \iff \mathbf{L}(\varrho_i) \uparrow \mathbf{L}(\varrho) \implies \mathbf{L}(\varrho) = \bigcup_i \mathbf{L}(\varrho_i)$$
$$\varrho_i \uparrow \varrho \iff \varrho_i^{-1}(b) \downarrow \varrho^{-1}(b) \iff \mathbf{L}(\varrho_i) \downarrow \mathbf{L}(\varrho) \implies \mathbf{L}(\varrho) = \bigcap_i \mathbf{L}(\varrho_i)$$

**Proof** Only the left implications are still to be proven. Moreover, we may assume **P** to be dense.
- If $\varrho_i \downarrow \varrho$, then $\varrho_i(a) \downarrow \varrho(a)$, hence $\varrho_i^{-1}(b)$ is strictly increasing by Corollary 6.11. It converges to some $t$. If $t < \varrho^{-1}(b)$, then $\varrho(t) < b$, hence $\varrho_i(t) < b$ for large $i$. Consequently, $t < \varrho_i^{-1}(b) \uparrow t$, which is impossible. Hence $\varrho_i^{-1}(b) \uparrow \varrho^{-1}(b)$.
- If $\varrho_i \uparrow \varrho$, then, as above, one sees that $\varrho_i^{-1}(b)$ is strictly decreasing, hence converges to some $t$. If $t > \varrho^{-1}(b)$, there is some $\tau \in \mathbf{P}_a$ with $t > \tau^{-1}(b) > \varrho^{-1}(b)$, by denseness. Now, $\varrho(a) = \lim \varrho_i(a) \leq \tau(a) < \varrho(a)$. Contradiction. **qed**

Now, we can define non-monotonous limits.

**Definition 10.4** We write for $\varrho, \varrho_i \in \mathbf{P}_a$
$$\varrho_i \to \varrho \iff \varrho_i(t) \to \varrho(t) \quad \text{for all } t \in \text{int } \mathbf{L}(\varrho)$$

Again, we assume that $t \in \mathbf{L}(\varrho_i)$ for almost all $i$ (i.e., all except for finitely many).

**Lemma 10.18** Let **P** be a positive motion set and let $\varrho, \varrho_i \in \mathbf{P}_a$. Then
$$\varrho_i \to \varrho \iff \varrho_i(s) \to \varrho(s) \quad \text{for some } s \in \mathbf{L}(\varrho)$$

**Proof** Let $t \in \text{int } \mathbf{L}(\varrho)$. We may assume that $s, t \in \mathbf{L}(\varrho_i)$ for all $i$.
- Assume that there are infinitely many $\varrho_i(s)$ larger than $\varrho(s)$. Then there is a strictly decreasing subsequence $(\varrho_{i'}(s))$. By Lemma 10.16, $\varrho_{i'}(t) \downarrow \varrho(t)$. As, for any $i'$, we have $\varrho_i(s) \leq \varrho_{i'}(s)$, hence $\varrho_i(t) \leq \varrho_{i'}(t)$ for large $i$, we have $\overline{\lim} \varrho_i(t) \leq \varrho(t)$.
- Assume that there are finitely many $\varrho_i(s)$ larger than $\varrho(s)$. Then $\varrho_i(s) \leq \varrho(s)$ for large $i$, hence $\varrho_i(t) \leq \varrho(t)$ for large $i$. Hence we have $\overline{\lim} \varrho_i(t) \leq \varrho(t)$.

Thus, we have always $\overline{\lim} \varrho_i(t) \leq \varrho(t)$. Completely analogously, one shows $\underline{\lim} \varrho_i(t) \geq \varrho(t)$, giving $\varrho_i(t) \to \varrho(t)$. **qed**

**Proposition 10.19** Let **P** be a positive motion set and let $\varrho, \sigma, \varrho_i, \sigma_i \in \mathbf{P}_a$.
If, moreover, $\varrho_i \to \varrho$, $\sigma_i \to \sigma$ and $\varrho_i \leq \sigma_i$ for large $i$, then $\lim \varrho_i \leq \lim \sigma_i$.

**Proof** By assumption, $\varrho_i(a) \leq \sigma_i(a)$, hence $\varrho(a) \leq \sigma(a)$. **qed**

### 10.4 Metric Structure

**Definition 10.5** Let **P** be a positive motion set. We define a metric $d$ on $\mathbf{P}_a$ by
$$d(\varrho_1, \varrho_2) \ := \ |\varrho_1(a) - \varrho_2(a)|$$

Obviously, $d$ defines a metric. Indeed, $d(\varrho_1, \varrho_2) = 0$ implies $\varrho_1(a) = \varrho_2(a)$, hence $\varrho_1 = \varrho_2$. Also, Definition 10.4 and Lemma 10.18 imply $\varrho_i \to \varrho$ iff $\varrho_i(a) \to \varrho(a)$. As $\varrho(a)$ never equals $b$, we have

**Proposition 10.20** Let **P** be a positive motion set. Then the convergence in the metric space $(\mathbf{P}_a, d)$ equals that of Definition 10.4.

In particular, we have

**Theorem 10.21** Let **P** be a positive motion set. Then the evaluation map
$$\begin{aligned} \mathbf{\Phi}: \ \mathbf{P}_a &\longrightarrow \mathbf{P}(a) \\ \varrho &\longmapsto \varrho(a) \end{aligned}$$
is an isomorphism of metric spaces.

## 11 Elementary Algebra on Motion Sets

For finite positive motion sets we already know that they are generated by a single element $\mu$. In the infinite case, this is of course not possible anymore (unless **P** is countable). Thus, let us go back again to our prime examples for positive motion sets, namely the action of $SO(2)$ by shifts on $S^1$, restricted to some non-trivial compact interval $I$. Now, any shift by $x$ can be seen as the multiplication of $e^{it} \in I \subseteq S^1$ with $e^{xA}$ where $A$ is an appropriate element of the Lie algebra $\mathfrak{so}(2) \cong i\mathbb{R}$. Appropriate means that $e^{xA}e^{it} = e^{i(x+t)}$, i.e., $A = i$. Now, we immediately see that any shift by $x$ is given by $\varrho^x$ with $\varrho = e^A$. Although this slightly abuses notation, the mapping $x \longmapsto \varrho^x$ fulfills all usual power laws. Even more, it is compatible with the discrete case: here, we have $x \in \mathbb{R}$; there, we have $x \in \mathbb{Z}$.

This motivates to search for fractional powers of $\varrho$ in infinite, i.e. dense motion sets. Since, for negative ones, this notion hardly makes sense, we will restrict ourselves to positive motion sets. We should even add a further assumption, namely the fullness of **P**. In fact, take the set $A$ of all rational numbers with finite triadic expansion, i.e., the numbers that can be written in the form $p/3^n$ with $p \in \mathbb{Z}$, $n \in \mathbb{N}$. Let now **P** consist of the operators on $[0,1]$ that are induced by shifts by lengths in $A$. Of course, **P** is a positive motion set. Also, it is dense, since $\mathbf{P}(0) = A \cap [0, 1)$ is dense in $[0, 1]$. Nevertheless, we cannot define a square root in **P** in the usual sense, i.e. as a inversion of taking the square. Indeed, if $\sigma$ is the shift by $s$, then $\sigma^2$ is the shift by $2s$. But, $\frac{1}{3}$ cannot by written as $2s$ for some $s \in A$, whence the shift by $\frac{1}{3}$ cannot have a square root in **P**. Therefore, we shall assume that **P** is even full.

For full positive motion sets **P**, we will indeed be able to construct fractional, even real powers of any of its elements. Let us get the idea, first for square roots. We will see in a moment, that the mapping $\varrho \longmapsto \varrho^2$ is continuous, provided the multiplicity of $\varrho$ is at least 2 (which is always true for some neighbourhood of $a$). Moreover, it is monotonous there, since $\varrho^2 \geq \varrho$ by $\varrho^2(a) \geq \varrho(a)$. Now, by fullness, taking squares maps the interval $[a, \varrho(a)]$ bijectively, even homeomorphically to the interval $[a, \varrho^2(a)]$. This now allows to define the square root as the inverse mapping to the squaring map above. Similarly, the $n$-th root can be defined.

Next, we have to admit also positive rational exponents. Naively, one just sets $\varrho^{\frac{m}{n}} := (\varrho^{\frac{1}{n}})^m$, but for which $m$ and $n$ is this well defined? Having a look at the prime example on $\mathbb{R}$ with $\varrho$ being the unit right-shift, we see that we can define $\varrho^x$ provided $x$ is smaller than the length of the interval $I$. But, how can we measure the length of $[a, b]$ in the general case? The idea is like for the length measurement by rulers of fixed length. We already know how to break the ruler in $n$ pieces of equals "length"; indeed, the square root $\sigma$ of $\varrho$ allows to divide the interval $[a, \varrho(a)]$ into two "aequilateral" intervals $[a, \sigma(a)]$ and $[\sigma(a), \sigma^2(a)]$. The $n$-th roots now do the job for any $n$. This way, we increase the precision of our measurement. Ultimately, we can refine the notion of multiplicity of $\varrho$ as the limit of $n\mathbf{m}(\varrho^{\frac{1}{n}})$. Thus, as long as $x$ is smaller than this fractional multiplicity $\mathbf{x}(\varrho)$, we can indeed define powers with rational exponents. Finally, continuity provides us with any positive real exponent, where positivity can be dropped upon admitting inverses. The crucial observation will ultimately be that all usual power laws transfer to the real exponents. In particular, the mapping $x \longmapsto \varrho^x$ is a homomorphism giving the desired identification of $[0, \mathbf{x}(\varrho)) \longrightarrow \mathbf{P}_a$ as a local sub-semigroup.

The structure of this section follows precisely the ideas above. We will round up with classifying also the general full motion sets, i.e., admit reflections at the end.

## 11.1 Multiplication

First, let us prove that multiplication is continuous in $\mathbf{P}_a$.

**Lemma 11.1** Let $\mathbf{P}$ be a positive motion set and $\varrho, \varrho_i, \sigma, \sigma_i \in \mathbf{P}_a$ with $\varrho_i \to \varrho$ and $\sigma_i \to \sigma$. Then the first of the following conditions implies the second one:
1. $\varrho_i \circ \sigma_i$ is right-moving for large $i$ and $\overline{\lim}_i \varrho_i(\sigma_i(a)) < b$.
2. $\varrho \circ \sigma$ is right-moving.

**Proof** We may assume that $\mathbf{P}$ is dense. Choose $r$ with $\overline{\lim}_i \varrho_i(\sigma_i(a)) < r < b$.
- First observe that $\overline{\lim}_i \sigma_i(a) < r$, hence $\sigma(a) < r$.
- Then, by denseness, there is some $\tau \in \mathbf{P}$ with $\sigma(a) < r < \tau(\sigma(a)) < b$.
- Corollary 6.8 implies $\sigma(a) \in \mathbf{L}(\tau)$ and $\tau \in \mathbf{P}_a$.
- As $\sigma(a) \in \operatorname{int} \mathbf{L}(\tau)$, we have $\sigma_i(a) \in \mathbf{L}(\tau)$ for large $i$.
- Now, obviously, $\varrho_i(\sigma_i(a)) < r < \tau(\sigma_i(a)) < b$ for large $i$.
- This implies $\varrho_i < \tau$ for large $i$ as $\sigma_i(a) \in \mathbf{L}(\varrho_i) \cap \mathbf{L}(\tau)$.
- Taking the limit, we have $\varrho \leq \tau$, hence $\operatorname{int} \mathbf{L}(\varrho) \supseteq \operatorname{int} \mathbf{L}(\tau) \ni \sigma(a)$.
- Together, we get $\sigma(a) \leq \varrho(\sigma(a)) \leq \tau(\sigma(a)) < b$. **qed**

**Proposition 11.2** Let $\mathbf{P}$ be a positive motion set and $\varrho, \varrho_i, \sigma, \sigma_i \in \mathbf{P}_a$. Then we have

$$\varrho_i \downarrow \varrho \ \wedge \ \sigma_i \downarrow \sigma \ \implies \ \varrho_i \bullet \sigma_i \downarrow \varrho \bullet \sigma$$
$$\varrho_i \to \varrho \ \wedge \ \sigma_i \to \sigma \ \implies \ \varrho_i \bullet \sigma_i \to \varrho \bullet \sigma$$
$$\varrho_i \uparrow \varrho \ \wedge \ \sigma_i \uparrow \sigma \ \implies \ \varrho_i \bullet \sigma_i \uparrow \varrho \bullet \sigma$$

in each of the two cases:
1. $\varrho_i \circ \sigma_i$ is right-moving for large $i$ and $\overline{\lim}_i \varrho_i(\sigma_i(a)) < b$.
2. $\varrho \circ \sigma$ is right-moving.

**Proof** Again, we may assume that $\mathbf{P}$ is dense. By Lemma 11.1, also $\varrho \circ \sigma$ is right-moving. Consequently, $\sigma(a) \in \operatorname{int} \mathbf{L}(\varrho)$. By denseness, there are some $\tau_k \in \mathbf{P}_a$ with $\tau_k \downarrow \sigma$ and $\tau_k(a) < \varrho^{-1}(b)$, hence $\tau_k(a) \in \operatorname{int} \mathbf{L}(\varrho)$ for all $k$. In particular, by monotonicity, we may assume $\tau_k(a) \in \mathbf{L}(\varrho_i)$ for all $i$. Fixing $k$ for a moment, we see that $\sigma_i(a) < \tau_k(a)$, hence $\varrho_i(\sigma_i(a)) < \varrho_i(\tau_k(a))$ for large $i$. This gives $\overline{\lim}_i \varrho_i(\sigma_i(a)) \leq \overline{\lim}_i \varrho_i(\tau_k(a)) = \varrho(\tau_k(a))$. As the left-hand side is independent from $k$, we get $\overline{\lim} \varrho_i(\sigma_i(a)) \leq \varrho(\sigma(a))$. Similarly, we see that $\underline{\lim} \varrho_i(\sigma_i(a)) \geq \varrho(\sigma(a))$, hence $\varrho_i(\sigma_i(a)) \to \varrho(\sigma(a))$. The assertions requiring monotonicity are now obvious, as concatenations preserve monotonicity; remember that anything takes place on left domains. **qed**

Sometimes, we cannot guarantee that $\sigma \circ \varrho$ is right-moving again, as $\overline{\lim}_i \varrho_i(\sigma_i(a))$ is $b$. Then, only the normal concatenation of the functions converges in the lower line.

**Lemma 11.3** Let $\mathbf{P}$ be a positive motion set, $\varrho, \varrho_i, \sigma, \sigma_i \in \mathbf{P}_a$ and $\varrho_i \circ \sigma_i$ right-moving. Then

$$\varrho_i \uparrow \varrho \ \wedge \ \sigma_i \uparrow \sigma \ \implies \ \varrho_i(\sigma_i(a)) \uparrow \varrho(\sigma(a)) \,.$$

In particular, $\varrho(\sigma(a))$ is well defined. Moreover, $a < \sigma(a) < \varrho(\sigma(a))$.

**Proof** By assumption, $\sigma_i(a) \in \mathbf{L}(\varrho_i) \subset \mathbf{L}(\varrho_j)$ for $i > j$. By closedness of domains, then $\sigma(a) \in \mathbf{L}(\varrho_j)$ for all $j$. On the other hand, $\mathbf{L}(\varrho_j) \downarrow \mathbf{L}(\varrho)$, whence $\sigma(a) \in \mathbf{L}(\varrho)$. Thus, $\varrho(\sigma(a))$ is well defined with $\varrho(\sigma(a)) > \sigma(a)$ as obviously $\varrho \neq \mathbf{1}$.. Moreover, $\sigma_i(a) \in \mathbf{L}(\varrho)$ as well, whence $\varrho(\sigma(a)) > \varrho(\sigma_i(a)) > \varrho_i(\sigma_i(a))$ for all $i$. If now $\overline{\lim}_i \varrho_i(\sigma_i(a)) < b$, then Lemma 11.1 and Proposition 11.2 show that $\varrho \circ \sigma$ is right-moving, hence $\varrho_i \circ \sigma_i = \varrho_i \bullet \sigma_i \uparrow \varrho \bullet \sigma = \varrho \circ \sigma$ in $a$. If, on the other hand, $\overline{\lim}_i \varrho_i(\sigma_i(a)) = b$, then $\varrho_i(\sigma_i(a)) < \varrho(\sigma(a))$ implies $b = \lim \varrho_i(\sigma_i(a)) = \varrho(\sigma(a))$. **qed**

The statements above can be generalized directly to multiple $\bullet$-products. One only has to extend the right-moving condition from the previously two functions to multiple functions. Indeed, if $\varrho_i \circ \sigma_i \circ \tau_i$ is right-moving, then $\varrho_i \circ (\sigma_i \bullet \tau_i)$ is right-moving as well as $\sigma_i \circ \tau_i$. Moreover, obviously, $\overline{\lim}_i \sigma_i(\tau_i(a)) \leq \overline{\lim}_i \varrho_i(\sigma_i(\tau_i(a))) < b$. Therefore, $\sigma_i \bullet \tau_i \to \sigma \bullet \tau$, and consequently, $\varrho_i \bullet (\sigma_i \bullet \tau_i) \to \varrho \bullet (\sigma \bullet \tau) = \varrho \bullet \sigma \bullet \tau$. Here, we have used that $\sigma_i \bullet \tau_i = \sigma_i \circ \tau_i$ in $a$ as well as that $\varrho \circ \sigma \circ \tau$ is right-moving (by assumption or by Lemma 11.1).

## 11.2 $n$-th Powers

The most important application for multiple products are integer powers $\varrho^n$.

**Proposition 11.4** Let $\mathbf{P}$ be a positive motion set. Then we have for all $\varrho, \varrho_i \in \mathbf{P}_a$ and $n \leq \mathbf{m}(\varrho)$

$$\varrho_i \to \varrho \implies \varrho_i^n \to \varrho^n$$

Let us summarize some important properties of taking the $n$-th power.

**Lemma 11.5** Let $\mathbf{P}$ be a positive motion set. Then we have for all $\varrho, \varrho_i \in \mathbf{P}_a$

$$\varrho_i \downarrow \varrho \implies \mathbf{m}(\varrho_i) \to \mathbf{m}(\varrho)$$

Moreover, unless $\varrho$ is the identity, we even have $\mathbf{m}(\varrho_i) = \mathbf{m}(\varrho)$ for large $i$.

**Proof** As the multiplicity is non-increasing (Corollary 7.8), $\mathbf{m}(\varrho_i)$ is non-descreasing and bounded by $\mathbf{m}(\varrho)$. Let $k < \mathbf{m}(\varrho)$. Then $\varrho^k(a) \in \mathrm{int}\,\mathbf{L}(\varrho)$, hence $\varrho^k(a) \in \mathrm{int}\,\mathbf{L}(\varrho_j)$ for some $j$ by Lemma 10.15. Now, $\varrho_i^k(a) \in \mathrm{int}\,\mathbf{L}(\varrho_j) \subseteq \mathrm{int}\,\mathbf{L}(\varrho_i)$ for large $i \geq j$. This shows $k < \mathbf{m}(\varrho_i)$ for large $i$ inductively.
- If $\varrho$ is not the identity, then $\mathbf{m}(\varrho)$ is finite, hence $\mathbf{m}(\varrho_i) = \mathbf{m}(\varrho)$ for large $i$.
- If $\varrho$ is the identity, then, as shown above, $\mathbf{m}(\varrho_i)$ is not bounded. As $\mathbf{m}(\varrho_i)$ is non-decreasing and $\mathbf{m}(\mathbf{1}) = \infty$, we get the proof. **qed**

**Lemma 11.6** Let $\mathbf{P}$ be a positive motion set. Moreover, let $\varrho, \varrho_i \in \mathbf{P}_a$ with $\varrho_i \uparrow \varrho$ and $n \leq \mathbf{m}(\varrho_i)$. Then $\mathbf{m}(\varrho_i)$ stabilizes at some $\mathbf{m}$ for large $i$ and we have the following implications

$$\begin{aligned} n = \mathbf{m} \text{ and } \varrho_i^n(a) \uparrow b &\implies \varrho_i^n(a) \uparrow \varrho_\circ^n(a) = b \text{ and } \mathbf{m}(\varrho) = \mathbf{m} - 1 \\ \text{otherwise} &\implies \varrho_i^n(a) \uparrow \varrho_\circ^n(a) = \varrho^n(a) \text{ and } \mathbf{m}(\varrho) = \mathbf{m} \end{aligned}$$

**Proof** As $\mathbf{m}(\varrho_i)$ is non-increasing, it stabilizes. Observe that $\varrho_i < \varrho$ implies $\mathbf{m} = \lim \mathbf{m}(\varrho_i) \geq \mathbf{m}(\varrho)$. The multiple-factor version of Lemma 11.3 shows that $\varrho_i^n(a) \uparrow \varrho_\circ^n(a)$. Moreover, it gives $a < \varrho(a) < \ldots < \varrho_\circ^{\mathbf{m}-1}(a) < \varrho_\circ^{\mathbf{m}}(a)$. This shows the relations for $\mathbf{m}(\varrho)$. The lower line follows taking also the snaking lemma into account. **qed**

## 11.3 $n$-th Roots

We now know that the exponentiation by $n$ is continuous provided the multiplicities of the involved mapping is at most $n$. This suggests to take that set as the domain for taking $n$-th roots.

**Definition 11.1** Let **P** be a positive motion set. Then we set for $n \in \mathbb{N}$
$$\mathbf{P}^{1/n} := \{\varrho \in \mathbf{P}_a \mid \mathbf{m}(\varrho) \geq n\}.$$

**Lemma 11.7** Let **P** be a positive motion set and $n \in \mathbb{N}$.
Then $\mathbf{P}^{1/n}$ is a semi-open[23] interval containing **1**.

**Proof** $\mathbf{P}^{1/n}$ contains **1**, since $\mathbf{m}(\mathbf{1}) = \infty$. It is an interval, since given $\varrho \in \mathbf{P}^{1/n}$, all $\sigma \leq \varrho$ fulfill $\mathbf{m}(\sigma) \geq \mathbf{m}(\varrho) \geq n$ by Lemma 11.6. To prove semi-openness, first observe that any interval in $\mathbf{P}_a$ is semi-open, provided **P** is finite. Thus, we may assume that $\mathbf{P}_a$ is dense. Assume that $\mathbf{P}^{1/n}$ is not semi-open. This means that $\mathbf{P}^{1/n}$ equals $[\mathbf{1}, \sigma]$ for some $\sigma$. By denseness, choose $\sigma_i \notin \mathbf{P}^{1/n}$ with $\sigma_i \downarrow \sigma$. If $\sigma$ is not the identity, Lemma 11.5 shows that $\mathbf{m}(\sigma_i)$ eventually equals $\mathbf{m}(\sigma)$, i.e., $\sigma_i \in \mathbf{P}^{1/n}$ for large $i$. If $\sigma$ is the identity, then $\mathbf{m}(\sigma_i) \to \infty$, whence also here $\sigma_i \in \mathbf{P}^{1/n}$. Contradiction. **qed**

**Proposition 11.8** Let **P** be a positive and full motion set. Then we have for $n \in \mathbb{N}_+$:
Taking the $n$-th power is a homeomorphism between $\mathbf{P}^{1/n}$ and $\mathbf{P}_a$.

Note that the proposition is in general not true for finite or dense motion sets.

**Proof** As **P** is full, $\mathbf{P}(a)$ equals $[a, b]$. Consequently, intervals in $\mathbf{P}_a$ are mapped to intervals in $[a, b]$. Therefore, to prove homeomorphy, we only have to check that taking the $n$-th power is a continuous bijection between $\mathbf{P}^{1/n}$ and $\mathbf{P}_a$. As continuity has already been proven in Proposition 11.4, we just have to show bijectivity.
- Injectivity
  Assume that $\varrho_1 < \varrho_2$ are elements in $\mathbf{P}^{1/n}$. Then, for $k < \mathbf{m}(\varrho_2) \leq \mathbf{m}(\varrho_1)$, we have
  $$\varrho_1^{k+1}(a) = \varrho_1(\varrho_1^k(a)) < \varrho_2(\varrho_1^k(a)) < \varrho_2(\varrho_2^k(a)) = \varrho_2^{k+1}(a)$$
  Here, we have used inductively that $\varrho_1^k(a) < \varrho_2^k(a) \in \mathbf{L}(\varrho_2) \subset \mathbf{L}(\varrho_1)$.
- Surjectivity
  From Lemma 11.7 we know that $\mathbf{P}^{1/n} = [\mathbf{1}, \sigma)$ for some $\sigma \in \mathbf{P}_a$. Choose some $\sigma_i \uparrow \sigma$. Since $\mathbf{m}(\sigma) < n \leq \mathbf{m}(\sigma_i)$, Lemma 11.6 shows that $\sigma_i^n(a) \uparrow \sigma_\circ^n(a) = b$. On the other hand, **P** is full, whence $\mathbf{P}^{1/n}(a)$ as an interval is connected. Consequently, its continuous image under taking the $n$-th power is connected again. As this image interval contains $a$ and allows to approximate $b$, it must equal $[a, b)$. **qed**

This allows us to define the $n$-th root on $\mathbf{P}_a$ just by inverting this bijection.

**Definition 11.2** Let **P** be a positive motion set, and let $\varrho$ and $\sigma$ be in $\mathbf{P}_a$. Then
$$\sigma \ldots n\text{-th root of } \varrho \iff \sigma \in \mathbf{P}^{1/n} \text{ and } \varrho \text{ equals } \sigma^n.$$

As usual we denote the $n$-th root of $\varrho$ by $\varrho^{\frac{1}{n}}$. By construction, we see that

**Proposition 11.9** Let **P** be a positive and full motion set.
Then each $\varrho \in \mathbf{P}_a$ has a unique $n$-th root.

Moreover, the first root of $\varrho$ is $\varrho$ itself. The roots fulfill the usual power law

**Lemma 11.10** Let **P** be a positive and full motion set. Then we have
$$\left(\varrho^{\frac{1}{nm}}\right)^m = \varrho^{\frac{1}{n}} \qquad \text{for all } \varrho \in \mathbf{P}_a \text{ and all } m, n \in \mathbb{N}_+.$$

---

[23]It means that $\mathbf{P}^{1/n}(a)$ is the intersection of $\mathbf{P}(a)$ with some interval $[a, c) \subseteq [a, b]$.

**Proof** Let $\sigma := \varrho^{\frac{1}{nm}} \in \mathbf{P}^{1/nm}$. First, observe that $\mathbf{m}(\sigma) \geq mn$ and Proposition 4.5 imply that $\varrho = \sigma^{mn} = (\sigma^m)^n$. Since $\mathbf{m}(\sigma) \geq mn \geq m$, Proposition 7.13 now implies

$$\mathbf{m}(\sigma^m) \;>\; \frac{\mathbf{m}(\sigma)}{m} - 1 \;\geq\; \frac{mn}{m} - 1 \;=\; n - 1$$

hence $\mathbf{m}(\sigma^m) \geq n$ and $\sigma^m \in \mathbf{P}^{1/n}$. Consequently, $\sigma^m = \varrho^{\frac{1}{n}}$. **qed**

Proposition 7.13 immediately gives

**Corollary 11.11** Let $\mathbf{P}$ be a positive and full motion set. Then we have for all $\mathbf{1} \neq \varrho \in \mathbf{P}_a$, $m, n \in \mathbb{N}_+$

$$\frac{\mathbf{m}(\varrho^{\frac{1}{n}})}{n} \;\leq\; \frac{\mathbf{m}(\varrho^{\frac{1}{nm}})}{mn} \;<\; \frac{\mathbf{m}(\varrho^{\frac{1}{n}}) + 1}{n}$$

## 11.4 Rational Powers

Next, we would like to admit (positive) rational exponents. As already announced, this requires a refinement of the notion of multiplicity.

**Lemma 11.12** Let $\mathbf{P}$ be a positive and full motion set. Moreover let $\varrho \in \mathbf{P}_a$. Then we have

$$\mathbf{m}(\varrho) \;\leq\; \mathbf{x}(\varrho) \;:=\; \sup_{n \in \mathbb{N}_+} \frac{\mathbf{m}(\varrho^{\frac{1}{n}})}{n} \;=\; \lim_{n \in \mathbb{N}_+} \frac{\mathbf{m}(\varrho^{\frac{1}{n}})}{n}.$$

In particular, $\mathbf{x}(\varrho)$ is positive. Moreover, it is finite iff $\varrho \neq \mathbf{1}$. Finally,

$$\frac{m}{n} < \mathbf{x}(\varrho) \;\implies\; \mathbf{m}(\varrho^{\frac{1}{n}}) \geq m\,.$$

**Definition 11.3** $\mathbf{x}(\varrho)$ is called **fractional multiplicity** of $\varrho$.

**Proof** For $\varrho = \mathbf{1}$, the statement is trivial; indeed, $\mathbf{x}(\mathbf{1})$ obviously equals $\infty$. Thus, let $\varrho \neq \mathbf{1}$ and

$$a_n \;:=\; \frac{\mathbf{m}(\varrho^{\frac{1}{n}})}{n}\,.$$

From Corollary 11.11 we get immediately $a_n \leq a_{nm} < a_n + \frac{1}{n}$ for $m, n \in \mathbb{N}_+$. Exchanging the rôles of $m$ and $n$, we see that $|a_n - a_m| < \max(\frac{1}{n}, \frac{1}{m})$. Hence, $(a_n)$ is Cauchy, proving the existence of the limit as well as finiteness. Moreover, if $\lim a_n < \sup a_n =: s$, then $s = a_N$ for some $N$. But, $s = a_N \leq a_{kN} \leq s$ implies $a_{kN} = s$ for all $k$, hence $\lim a_n = s$. Contradiction. Furthermore, observe that $\mathbf{x}(\varrho) \geq a_1 = \mathbf{m}(\varrho)$, which is positive. For the final implication use Corollary 11.11 to obtain

$$\mathbf{m}(\varrho^{\frac{1}{n}}) + 1 \;\geq\; n \lim_{k \to \infty} \frac{\mathbf{m}(\varrho^{\frac{1}{kn}})}{kn} \;=\; n\,\mathbf{x}(\varrho) \;>\; nx \;=\; m\,.$$

**qed**

**Definition 11.4** The $\frac{m}{n}$-**th power** of $\varrho \in \mathbf{P}$ is defined for $0 \leq \frac{m}{n} < \mathbf{x}(\varrho)$ with $m, n \in \mathbb{N}$ by

$$\varrho^{\frac{m}{n}} \;:=\; \left(\varrho^{\frac{1}{n}}\right)^m$$

**Proposition 11.13** Let $\mathbf{P}$ be a positive and full motion set.
Then $\varrho^x$ is well defined for all $\varrho \in \mathbf{P}_a$ and all rational $0 \leq x < \mathbf{x}(\varrho)$.

**Proof** Let $x = \frac{m}{n}$. Lemma 11.12 gives $\mathbf{m}(\varrho^{\frac{1}{n}}) \geq m$, whence $\varrho^{\frac{m}{n}}$ exists. Similarly, we see that

$$\varrho^{\frac{mk}{nk}} \;\equiv\; \left(\varrho^{\frac{1}{nk}}\right)^{mk} \;=\; \left[\left(\varrho^{\frac{1}{nk}}\right)^k\right]^m \;=\; \left[\varrho^{\frac{1}{n}}\right]^m \;=\; \varrho^{\frac{m}{n}}\,,$$

by Lemma 11.10. Choosing co-prime $m$ and $n$, this proves well-definedness. **qed**

**Proposition 11.14** Let $\mathbf{P}$ be a positive and full motion set. Moreover, let $\varrho \in \mathbf{P}_a$.
Then we have for all rational $0 \leq x, y, x_k < \mathbf{x}(\varrho)$
1. If $x + y < \mathbf{x}(\varrho)$, then $\varrho^x \circ \varrho^y$ is right-moving and
$$\varrho^{x+y} = \varrho^x \bullet \varrho^y$$
2. $$x \leq y \implies \varrho^x \leq \varrho^y$$
3. $$x_k \downarrow 0 \implies \varrho^{x_k} \searrow \mathbf{1}$$
4. $$\varrho^x = \mathbf{1} \iff \varrho = \mathbf{1} \text{ or } x = 0.$$

**Proof** For the first two assertions, we may write $x = m_1/n$ and $y = m_2/n$ for $m_1, m_2, n \in \mathbb{N}$, and let $\sigma := \varrho^{1/n}$. If now $x \leq y$, i.e. $m_1 \leq m_2 \leq \mathbf{m}(\sigma)$, we get $\sigma^{m_1}(a) \leq \sigma^{m_2}(a)$, from the right-movement definition. Hence $\varrho^x = \sigma^{m_1} \leq \sigma^{m_2} = \varrho^y$ by Proposition 6.2. If now $x + y < \mathbf{x}(\varrho)$, i.e. $m_1 + m_2 \leq \mathbf{m}(\sigma)$, we get $\varrho^{x+y} = \sigma^{m_1+m_2} = \sigma^{m_1} \bullet \sigma^{m_2} = \varrho^x \bullet \varrho^y$ and the right-moving property. For the third assertion, observe that $\varrho^{x_k}$ is monotonously decreasing, hence converging to some $\tau \in \mathbf{P}_a$. Since the multiplicity is non-increasing, we have $\mathbf{m}(\tau) \geq \mathbf{m}(\varrho^{\frac{1}{n}}) \geq n\mathbf{m}(\varrho) \geq n$, for all $n$, hence $\mathbf{m}(\tau) = \infty$. Consequently, $\tau = \mathbf{1}$. For the final line, only the implication is nontrivial. Indeed, if $\varrho^x = \mathbf{1}$ with $\frac{1}{n} \leq x$ for some $n$, then $\varrho^{1/n} = \mathbf{1}$ by monotonicity, hence $\varrho = \mathbf{1}$. **qed**

**Corollary 11.15** Let $\mathbf{P}$ be a positive and full motion set. Moreover, let $\mathbf{1} \neq \varrho \in \mathbf{P}_a$. Then
$$\sup\{\varrho^x(a) \mid 0 < x < \mathbf{x}(\varrho) \text{ rational}\} = b.$$

**Proof** Denote the supremum by $t$ and assume $t < b$. Then there are $\sigma, \tau \in \mathbf{P}_a$ with $t < \sigma(a) < \tau(a)$. As $\sigma^{-1} \bullet \tau$ is in $\mathbf{P}_a$, but not the identity, we have $\upsilon := \varrho^{\frac{1}{n}} < \sigma^{-1} \bullet \tau$ for all large $n$. Choose sufficiently large $m$ and $n$, such that additionally
$$x := \frac{m}{n} < \mathbf{x}(\varrho) < \frac{m}{n} + \frac{1}{n}.$$
Now, $\sigma \circ \sigma^{-1} \bullet \tau$ is right-moving, by $\sigma < \tau$ and Lemma 5.11. Since $\varrho^x < \sigma$ and $\upsilon < \sigma^{-1} \bullet \tau$, we see from Lemma 7.5 that also $\varrho^x \circ \upsilon$ is right-moving. Moreover, $\varrho^x \bullet \upsilon$ is well defined and smaller than $\sigma \bullet \sigma^{-1} \bullet \tau = \tau$. From $\varrho^x = \upsilon^m$, we see that $\upsilon^m \circ \upsilon$ is right-moving. Observe that by Corollary 11.11 and Lemma 11.12, we have $\mathbf{m}(\upsilon) \geq m$, hence $\upsilon_\circ^m$ is right-moving. Now, Lemma 7.11 shows that also $\upsilon_\circ^{m+1}$ is right-moving, implying $\mathbf{m}(\upsilon) \geq m + 1$. This, in turn, shows
$$\mathbf{x}(\varrho) \geq \frac{\mathbf{m}(\varrho^{\frac{1}{n}})}{n} = \frac{m}{n} + \frac{1}{n} > \mathbf{x}(\varrho),$$
giving the desired contradiction. **qed**

## 11.5 Positive Real Powers

Using continuity, we now admit even real exponents.

**Definition 11.5** The $x$-**th power** of $\varrho \in \mathbf{P}$ is defined for $0 \leq x < \mathbf{x}(\varrho)$ by
$$\varrho^x := \lim_{k \to \infty} \varrho^{x_k} \quad \text{for any rational sequence } x_k \to x.$$

Of course, we have tacitly assumed that $0 \leq x_k \leq \mathbf{x}(\varrho)$.

**Lemma 11.16** Let $\mathbf{P}$ be a positive and full motion set.
Then $\varrho^x$ is well defined for all $\varrho \in \mathbf{P}_a$ and $0 \leq x < \mathbf{x}(\varrho)$.

Thus, in particular, Definition 11.4 and Definition 11.5 are consistent.

**Proof** Choose first two rational sequences with $y_k \nearrow x$ and $z_k \searrow x$. By homeomorphy of $\mathbf{P}_a$ and $\mathbf{P}(a)$, we see that $\varrho^{y_k} \nearrow \sigma$ and $\varrho^{z_k} \searrow \tau$ for some $\sigma, \tau \in \mathbf{P}_a$; for this, note that $y_k < y < \mathbf{x}(\varrho)$ for some $y \in \mathbb{Q}$. As $y_k \leq z_k$, we have $\varrho^{y_k} \leq \varrho^{z_k}$, hence $\sigma \leq \tau$. Assume $\sigma < \tau$. Lemma 5.11, implies that $\sigma^{-1} \bullet \tau$ is in $\mathbf{P}_a$ again. As it is obviously not the identity, Proposition 11.14$_3$. provides us with some rational $0 < q < \mathbf{x}(\varrho) - x$ fulfilling $\varrho^q < \sigma^{-1} \bullet \tau$. As now $z_k < y_k + q$ for some large $n$, we get the contradiction by

$$\varrho^{z_k} \;<\; \varrho^{y_k+q} \;=\; \varrho^{y_k} \bullet \varrho^q \;\leq\; \sigma \bullet \varrho^q \;<\; \sigma \bullet \sigma^{-1} \bullet \tau \;=\; \tau \;\leq\; \varrho^{z_k}.$$

Here, the first inequality comes from Proposition 11.14$_2$., the second and the third one from Lemma 7.5 together with Lemma 5.11, and the first equality comes from Proposition 11.14$_1$.. To finish the proof, let $x_k \to x$ be a rational sequence. Then, we find rational sequences $y_k \nearrow x$ and $z_k \searrow x$ with $y_k \leq x_k \leq z_k$. By the arguments above, $\varrho^{y_k}$ and $\varrho^{z_k}$ converge to the same limit, whence also $\varrho^{x_k}$ must do it giving well-definedness. **qed**

**Proposition 11.17** Let $\mathbf{P}$ be a positive and full motion set. Moreover, let $\varrho \in \mathbf{P}_a$.
Then we have for all real $0 \leq x, y, x_k < \mathbf{x}(\varrho)$
1. If $x + y < \mathbf{x}(\varrho)$, then $\varrho^x \circ \varrho^y$ is right-moving and

$$\varrho^{x+y} \;=\; \varrho^x \bullet \varrho^y$$

2. $$x \leq y \;\Longrightarrow\; \varrho^x \leq \varrho^y$$

3. $$x_k \to x \;\Longrightarrow\; \varrho^{x_k} \to \varrho^x$$

4. $$\varrho^x = \mathbf{1} \;\Longleftrightarrow\; \varrho = \mathbf{1} \text{ or } x = 0.$$

5. $$x_k \uparrow \mathbf{x}(\varrho) \;\Longrightarrow\; \varrho^{x_k}(a) \uparrow b$$

In particular, $\mathbf{P}_a$ is commutative as long as the product is well defined and in $\mathbf{P}_a$ again.

**Proof** 1. Choose rational $x_k \nearrow x$ and $y_k \nearrow y$. Then $\varrho^{x_k}(\varrho^{y_k}(a)) = \varrho^{x_k+y_k}(a) < b$ by Proposition 11.14$_1$.. Moreover, $\overline{\lim}_k \varrho^{x_k+y_k}(a) \leq \varrho^{x+y}(a) < b$ by monotony. Thus, by Lemma 11.1, $\varrho^x \circ \varrho^y$ is right-moving and we have by Proposition 11.2

$$\varrho^{x+y} \;=\; \lim \varrho^{x_k+y_k} \;=\; \lim \varrho^{x_k} \bullet \varrho^{y_k} \;=\; \lim \varrho^{x_k} \bullet \lim \varrho^{y_k} \;=\; \varrho^x \bullet \varrho^y.$$

2. This follows from the corresponding relation in $\mathbb{Q}$ (see Proposition 11.14$_2$.), as limits preserve the ordering.
3. Choose monotonous rational sequences $y_k, z_k \to x$ with $y_k \leq x_k \leq z_k$. This now gives $\varrho^x \leftarrow \varrho^{y_k} \leq \varrho^{x_k} \leq \varrho^{z_k} \to \varrho^x$, hence $\varrho^{x_k} \to \varrho^x$.
4. Here, the proof is verbatim the same as that for Proposition 11.14$_4$.
5. This is a simple consequence of Corollary 11.15 and of monotonicity. **qed**

**Corollary 11.18** Let $\mathbf{P}$ be a positive and full motion set. Then the mapping

$$\varphi : \;\; [0, \mathbf{x}(\varrho)) \;\longrightarrow\; [a, b)$$
$$x \;\longmapsto\; \varrho^x(a)$$

is a homeomorphism for any $\mathbf{1} \neq \varrho \in \mathbf{P}_a$.

**Proof** As continuity follows from the continuity of $x \longmapsto \varrho^x$ and that of the evaluation mapping $\sigma \longmapsto \sigma(a)$, it is sufficient to prove bijectivity. Injectivity follows, as $\varrho^x(a) = \varrho^y(a)$ with $x \leq y$ implies $\varrho^x = \varrho^y = \varrho^{x+(y-x)} = \varrho^x \bullet \varrho^{y-x}$, hence $\varrho^{y-x} = \mathbf{1}$ giving $y - x = 0$ by Proposition 11.17$_4$.. Surjectivity follows from continuity, since $a = \varrho^0(a)$ and $b$ is the limit of $\varrho^{x_k}(a)$ for some appropriate sequence $(x_k)$ by Corollary 11.15. **qed**

This, in particular, proves Theorem 3.6 in the full case. Indeed, let there $\mathbf{T} := [0, \mathbf{x}(\varrho))$ for some $1 \neq \varrho \in \mathbf{P}_a$. Obviously, $\mathbf{T}$ as a subinterval starting at 0 is a local sub-semigroup of $\mathbb{R}$. The mapping $\varphi(x) := \varrho^x(a)$ is a homeomorphism between $\mathbf{T}$ and $\mathbf{P}(a) = [a, b)$. The homomorphy relations have been proven in Proposition 11.17. Finally, the coincidence of $\varrho^{x+y}(a)$ and $\varrho^x(\varrho^y(a))$ is obvious.

## 11.6 Real Powers

Corollary 11.18 provides us with a complete characterization of $\mathbf{P}_a$. It now remains $\mathbf{P}_b$. For this, recall that the inverse of any element in $\mathbf{P}_a$ is an element in $\mathbf{P}_b$. It may now happen that the intersection of $\mathbf{P}_a$ and $\mathbf{P}_b$ is nontrivial, hence contains more than the identity. Let us

**Definition 11.6** The $x$-**th power** of $\varrho \in \mathbf{P}_a$ is defined for $-\mathbf{x}(\varrho) \leq x < 0$ by
$$\varrho^x := (\varrho^{-x})^{-1}.$$

Observe that now $\varrho^x \in \mathbf{P}_b$ for all $-\mathbf{x}(\varrho) < x \leq 0$.

**Proposition 11.19** Let $\mathbf{P}$ be a positive and full motion set. Then we have for all $\varrho \in \mathbf{P}_a$
$$|x|, |y|, |x+y| < \mathbf{x}(\varrho) \implies \varrho^{x+y} = \varrho^x \bullet \varrho^y$$

**Proof** We only prove the two typical cases. In both cases we reduce the problem to the statement for non-negative exponents proven above in Proposition 11.17.
- For $x, y \leq 0$ observe that $\varrho^{-(y+x)} = \varrho^{-y} \bullet \varrho^{-x}$, whence we get the equation by inversion.
- If $x < 0 < y$ and $x + y \geq 0$, then $\varrho^y = \varrho^{-x} \bullet \varrho^{x+y}$. $\varrho^x \bullet \varrho^y$ is well defined as $-x \leq y$. Moreover, $a$ is snaking along $(\varrho^{-x})^{-1} \circ \varrho^{-x} \circ \varrho^{x+y}$, whence $\varrho^x \bullet \varrho^y = \varrho^x \bullet \varrho^{-x} \bullet \varrho^{x+y} = \varrho^{x+y}$. **qed**

**Proposition 11.20** Let $\mathbf{P}$ be a positive and full motion set. Moreover, let $1 \neq \varrho \in \mathbf{P}_a$.
Then there is some $\mathbf{u}(\varrho) \geq 0$, such that
$$\mathbf{P}_a \cap \mathbf{P}_b = \{\mathbf{1}\} \cup \{\varrho^x \mid \mathbf{u}(\varrho) < x < \mathbf{x}(\varrho)\}.$$
Moreover, we have $\varrho^x(b) \downarrow a$ for $x \downarrow \mathbf{u}(\varrho)$, if $\mathbf{P}_a \cap \mathbf{P}_b$ is nontrivial.

Observe that $\mathbf{u}(\varrho)$ is unique if $\mathbf{P}_a \cap \mathbf{P}_b$ is nontrivial. In the trivial case we may choose any number $\mathbf{u}(\varrho) \geq \mathbf{x}(\varrho)$.

**Proof** The statement is trivial for trivial $\mathbf{P}_a \cap \mathbf{P}_b$. Just define $\mathbf{u}(\varrho) := \mathbf{x}(\varrho)$. Thus, let $\mathbf{P}_a \cap \mathbf{P}_b$ be nontrivial and define $\mathbf{u} := \mathbf{u}(\varrho) := \inf\{x \mid \varrho^x \in \mathbf{P}_b, 0 < x < \mathbf{x}(\varrho)\}$.
- By assumption, $0 \leq \mathbf{u} < \mathbf{x}(\varrho)$ is well defined.
- For $\mathbf{u} < x < \mathbf{x}(\varrho)$, choose $\mathbf{u} \leq y < x$ with $\varrho^y \in \mathbf{P}_b$. From $\varrho^y < \varrho^x$ by Proposition 11.17$_2$., we get $\mathbf{L}(\varrho^y) \supset \mathbf{L}(\varrho^x)$ by Proposition 6.2, hence $b \in \mathbf{R}(\varrho^y) \subset \mathbf{R}(\varrho^x)$ by Proposition 6.15. This gives $\varrho^x \in \mathbf{P}_b$.
- For $x_k \downarrow \mathbf{u}$, we have $\varrho^{x_k}(b) \downarrow t$ for some $t < b$ by monotonicity (again Proposition 6.15). If $\mathbf{u} = 0$, then $\varrho^{x_k}(b) < \varrho^{x_k}(a) \to \varrho^0(a) = a$, hence $\varrho^{x_k}(b) \to a$ as well. Pointwise properness, however, shows $a = b$. Therefore, $\mathbf{u} > 0$.
- Assume now $a < t$ and $\mathbf{u} < x$. Choose $s$ and some $0 < y < \mathbf{x}(\varrho)$ with $s \in \mathbf{L}(\varrho^y)$ and $a < s < t < \varrho^x(b) = \varrho^y(s)$. As $\varrho^y(a) < \varrho^y(s) = \varrho^x(b) < \varrho^x(a)$, we have $\varrho^y < \varrho^x$. Consequently, $\varrho^{-y} \bullet \varrho^x \in \mathbf{P}_a$ by Lemma 5.11. On the other hand, also $b$ is snaking along $\varrho^{-y} \circ \varrho^x$, whence $[\varrho^{-y} \bullet \varrho^x](b) = s < t$. Thus, $t$ cannot be the infimum. Contradiction.
- If $\varrho^{\mathbf{u}} \in \mathbf{P}_b$, then $\varrho^{\mathbf{u}}(b) < \varrho^{x_k}(b) \downarrow a$ for any $x_k \downarrow \mathbf{u}$, hence $\varrho^{\mathbf{u}}(b) = a$. This is impossible. **qed**

If $\varrho$ is clear from the context, we may simply write $\mathbf{u}$ instead of $\mathbf{u}(\varrho)$ and $\mathbf{x}$ instead of $\mathbf{x}(\varrho)$.

**Proposition 11.21** Let $\mathbf{P}$ be a positive and full motion set. Moreover, let $\mathbf{1} \neq \varrho \in \mathbf{P}_a$. Then
$$\varrho^x = \varrho^y \iff |x - y| \text{ equals } 0 \text{ or } \mathbf{u}(\varrho) + \mathbf{x}(\varrho)$$

**Proof** For $x$ and $y$ not having opposite sign, it suffices to show that $\varrho^x = \varrho^y$ iff $x = y$.
- If $x, y \geq 0$, then both elements are in $\mathbf{P}_a$, hence $x = y$ by Corollary 11.18.
- If $x, y \leq 0$, then the inverses of both elements are in $\mathbf{P}_a$ giving $-x = -y$.

For $x$ and $y$ having opposite sign, we have to show that $\varrho^x = \varrho^y$ iff $|x - y| = \mathbf{x}(\varrho) + \mathbf{u}(\varrho)$. We may assume $x < 0 < y$ and that $\mathbf{P}_a \cap \mathbf{P}_b$ is nontrivial as otherwise neither side can be fulfilled.
- If $\varrho^x = \varrho^y$, then $\mathbf{u} < -x, y < \mathbf{x}$. Choose positive $z_k \uparrow \mathbf{x} - y$. Then, by homomorphy, $\varrho^{x+z_k}(a) = \varrho^{y+z_k}(a) \uparrow b$. As $x + z_k < 0$, even $x + z_k < -\mathbf{u}$, hence $t := y - x \geq \mathbf{x} + \mathbf{u}$. Assume that the inequality is strict. Then $\varrho^{\mathbf{x}-t} \in \mathbf{P}_a$ by the definition of $\mathbf{u}$. Thus, for large $k$, we have $\varrho^{\mathbf{x}-t}(a) < \varrho^{y+z_k}(a) = \varrho^{y-t+z_k}(a) < b$. Now, Lemma 5.11 implies that $\varrho^{y+z_k-\mathbf{x}} \in \mathbf{P}_a$. But, $y + z_k - \mathbf{x} \uparrow 0$ contradicts the definition of $\mathbf{u}$. Thus, $t = \mathbf{x} + \mathbf{u}$.
- If $y - x = t$, choose $r < 0 < s$ with $\varrho^r = \varrho^s$, which is possible as $\mathbf{P}_a \cap \mathbf{P}_b$ is nontrivial. As shown above, $s - r = t = y - x$. Now, $\varrho^x = \varrho^r \bullet \varrho^{x-r} = \varrho^s \bullet \varrho^{y-s} = \varrho^y$ by homomorphy. **qed**

## 11.7 Classification Result – Positive

**Theorem 11.22 Classification of Full Positive Motion Sets**
Let $\mathbf{P}$ be a positive and full motion set. Moreover, let $\mathbf{1} \neq \varrho \in \mathbf{P}_a$. Then $\mathbf{P}$ is isomorphic to the motion set induced by all translations on $S^1$ restricted to some subinterval of length $\mathbf{x}(\varrho)$, whereas $S^1$ has circumference $\mathbf{x}(\varrho) + \mathbf{u}(\varrho)$.

**Proof** The mapping $\boldsymbol{\varphi}$ from Corollary 11.18 induces a homeomorphism between $I := [0, \mathbf{x}(\varrho)] \subseteq S^1$ and $[a, b]$. From $\boldsymbol{\varphi}(x) = \varrho^x(a) = \varrho^x(\boldsymbol{\varphi}(0))$ for $x \in [0, \mathbf{x}(\varrho))$, we get $x = [\boldsymbol{\varphi}^{-1} \circ \varrho^x \circ \boldsymbol{\varphi}](0)$. Now, $\varrho^{x+y}(a) = \varrho^x(\varrho^y(a))$ implies
$$R_x(y) \equiv x + y = [\boldsymbol{\varphi}^{-1} \circ \varrho^{x+y} \circ \boldsymbol{\varphi}](0)$$
$$= [[\boldsymbol{\varphi}^{-1} \circ \varrho^x \circ \boldsymbol{\varphi}] \circ [\boldsymbol{\varphi}^{-1} \circ \varrho^y \circ \boldsymbol{\varphi}]](0) = [\boldsymbol{\varphi}^{-1} \circ \varrho^x \circ \boldsymbol{\varphi}](y),$$
hence $R_x = \boldsymbol{\varphi}^{-1} \circ \varrho^x \circ \boldsymbol{\varphi}$ on $[0, \mathbf{x}(\varrho) - x]$, at least after completion.
- If $\mathbf{P}_a \cap \mathbf{P}_b$ is trivial, then the domain of $R_x$ has the single component $[0, \mathbf{x}(\varrho) - x]$, because the circumference at least doubles the length of the interval. The same applies to $\varrho^x$. One easily checks that $\varrho^x(\boldsymbol{\varphi}(y)) = \varrho^{x+y}(a) \uparrow b$ for $y \uparrow \mathbf{x}(\varrho) - x$, hence dom $\varrho^x = \mathbf{L}(\varrho^x) = \boldsymbol{\varphi}[0, \mathbf{x}(\varrho) - x]$. This shows $R_x = \boldsymbol{\varphi}^{-1} \circ \varrho^x \circ \boldsymbol{\varphi}$ everywhere. The same applies to negative $x$ corresponding to the inverses.
- If $\mathbf{P}_a \cap \mathbf{P}_b$ is nontrivial, the statement on $\mathbf{L}(R_x)$ is as above. Observe that both $\mathbf{L}(R_x)$ and $\mathbf{L}(\varrho^x)$ have a single component for $0 < x \leq \mathbf{u}(\varrho)$, by construction. Hence, $R_x = \boldsymbol{\varphi}^{-1} \circ \varrho^x \circ \boldsymbol{\varphi}$ everywhere for these $x$, and for $x$ replaced by $-x$ as well. Let now $x > \mathbf{u}(\varrho)$ and $y \in \mathbf{R}(R_x)$. Then $R_x y \in \mathbf{L}(R_{-x})$, hence $y = R_{-x}(R_x(y)) = [\boldsymbol{\varphi}^{-1} \circ \varrho^{-x} \circ \boldsymbol{\varphi}](R_x(y))$. Altogether $R_x = \boldsymbol{\varphi}^{-1} \circ \varrho^x \circ \boldsymbol{\varphi}$ everywhere. Note that we have used that, by choice of the circumference, the change of the number of domain components occurs for $R_x$ and $\varrho^x$ at the same $x$, namely $|x| = \mathbf{u}(\varrho)$. This gives the proof. **qed**

## 11.8 Reflections

So far, we have only considered full and *positive* motion sets. Let us now consider general motion sets $\mathbf{P}$. The aim is to show that any such set is isomorphic to some restriction of $O(2)$ acting on $S^1$. For this, we first show that $\mathbf{P}^-$ contains a reflection defined on full $I$. Recall that $\mathbf{F}_a$ collects all fixed points in the left domains of all reflections; similarly $\mathbf{F}_b$ comprises the right fixed points. $\mathbf{F}$ equals the union of both sets.

**Lemma 11.23**  Let $\mathbf{P}$ be an infinite motion set with non-empty $\mathbf{P}_-$.
Then neither $\mathbf{F}_a$ nor $\mathbf{F}_b$ are empty.

**Proof**  Assume that $\mathbf{F}_b$ is empty. By Theorem 10.13, $\mathbf{F} = \mathbf{F}_a$ is dense. Hence, there are $\mathbf{x}_i \uparrow b$ in $\mathbf{F}_a$. Let $a < t < b$ for some $t$. Then $\mathbf{x}_i > t$ for large $i$, whence the corresponding reflections $\sigma_i$ fulfill $\sigma_i(a) > \sigma_i(t) > \sigma_i(\mathbf{x}_i) = \mathbf{x}_i \uparrow b$, hence $\lim \varrho_i(a) = \lim \varrho_i(t)$ giving $a = t$. Contradiction. **qed**

**Lemma 11.24**  Let $\mathbf{P}$ be a motion set with full $\mathbf{P}^+$ and non-empty $\mathbf{P}^-$.
Then $\mathbf{P}^-$ contains some reflection with domain $[a,b]$.

**Proof**  We know from Corollary 6.18 that the fixed point set $\mathbf{F}$ equals $(a,b)$. Moreover, it can be written as a union $\mathbf{F}_a \cup \mathbf{F}_b$ having at most a single-element intersection and its elements fulfilling $\mathbf{x}_a \leq \mathbf{x}_b$. Hence, $\mathbf{F}_a$ and $\mathbf{F}_b$ are intervals. As both intervals are non-empty, we have $\sup \mathbf{F}_a = \mathbf{x} = \inf \mathbf{F}_b$. As $\mathbf{x} \in \mathbf{F}$, we may assume $\mathbf{x} \in \mathbf{F}_a$. If $\mathbf{L}(\sigma) = [a,b]$, we are done. Thus, let us assume $\mathbf{L}(\sigma) = [a,t]$ with $t < b$. Then choose some $\varrho \in \mathbf{P}_t^+$ with $t < \varrho(t) < b$, hence $\varrho \in \mathbf{P}_a$. As

$$\mathbf{L}(\varrho \bullet \sigma) \;=\; (\varrho \bullet \sigma)\mathbf{L}(\varrho \bullet \sigma) \;\supseteq\; (\varrho \circ \sigma)\mathbf{L}(\varrho \circ \sigma) \;=\; (\varrho \circ \sigma)\mathbf{L}(\sigma) \;=\; [\varrho(a), \varrho(t)],$$

we have $\mathbf{L}(\varrho \bullet \sigma) \supseteq [a, \varrho(t)] \supset [a,t] = \mathbf{L}(\sigma)$. Hence, the fixed point to $\varrho \bullet \sigma$ is in $\mathbf{F}_a$ by construction, but, by Proposition 6.4, strictly larger than $\mathbf{x}$, which is impossible. Hence, $\mathrm{dom}\,\sigma = [a,b]$. **qed**

For any general motion set $\mathbf{P}$, we may assume by Theorem 11.22 that the positive part $\mathbf{P}^+$ is the restriction of $SO(2)$ to some compact subinterval $I$ of $S^1$. We also may assume that the circumference of $S^1$ is $2\pi$. Our aim is to show that then $\mathbf{P}$ equals the restriction of $O(2)$ to that subinterval. For this, let $\sigma_0$ denote the reflection having full domain $I$. We may finally assume that $I = [-b, b]$.

- $\sigma_0$ is the reflection in $0$.

  For this, consider the right-shift $\varrho$ by $b$. Then $\varrho \circ \sigma_0 \circ \varrho$ maps $0$ via $b$ and $-b$ to $0$. Hence $0$ is snaking. Indeed, between $b$ and $-b$, an odd number of reflections, namely a single one, is used. Thus, $\sigma_0 \bullet \varrho \bullet \sigma_0 \bullet \varrho$ is well defined as well as $\sigma_0 \bullet \varrho$. The latter one is obviously a reflection, whence the former one is even the identity. In particular, $\sigma_0(0) = [\varrho \bullet \sigma_0 \bullet \varrho](0) = [\varrho \circ \sigma_0 \circ \varrho](0) = 0$. Thus, $\sigma_0$ is the reflection in $0$.

- $\sigma_0$ equals $-\mathbf{1}$.

  For this, let $\varrho_x$ be the right-shift by $x$. Then we see as above that $\sigma_0 \circ \varrho_x \circ \sigma_0 \circ \varrho_x$ maps $-x$ via $0$, $0$ and $x$ to $\sigma_0(x)$. This shows that $-x$ is snaking at least for $|x| \neq b$. As above, we have $\sigma_0 \bullet \varrho_x \bullet \sigma_0 \bullet \varrho_x = \mathbf{1}$. This shows $\sigma_0 = -\mathbf{1}$ on $(-b, b)$, hence the claim by continuity.

- $\sigma_\mathbf{x} := \varrho_\mathbf{x} \bullet \sigma_0 \bullet \varrho_{-\mathbf{x}}$ has fixed point $\mathbf{x}$ for $|\mathbf{x}| < b$.

  This is a direct consequence of Proposition 6.17.

- $\sigma_\mathbf{x} = \sigma_\mathbf{y} \iff \mathbf{x} - \mathbf{y} \in \pi\mathbb{Z}$.

  For the implication assume $\mathbf{x} < \mathbf{y}$. Then $\mathbf{x}$ is a left fixed point and $\mathbf{y}$ is a right fixed point. In particular, $\varrho := \sigma_\mathbf{x} \bullet \sigma_0 \in \mathbf{P}_{-b}$ with $\varrho(-b) = \sigma_\mathbf{x}(b) = 2\mathbf{x} - b$. This means that $\varrho$ is the shift by $2\mathbf{x} \bmod 2\pi$. Similarly, $\varrho(b) = 2\mathbf{y} + b$, hence $\varrho$ is the shift by $2\mathbf{y} \bmod 2\pi$.

  For the reversed implication again assume $\mathbf{x} < \mathbf{y}$. Then $\mathbf{x} < 0$ as otherwise $\mathbf{y} - \mathbf{x} < b < \pi$. Hence, $\varrho_{2\mathbf{x}} \bullet \sigma_0$ is well defined and fixes $\mathbf{x}$; hence, it equals $\sigma_\mathbf{x}$. Similarly, $\sigma_\mathbf{y} = \varrho_{2\mathbf{y}} \bullet \sigma_0$. By assumption, $\varrho_{2\mathbf{x}} = \varrho_{2\mathbf{y}}$ giving the claim from Proposition 11.21.

- $\sigma_{\mathbf{x}}(t) = 2\mathbf{x} - t$ for $t \in [2\mathbf{x} - b, b]$ and $\mathbf{x} \geq 0$

  First observe that $\varrho_{\mathbf{x}} \circ \sigma_0 \circ \varrho_{-\mathbf{x}}$ is defined on $I \cap (I + \mathbf{x}) \cap (I + 2\mathbf{x})$. For $\mathbf{x} > 0$, this comprises the interval $J$ from $2\mathbf{x} - b$ to $b$. As $J$ is nontrivial and as $[\varrho_{\mathbf{x}} \circ \sigma_0 \circ \varrho_{-\mathbf{x}}](t) = -t + 2\mathbf{x}$ thereon, we get the claim.

- $\sigma_{\mathbf{x}}(t) = 2\mathbf{x} - t$ for $t \in [-b, 2\mathbf{x} + b]$ and $\mathbf{x} \leq 0$

  As above.

It is now clear that $\sigma_{\mathbf{x}}$ is the restriction to $I$ of the reflection on the diameter through $\mathbf{x}$. In particular, we have shown that with $\mathbf{x}$ also $\mathbf{x} \pm \pi$ is a fixed point of $\sigma_{\mathbf{x}}$ whenever $\mathbf{x} \pm \pi$ is in the interior of the interval $[-b, b]$ again.

## 11.9 Classification Result – General

Altogether, we have

**Theorem 11.25  Classification of Full Motion Sets**

Let $\mathbf{P}$ be a full motion set.
Then there is a nontrivial interval $I \subseteq S^1$ such that $\mathbf{P}$ is isomorphic to
- the restriction of $SO(2)$ to $I$, if $\mathbf{P}$ is positive;
- the restriction of $O(2)$ to $I$, if $\mathbf{P}$ is not positive.

More concretely, this isomorphy is given iff the length of the interval $I$ is

$$\text{smaller than } \pi, \qquad \text{if } \mathbf{P}_a^+ \cap \mathbf{P}_b^+ \text{ is trivial};$$

$$2\pi \frac{\mathbf{x}(\varrho)}{\mathbf{x}(\varrho) + \mathbf{u}(\varrho)}, \qquad \text{if } \mathbf{P}_a^+ \cap \mathbf{P}_b^+ \text{ is trivial}.$$

# 12 Conclusions

In this paper we have derived an explicit classification of the symmetries of analytic paths, provided the acting group is analytic and pointwise proper. We have seen that each path is either a Lie path, i.e., part of an integral curve of a fundamental vector field, or a brick path, i.e., part of a concatenation of translates of a free segment. Such a free segment is either preserved by an element of the symmetry group or mapped to another path sharing at most finitely many points with the original one. In contrast to the limitations in [6], this classification applies not only to the action of the translations on $\mathbb{R}^3$, but also to that of rotations around the origin and even to the full Euclidean group. In fact, one easily checks that both actions are pointwise proper. This will very much ease the investigation of symmetric distributional connections in the spherically symmetric situation (acting group $SO(3)$ or its cover $SU(2)$) and the homogeneous isotropic scenario (full Euclidean group or full connected component acting).

Our classification might even open the road to group actions that are not pointwise proper. Indeed, as mentioned in the beginning, the dilations acting on $\mathbb{R}^n$ do not fulfill that requirement. Nevertheless, we can decompose $\mathbb{R}^n$ into two invariant parts, namely the origin and the punctured $\mathbb{R}^n$, where the action *is* pointwise proper. Hence, any path that does not touch both parts is either a brick or a Lie path. The only paths not covered by this statement are those running through the origin. To classify them by hand should be feasible. Nevertheless, a general classification appears very difficult as the structure of non-proper actions is much more involved.

The classification theorem for paths provides us with a separation between Lie and brick paths. However, such a separation is just an intermediate step in the investigation of homomorphisms on the path groupoid. One very important question is to distinguish within the respective realms of Lie paths and of brick paths. In particular, can we classify the free segments arising? Can we classify the Lie paths arising? The latter problem seem more feasible. Indeed, assume that we are given two paths $\gamma_i(t) := e^{tA_i}x_i$ with $A_i$ in the Lie algebra $\mathfrak{g}$ of the symmetry group and $x_i \in M$. When do translates of $\gamma_1$ and $\gamma_2$ have nontrivial overlap? Well, $\varphi_g \circ \gamma_1 = \gamma_2 \circ \varrho$ on some dom $\varrho$ leads to

$$e^{(t-t_0)\mathrm{Ad}_g A_1} z \;\; = \;\; e^{(\varrho(t)-\varrho(t_0))A_2} z \qquad \text{with } z = e^{\varrho(t_0)A_2} x_2 = g e^{t_0 A_1} x_1$$

for some $t_0 \in \text{dom } \varrho$. Moreover, we should also consider the option that a subpath of $\gamma_1$ equals the inverse of a subpath of $\gamma_2$. Starting there with some fixed $z$, we should now merge the equation above as that for the inverse into an equivalence relation on $\mathfrak{g}$ and then describe these classes. For the case of certain subgroups of the cover of the Euclidean group in $\mathbb{R}^3$, this has been done by Hanusch [6] in a lengthy case-by-case analysis. It would be very helpful to have some more general statements in this respect.

Brick paths, at least, can be classified w.r.t. the invariance groups $G_\gamma$. Hanusch [6] has already started into that direction. He distinguished between so-called "symmetric" and "non-symmetric" paths. The former ones have non-trivial stabilizer $G_\gamma$. He also used that the stabilizer of $\gamma$ coincides with any of its nontrivial subpaths. This is very relevant for the definition of homomorphisms. The next step now should be the classification of the occurring stabilizers. Of course, a classification seems reasonable only up to conjugation, similarly to the usual definition of orbit types in transformation group theory. Observe, however, that there are usually less "path types" than orbit type. For instance, the rotation group acting on $\mathbb{R}^n$ keeps the origin invariant, but no analytic path. In other words, the orbit type $[O(n)]$ is not a path type. In general, we expect this to be a typical feature.

## Acknowledgments

The author thanks Maximilian Hanusch for numerous discussions. The author is also grateful to Matthias Schmidt for useful information on initial manifolds.